\documentclass[3p,times]{elsarticle}

\usepackage{amssymb}
\usepackage{amsmath}
\usepackage{soul,xcolor}
\usepackage{caption}
\usepackage{subcaption}
\usepackage{graphicx}
\usepackage{tikz}
\usepackage{tkz-euclide}
\usepackage{tikz-dimline}
\usepackage{pgfplots}
\usetikzlibrary{patterns, pgfplots.fillbetween, bending, shapes}
\usepackage{color}
\usepackage{hyperref}
\usepackage[ruled,vlined]{algorithm2e}
\usepackage{comment}
\usepackage{listings}
\usepackage[dvipsnames]{xcolor}

\pgfdeclareverticalshading{distance}{100pt}{
  rgb(0pt)=(1,0.8,0.8); 
  rgb(5pt)=(1,1,1); 
  rgb(50pt)=(0.2,0.2,1) 
}

\definecolor{codebg}{rgb}{0.97, 0.97, 0.97}

\lstdefinestyle{kratos}{
  language=Python,
  basicstyle=\ttfamily\scriptsize,       
  backgroundcolor=\color{codebg},        
  keywordstyle=\color{blue}\bfseries,    
  commentstyle=\color{gray},             
  stringstyle=\color{red},               
  numbers=left,                          
  numberstyle=\tiny\color{gray},         
  stepnumber=1,
  numbersep=5pt,
  showstringspaces=false,
  breaklines=true,
  frame=single,
  captionpos=b,
  tabsize=4,
  morekeywords={Parameters, Model}
}

\usetikzlibrary{shapes.misc}
\tikzset{cross/.style={cross out, draw=black, minimum size=2*(#1-\pgflinewidth), inner sep=0pt, outer sep=0pt}, cross/.default={1pt}}

\newdefinition{definition}{Definition}
\newdefinition{rem}{Remark}

\newcommand{\ignore}[1]{}

\journal{Engineering with Computers}

\begin{document}

\begin{frontmatter}



\title{A Non-intrusive Approach for the Imposition of Strong Dirichlet Boundary Conditions in Unfitted Boundary Meshes}

\author[tum]{Juan Ignacio Camarotti}
\ead{juan.camarotti@tum.de}
\author[tum]{Ricky Aristio}
\ead{ricky.aristio@tum.de}
\author[deca,cimne]{Riccardo Rossi}
\ead{rrossi@cimne.upc.edu}
\author[deca,cimne]{Rub\'en Zorrilla}
\ead{rzorrilla@cimne.upc.edu}
\author[tum]{Roland Wüchner}
\ead{wuechner@tum.de}

\address[tum]{Chair of Structural Analysis, Technical University of Munich, Arcisstr. 21, 80333 München, Germany}
\address[deca]{Departament d'Enginyeria Civil i Ambiental, Universitat Polit\`{e}cnica de Catalunya, Barcelona, 08034, Spain}
\address[cimne]{International Center for Numerical Methods in Engineering (CIMNE), Barcelona, 08034, Spain}
\cortext[ca]{Corresponding author}

\begin{abstract}
The enforcement of essential boundary conditions is a fundamental challenge in unfitted boundary methods. This paper presents a non-intrusive, black-box strategy for imposing such conditions in unfitted meshes. The approach is intended for situations where the user does not have access to the solver's source code or its mathematical formulation, which is often the case when using commercial software. The proposed algorithm allows solvers originally designed for body-fitted meshes to be used in unfitted cases, provided that four conditions are satisfied: (i) the solver must support user customization by means of scripting, (ii) allow the imposition of Dirichlet boundary conditions at the node level through scripting, (iii) permit the deactivation of elements outside the physical domain, and (iv) provide access to the solution gradient within active elements. The last condition can also be satisfied by externally
reconstructing the gradient from nodal values and connectivity information, provided the element formulation is
known, making it optional in practice. These requirements are very fair demands and are satisfied by the vast majority of production-ready, possibly commercial, codes. In the current work, we show the application of this non-intrusive algorithm in the context of the Finite Element Method (FEM) and Isogeometric Analysis (IGA) discretizations, demonstrating optimal $L^2$-norm error convergence. This is demonstrated using the Kratos Multiphysics code (release \texttt{v10.1}) \emph{from the user API, simply leveraging the capabilities mentioned above.}
\end{abstract}




\end{frontmatter}


\section{Introduction}
\label{sec:introduction}

\begin{figure}[h]
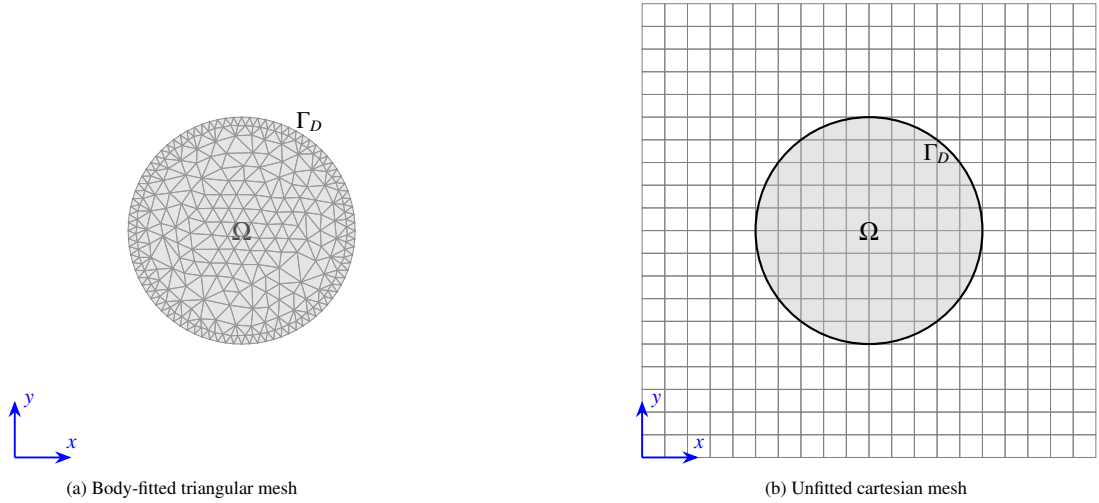

    \centering
    \begin{subfigure}{0.45\textwidth}
        \centering

        \caption{Body-fitted triangular mesh}
        \label{fig:body_fitted}
    \end{subfigure}
    \hfill
    \begin{subfigure}{0.45\textwidth}
        \centering
        %
        \caption{Unfitted cartesian mesh}
        \label{fig:unfitted}
    \end{subfigure}
    \caption{Comparison between body-fitted and unfitted meshes for a circular domain. (a) The body-fitted case, where the mesh conforms to the circular boundary using triangular elements. (b) The unfitted case, where the circular domain is embedded in a structured background mesh, allowing for simpler meshing but requiring special techniques for boundary condition enforcement.}
    \label{fig:mesh_comparison}
\end{figure}

Unfitted or embedded boundary methods have gained significant importance in computational mechanics, offering an efficient alternative to body-fitted (Fig.~\ref{fig:body_fitted}) approaches for handling complex geometries as well as for arbitrarily large boundary motions. These methods embed the physical boundary within a structured background mesh (Fig.~\ref{fig:unfitted}), removing the need for conformal meshing. This characteristic is particularly advantageous in scenarios involving evolving geometries such as crack propagation \cite{Fries2003XFEM, Yazid2009, Nagaraja2019}, shape optimization \cite{Burman2018, Edke2010, MESMER2024}, fluid-structure interaction \cite{Zorrilla2020, Schott2019, BOILEVINKAYL2019744}, additive manufacturing \cite{Neiva2020}, and biomedical applications \cite{WU2019112556, Xu2018, Hsu2015, Kamensky2015, Kamensky2017, Nitti2020, ZORRILLA_SOUDAH_2024}, where re-meshing introduces substantial computational costs. 

Despite their geometric flexibility, unfitted methods present significant challenges related to the integration over cut elements, the enforcement of boundary conditions along immersed interfaces, and the stability of the formulation in the presence of small-cut cells \cite{dePrenter2023}. Among these, the imposition of essential (Dirichlet) boundary conditions is particularly problematic, as the lack of alignment between mesh nodes and the physical boundary prevents direct enforcement. \textcolor{black}{Variational approaches, such as Nitsche’s method \cite{Nitsche1971, Juntunen2009}, Lagrange multipliers \cite{Babuska1973, Brezzi1974}, and penalty formulations \cite{Babuska1973Penalty, Hughes1987}}, impose these conditions weakly but often require careful tuning of penalty parameters and may lead to ill-conditioned systems. Alternatively, some methods aim for strong enforcement by modifying basis functions so they become interpolatory along the boundary, as in WEB-splines \cite{Hollig2001, Hollig2005} and i-splines \cite{Sanches2011}. While these techniques avoid modifying the weak form, they introduce significant algorithmic and analytical complexities due to the need for constructing specialized basis functions. Notably, none of the methods reviewed in \cite{dePrenter2023} support a black-box strategy for Dirichlet boundary condition enforcement, an increasingly important limitation in scenarios where users lack access to the solver internals.

Several prominent embedded boundaries methodologies have been developed, including the Finite Cell Method (FCM) \cite{Parvizian2007, Duster2008_FCM, Rank2012,Schillinger2012,Schillinger2012_FCM,Ruess2013}, the extended Finite Element Method (XFEM) \cite{foucard2015x}, the Cut Finite Element Method (CutFEM) \cite{ Zorrilla2020, Burman2015_CutFEM, MASSING2018262, Winter_2018}, Immersed Boundary Method (IBM)\cite{Nitti2020,peskin1977numerical}, Immersogeometric Analysis (IMGA) \cite{Kamensky2015}, Isogeometric B-Rep Analysis (IBRA)\cite{Breitenberger2015,Teschemacher2018,Teschemacher2022,Messmer2022, Messmer2024}, the Shifted Boundary Method (SBM) \cite{Main2018a, Main2018b, Antonelli2024, zorrilla2024sbm, COLLINS2023116301}, each addressing specific challenges in accurate integration schemes, enforcement of boundary conditions and maintaining numerical stability.

\textcolor{black}{Each of the existing methodologies approaches the challenges mentioned above differently, but a common numerical difficulty across most immersed methods is the so-called \emph{small cut cell problem}, which occurs when only a small fraction of an element lies inside the physical domain. This situation can lead to severe ill-conditioning of the system matrix, and different methods address it through different stabilization strategies.
The Finite Cell Method (FCM) \cite{Parvizian2007, Duster2008_FCM} embeds the physical domain into a larger, structured computational mesh and distinguishes physical and fictitious regions using an indicator function. It enables high-order analysis on non-boundary-fitted meshes and typically enforces boundary conditions weakly (e.g., via Nitsche’s method). Accurate integration near the boundary is achieved through advanced quadrature schemes, but the method can suffer from ill-conditioning in cells with small physical volume fractions.
The extended Finite Element Method (XFEM) enriches the standard finite element space with additional functions to capture discontinuities introduced by immersed boundaries, such as jumps or singularities. While this allows for accurate representation of complex features without mesh conformity, the method introduces extra degrees of freedom (DOFs), increasing computational cost. In both FCM and XFEM, the small cut cell problem can be effectively alleviated using the \emph{Eigenvalue Stabilisation Technique} pioneered by Löhnert \cite{Loehnert2014EigenStabXFEM}, which regularizes ill-conditioned element stiffness matrices via eigenmode filtering \cite{Garhuom2022EigenvalueFCM, EISENTRAGER2024129}.
The family of Immersed Boundary Methods (IBM), originally developed for fluid dynamics \cite{peskin1977numerical, Nitti2020}, represents boundaries within a fixed Cartesian grid and applies boundary conditions via additional body forces in the governing equations. While this method simplifies mesh generation and allows for large deformations, it often suffers from mass conservation issues and difficulties in accurately enforcing boundary conditions at the interface.
IBRA, on the other hand, employs the CAD model’s boundary representation (B-Rep) along with isogeometric basis functions to approximate solution fields, reducing pre-processing time but suffering from computational challenges associated with integrating trimmed elements and conditioning issues due to small cut cells.
The Shifted Boundary Method (SBM) addresses boundary enforcement by shifting the physical boundary onto a surrogate location while modifying the imposed boundary conditions, effectively bypassing small cut-cell instabilities but introducing other challenges, such as the non-trivial modification of boundary conditions and a geometrically non-obvious treatment of the surrogate boundary \cite{dePrenter2023}.
CutFEM classifies elements as either fully inside the physical domain or intersected by the boundary. For cut elements, it employs sub-cell integration to accurately compute contributions over the physical region. Boundary conditions are enforced variationally, often using Nitsche’s method, and the method incorporates \emph{ghost penalty stabilization} \cite{Burman2010} to address issues related to small cut elements and ensure numerical stability.}

Despite this rich landscape of methods, explicit (strong-like) imposition of Dirichlet boundary conditions in unfitted meshes, without modifying the weak form or the basis functions used for analysis, has received limited attention, particularly in black-box settings where the solver’s internals are inaccessible. Motivated by this gap, we propose a physics-agnostic, non-intrusive iterative approach that strongly enforces Dirichlet boundary conditions in unfitted meshes without requiring access to the solver’s formulation or source code. This is especially relevant in modern computational practice, where black-box solvers such as commercial software packages are widely used and cannot be altered by the end user.

\ignore{Despite these advancements, a method that enforces explicitly Dirichlet BCs in unfitted meshes has received little attention. We propose a physics-agnostic, non-intrusive iterative approach that exlicitly imposes Dirichlet BCs without the need for penalty parameters or weak imposition techniques. Our method reformulates boundary condition enforcement as an $L^2$-norm error minimization problem, augmented by a stabilization term that approximates the normal gradient within trimmed elements. Notably, the stabilization mechanism is computed over the entire intersected elements, which implies that these elements no longer participate in the resulting linear system derived from the weak form of the governing equation (note that this idea is conceptually similar to the treatment used in the SBM). This approach mitigates ill-conditioning and eliminates some of the complexities required by variational methods, such as the need for fine tuning or weak form modification.}

The proposed method iteratively refines the BC imposition by minimizing the $L^2$-norm error between the numerical and prescribed boundary values. The stabilization mechanism ensures smooth normal gradient transitions within intersected and fully active elements. This is achieved through interpolation techniques such as Radial Basis Functions (RBF) and Moving Least Squares (MLS), which provide a robust approximation of gradient fields.

Our approach presents several potential benefits over existing unfitted methods.  Its physics-agnostic and non-intrusive nature ensures broad applicability across various engineering fields, seamlessly integrating with existing black-box body-fitted solvers without requiring fundamental modifications. Furthermore, it eliminates the reliance on penalty parameters, thereby removing the need for empirical tuning, and enhances system conditioning, mitigating numerical instabilities associated with small cut cells. Nevertheless, some challenges persist, including potential less accuracy in the solution field due to the gradient approximation (especially in high-order geometries), increased computational cost due to the iterative enforcement process, and a dependency on interpolation schemes, which must be carefully chosen to achieve an optimal balance between accuracy and stability.

The remainder of the paper is structured as follows: 
In Section~\ref{sec:poisson_problem}, we introduce the governing equations and boundary conditions of the Poisson problem, which serves as the PDE for validating the method, and review its weak formulation.
In Section \ref{sec:strong_DC_BCs_L2_Norm}, we present the concept of enforcing strong Dirichlet boundary conditions as an $L^2$-error minimization problem. Furthermore, we show that applying this approach to unfitted meshes results in a singular minimization problem. 
In Section \ref{sec:stab_L2_functional_unfitted}, we show that this singularity issue can be mitigated by incorporating an additional stabilization term into the original functional, which is based on the solution normal gradient within the trimmed elements. In addition, we outline the resulting solution algorithm and describe the classification of elements within this framework.
Different interpolation techniques for approximating the normal gradient within trimmed elements, such as Radial Basis Functions (RBF) and Moving Least Squares (MLS), are proposed in Section \ref{sec:normal_gradient_approximation}.
In Section~\ref{sec:numerical_solution_poisson_problem}, we assess the validity of the proposed approach by numerically solving the Poisson problem using FEM and IGA discretizations.
In addition, we demonstrate how the proposed method can be integrated into an existing black-box solver by implementing it in the Kratos Multiphysics \cite{dadvand2010, dadvand2013} framework (release \texttt{v10.1}) without modifying the underlying solver algorithms. This practical implementation confirms that the method can be deployed non-intrusively on top of production-ready, body-fitted solvers.
Finally, in Section \ref{sec:conclusions}, we provide concluding remarks and discuss possible future applications of this method.

\section{Governing equations of Poisson's problem}
\label{sec:poisson_problem}

In this paper, we consider the numerical solution of the Poisson equation. 
Let $\Omega \subset \mathbb{R}^{n}$ be a bounded computational domain, where $n=2,3$ represents the spatial dimension. The domain boundary, $\partial \Omega = \Gamma$, consists of two disjoint subsets: the Dirichlet boundary $\Gamma_D$ and the Neumann boundary $\Gamma_N$, satisfying $\Gamma_D \cup \Gamma_N = \Gamma$ and $\Gamma_D \cap \Gamma_N = \emptyset$. The strong form of the Poisson equation is given by:
\begin{subequations} 
    \label{eq:strong_form} 
    \begin{align}
        -\nabla^2\phi &= \sigma \quad \quad \quad \quad \; \text{in } \Omega, \label{eq:poisson} \\
        \phi &= \phi_D \;\; \;  \quad \quad \;\;\; \text{on } \Gamma_D, \label{eq:dirichlet} \\
        \nabla \phi \cdot \mathbf{n} &= \frac{\partial \phi}{\partial \mathbf{n}}=t_n \quad \:\;\; \text{on } \Gamma_N. \label{eq:neumann}
    \end{align}
\end{subequations}
Here, $\phi$ is the primary variable, $\phi_D$ represents the prescribed value on the Dirichlet boundary $\Gamma_D$, $t_N$ denotes the imposed normal gradient on the Neumann boundary $\Gamma_N$, and $\sigma$ acts as the forcing or source term.
When the forcing term vanishes, i.e., $\sigma = 0$ throughout the entire domain $\Omega$, the equation simplifies to the Laplace equation:
\begin{equation}  
    \label{eq:laplace_equation}  
    \nabla^2 \phi = 0  \quad \text{in } \Omega.
\end{equation}
Finally, in Eq.~\ref{eq:neumann}, $\mathbf{n}$ denotes the outward-pointing normal vector to the boundary $\Gamma$.

\subsection{Weak form of the Poisson's problem}

The Sobolev space \( H^1(\Omega) \) is a fundamental function space in the weak formulation of partial differential equations. To define it, we first introduce the space of square-integrable functions, denoted by \( L^2(\Omega) \), which is defined as

\begin{equation}
    L^2(\Omega) := \left\{ v: \Omega \to \mathbb{R} \mid \int_{\Omega} |v|^2 \, d\Omega < \infty \right\}
\end{equation}

\noindent
This means that a function \( v \) belongs to \( L^2(\Omega) \) if its square is integrable over the domain \( \Omega \). 

Building on this, the Sobolev space \( H^1(\Omega) \) extends \( L^2(\Omega) \) by also requiring the first-order weak derivatives of \( v \) to be in \( L^2(\Omega) \). Formally, it is defined as

\begin{equation}
    H^1(\Omega) := \left\{ v \in L^2(\Omega) \mid \nabla v \in L^2(\Omega) \right\}
\end{equation}

\noindent 
where \( \nabla v \) represents the weak gradient of \( v \). This ensures that functions in \( H^1(\Omega) \) not only belong to \( L^2(\Omega) \) but also have weak derivatives that are square-integrable, providing a certain degree of smoothness.

\ignore{For problems involving Dirichlet boundary conditions,} We introduce a special subspace of \( H^1(\Omega) \), known as \( H^1_0(\Omega) \). This space is defined as

\begin{equation}
    H^1_0(\Omega) := \left\{ v \in H^1(\Omega) \mid v = 0 \text{ on } \Gamma_D \right\}
\end{equation}

\noindent
This definition implies that \( H^1_0(\Omega) \) consists of functions in \( H^1(\Omega) \) that vanish on the Dirichlet boundary \( \Gamma_D \). It serves as the natural function space for test functions in the weak formulation when Dirichlet boundary conditions are imposed.

Introducing the notation
\begin{equation}
\label{eq:bilinear_form_omega_gamma}
    (\alpha, \beta)_\Omega = \int_{\Omega} \alpha \beta \, d\Omega \quad, \quad 
    \langle \alpha, \beta\rangle_\Gamma = \int_{\Gamma} \alpha \beta \, d\Gamma
\end{equation}
and applying Green's theorem allows to write the weak formulation for the Poisson problem as: find \( \phi \in H^1(\Omega) \) such that for all \( w \in H^1_0(\Omega) \),
\begin{equation}
    \label{eq:final_weak_form}
    (\nabla \phi, \nabla w)_\Omega = (\sigma, w)_\Omega + \langle t_N, w\rangle_{\Gamma_N}
\end{equation}

\section{Imposition of strong-like Dirichlet BCs as an $L^2$-norm error minimization problem}
\label{sec:strong_DC_BCs_L2_Norm}

\ignore{
Again, we consider a general Partial Differential Equation defined on a domain $\Omega \subset \mathbb{R}^n$ with boundary $\partial \Omega = \Gamma_D \cup \Gamma_N$, where $\Gamma_D$ and $\Gamma_N$ are the Dirichlet and Neumann boundaries, respectively
\begin{subequations} 
    \label{eq:strong_form_arb} 
    \begin{align}
        \mathcal{L} (\phi) &= f \quad \text{in } \Omega, \label{eq:diff_oper} \\
        \phi &= \phi_D \quad \text{on } \Gamma_D, \label{eq:dirichlet_diff_op} \\
        \nabla \phi \cdot \mathbf{n} &= \frac{\partial \phi}{\partial \mathbf{n}}=t_n \quad \text{on } \Gamma_N. \label{eq:neumann_diff_op}
    \end{align}
\end{subequations}
}

For this section, we refer to the Poisson problem introduced in Section~\ref{sec:poisson_problem}, defined on a domain $\Omega \subset \mathbb{R}^n$ with boundary $\partial \Omega = \Gamma_D \cup \Gamma_N$, where $\Gamma_D$ and $\Gamma_N$ denote the Dirichlet and Neumann boundaries, respectively.

The Dirichlet boundary condition $\phi = \phi_D$ on $\Gamma_D$ can be interpreted as the minimization of the $L^2$-norm error between the field $\phi$ and the imposed value $\phi_D$. The corresponding functional representing the error is defined as
\begin{equation}
    \label{eq:dirichlet_L2_norm}
    \psi(\phi) = \frac{1}{2} \int_{\Gamma} \left( \phi(x) - \phi_D(x) \right)^2 \, d\Gamma
\end{equation}
This equation defines a least-squares fit problem and constitutes a projection of the Dirichlet boundary condition onto the discretized function space. It serves as the foundation for the auxiliary algebraic system used to impose the boundary conditions in our method.

\noindent In a FEM-like approach, the solution is approximated by
\begin{equation}
    \label{eq:FEM_approx}
    \phi_h(x) = \sum_{j=1}^{n} N_j(x) \phi_j = \mathbf{N}(\mathbf{x}) \boldsymbol{\phi}
\end{equation}
where $N_j(x)$ are the finite element shape functions, $\phi_j$ are the corresponding nodal values and $n$ represents the number of nodes (or control points in IGA). 

Substituting the approximation \ref{eq:FEM_approx} into the error functional \ref{eq:dirichlet_L2_norm} yields
\begin{equation}
    \psi(\boldsymbol{\phi}) = \frac{1}{2} \int_{\Gamma_D} \left( \sum_{j=1}^{n_{\Gamma_D}} N_j(x) \phi_j - \phi_D(x) \right)^2 \, d\Gamma.
\end{equation}
\textcolor{black}{
\begin{rem}
    In Eq.~\ref{eq:FEM_approx}, the index $n$ refers to the total number of nodes (or control points) in the discretized domain. However, only those nodes whose associated shape functions satisfy $N_j(x) \neq 0$ on the Dirichlet boundary $\Gamma_D$ actually contribute to the integral in Eq.~\ref{eq:dirichlet_L2_norm}. As a result, the effective linear system involves only a subset of these nodes. We denote this reduced number by $n_{\Gamma_D}$.
\end{rem}
}

To determine the degrees of freedom $\phi_i$, we take the derivative of $\psi(\boldsymbol{\phi})$ with respect to each $\phi_i$ and set it equal to zero:
\begin{equation}
    \label{eq:variation}
    \frac{\partial \psi}{\partial \phi_i} = \int_{\Gamma_D} \left( \sum_{j=1}^{n_{\Gamma_D}} N_j(x) \phi_j - \phi_D(x) \right) N_i(x) \, d\Gamma = 0\quad i=1,\ldots,n_{\Gamma_D}.
\end{equation}

\noindent
This set of equations leads to the linear system
\begin{equation}
    \label{eq:linear_system}
    \sum_{j=1}^{n_{\Gamma_D}} \left( \int_{\Gamma_D} N_i(x) N_j(x) \, d\Gamma \right) \phi_j = \int_{\Gamma_D} \phi_D(x) N_i(x) \, d\Gamma,\quad i=1,\ldots,n_{\Gamma_D}.
\end{equation}

\noindent
In matrix form, the linear system can be written as
\begin{equation}
    \mathbf{M}_{\Gamma_D}\boldsymbol{\phi} = \mathbf{b},
    \label{eq:original_linear_system}
\end{equation}
where the mass matrix $\mathbf{M}_{\Gamma_D}$ and the right-hand side vector $\mathbf{b}$ have entries defined by
\begin{align}
    \left[\mathbf{M}_{\Gamma_D}\right]_{ij} &= \int_{\Gamma_D} N_i(x) N_j(x) \, d\Gamma, \\
    b_i &= \int_{\Gamma_D} \phi_D(x) N_i(x) \, d\Gamma.
\end{align}

This system is solved independently of the weak form of the original PDE and is used solely to compute boundary values for subsequent enforcement.

In the following part we will demonstrate that, for unfitted discretizations, the resulting linear system is ill-posed, leading to infinite solutions.

\subsection{Singularity problem for unfitted discretizations}

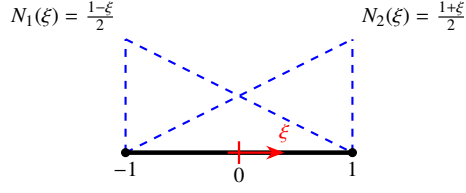
\begin{figure}[h]
    \centering
    \begin{tikzpicture}[scale=1.5]
        \draw[dashed, thick, blue] (-1,1) -- (1,0);
        \draw[dashed, thick, blue] (-1,0) -- (1,1);
        \draw[dashed, thick, blue] (-1,0) -- (-1,1);
        \draw[dashed, thick, blue] (1,0) -- (1,1);

       \draw[-, line width=0.60mm, color=black] (-1,0) -- (1,0);
        \draw[-{Stealth[red,flex]}, line width=0.30mm, color=red] (-0.1,0) -- (0.4,0) node[above] {\footnotesize $\xi$};
        \draw[-, line width=0.30mm, color=red] (0,-0.1) -- (0,0.1);

        \filldraw[black] (-1,0) circle (1pt);
        \filldraw[black] (1,0) circle (1pt);

        \node[below] at (-1,0) {\footnotesize $-1$};
        \node[below] at (1,0) {\footnotesize $1$};
        \node[below] at (0,-0.05) {\footnotesize $0$};

        \node[black, above left] at (-1,1) {\footnotesize $N_1(\xi) = \frac{1-\xi}{2}$};
        \node[black, above right] at (1,1) {\footnotesize $N_2(\xi) = \frac{1+\xi}{2}$};

    \end{tikzpicture}
    \caption{2-node linear element in natural coordinates space $[-1,1]$ with corresponding shape functions.}
    \label{fig:fem_linear_element}
\end{figure}

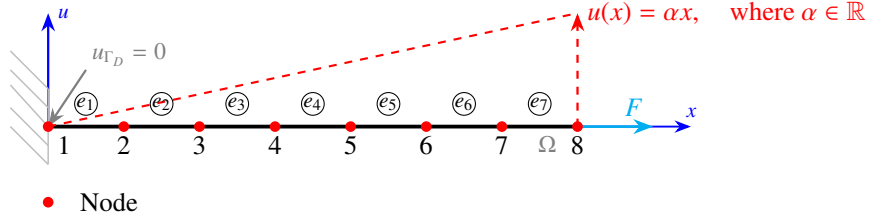
\begin{figure}[h!]
    \centering
    \begin{tikzpicture}[scale=1.0]
        \centering
        
        \draw[-{Stealth[blue,flex]}, line width = 0.25mm, color=blue] (0,0.0)--(8.5,0.0) node[above]{\footnotesize$x$};
        \draw[-{Stealth[blue,flex]}, line width = 0.30mm, color=blue] (0.0,0.0)--(0.0,1.5) node[right]{\footnotesize$u$};
        \draw[-{Stealth[red,flex]}, line width = 0.30mm, color=red, dashed]  (7,0) -- (7,1.5) node[anchor=west] {$u(x)=\alpha x, \quad \text{where } \alpha \in \mathbb{R}$};  
        \draw[-{Stealth[gray,flex]}, line width=0.30mm, color=gray]  
        (0.5,0.75) -- (0,0) node[pos=0.1, anchor=south west] {\small $u_{\Gamma_D}=0$};
        
        \draw[ line width = 0.50mm, color=black] (0,0) -- (7,0);
        
        \draw[-{Stealth[cyan,flex]}, line width = 0.35mm, color=cyan]  (7,0) -- (8.0,0) node[anchor=south east] {$F$};
        
        \draw[thick,red, dashed] (0,0) -- (7.0,1.5);
        
        \draw[ line width = 0.20mm, color=lightgray] (0,-0.5) -- (0,0.5);
         \draw[ line width = 0.20mm, color=lightgray] (0,0.0) -- (-0.5,0.55);
         \draw[ line width = 0.20mm, color=lightgray] (0,-0.5) -- (-0.5,0);
         \draw[ line width = 0.20mm, color=lightgray] (0,0.5) -- (-0.5,1);
         \draw[ line width = 0.20mm, color=lightgray] (0,-0.25) -- (-0.5,0.25);
         \draw[ line width = 0.20mm, color=lightgray] (0,0.25) -- (-0.5,0.75);
        
        \foreach \x in {1,2,...,8} {
            \fill[red] (\x-1,0) circle (2pt);
        }
        
        \foreach \i in {1,...,7} {
            \pgfmathsetmacro{\xval}{\i -0.5}
            \draw (\xval,0.3) circle (0.15) node {\footnotesize $e_{\i}$};
        }
        
        \node[anchor=north west] at (0,0) {1};
        \node[anchor=north] at (1,0) {2};
        \node[anchor=north] at (2,0) {3};
        \node[anchor=north] at (3,0) {4};
        \node[anchor=north] at (4,0) {5};
        \node[anchor=north] at (5,0) {6};
        \node[anchor=north] at (6,0) {7};
        \node[anchor=north] at (7,0) {8};
        
        \fill[red] (0,-1) circle (2pt);
        \node[anchor=west] at (0.3,-1) {Node};
        \node[anchor=north, color=gray] at (6.6,0.0) {\small $\Omega$};
    \end{tikzpicture}
    \caption{Body-fitted discretization of a bar structure subjected to an external force $F$ at its free end. 
    The bar is fixed at the left support and undergoes a displacement $u(x)$ due to the applied load. 
    The discretization is achieved using a structured mesh, where boundary nodal points (marked in red) conform to the geometry of the bar. 
    The horizontal axis represents the spatial coordinate $x$, while the vertical axis shows the displacement field $u(x)$ for illustrative purposes.}
    \label{fig:bar_structure_body_fitted}
\end{figure}

\begin{figure}[h!]
    \centering
    \begin{tikzpicture}[scale=1.0]
        \centering
        
        \draw[-{Stealth[blue,flex]}, line width = 0.25mm, color=blue] (-0.5357,0.0)--(8.5,0.0) node[above]{\footnotesize$x$};
        \draw[-{Stealth[blue,flex]}, line width = 0.30mm, color=blue] (0.0,0.0)--(0.0,1.5) node[right]{\footnotesize$u$};
        \draw[-{Stealth[red,flex]}, line width = 0.30mm, color=red, dashed]  (7,0) -- (7,1.5) node[anchor=west] {$u(x) = \alpha x + \beta, \quad \text{where } \alpha, \beta \in \mathbb{R}$;
};
        \draw[-{Stealth[black,flex]}, line width = 0.30mm, color=black] ;
        \draw[-{Stealth[gray]}, line width=0.30mm, color=gray]  
        (1.2,1) -- (0,0) node[pos=0, anchor=south, color=gray] {\small $u_{\Gamma_D}=0$};

        \draw[ line width = 0.50mm, color=black] (0,0) -- (7,0);
        
          \draw[-{Stealth[cyan,flex]}, line width = 0.35mm, color=cyan]  (7,0) -- (8.0,0) node[anchor=south east] {$F$};
        
        \draw[thick,red, dashed] (0,0) -- (7.0,1.5);
        
        \draw[ line width = 0.20mm, color=lightgray] (0,-0.5) -- (0,0.5);
         \draw[ line width = 0.20mm, color=lightgray] (0,0.0) -- (-0.5,0.55);
         \draw[ line width = 0.20mm, color=lightgray] (0,-0.5) -- (-0.5,0);
         \draw[ line width = 0.20mm, color=lightgray] (0,0.5) -- (-0.5,1);
         \draw[ line width = 0.20mm, color=lightgray] (0,-0.25) -- (-0.5,0.25);
         \draw[ line width = 0.20mm, color=lightgray] (0,0.25) -- (-0.5,0.75);
        
        \foreach \i in {0,...,7} {
            \pgfmathsetmacro{\xval}{-0.5357 + \i * (6.9617 - (-0.5357)) / 7}
            \fill[red] (\xval,0) circle (2pt);
        }

        \foreach \i in {1,...,7} {
            \pgfmathsetmacro{\xval}{-0.5357 + (\i - 0.5) * (6.9617 - (-0.5357)) / 7}
            \draw (\xval,0.3) circle (0.15) node {\footnotesize $e_{\i}$};
        }
        
        \node[anchor=north west] at (-0.5357,0) {1};
        \node[anchor=north] at (0.5357,0) {2};
        \node[anchor=north] at (1.6067,0) {3};
        \node[anchor=north] at (2.6777,0) {4};
        \node[anchor=north] at (3.7487,0) {5};
        \node[anchor=north] at (4.8197,0) {6};
        \node[anchor=north] at (5.8907,0) {7};
        \node[anchor=north] at (6.9617,0) {8};
        
        \fill[red] (0,-1) circle (2pt);
        \node[anchor=west] at (0.3,-1) {Node};
        \node[anchor=north, color=gray] at (6.4,0.0) {\small $\Omega$};
    \end{tikzpicture}
    \caption{Unfitted discretization of a bar structure subjected to an external force $F$ at its free end. 
    The bar is fixed at the left support and undergoes a displacement $u(x)$ due to the applied load. 
    The discretization is achieved using a structured mesh, where boundary nodal points (marked in red) do not conform to the geometry of the bar. 
    The horizontal axis represents the spatial coordinate $x$, while the vertical axis shows the displacement field $u(x)$ for illustrative purposes.}
    \label{fig:bar_structure_unfitted}
\end{figure}
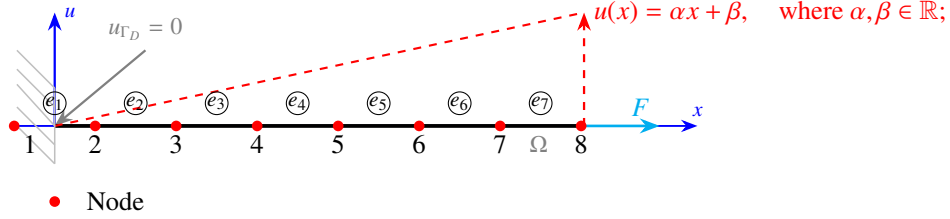

 In this section, we analyze a simplified example involving two different FEM discretizations of a 1D bar structure using linear elements (Fig.~\ref{fig:fem_linear_element}): a body-fitted discretization (Fig.~\ref{fig:bar_structure_body_fitted}) and an unfitted discretization (Fig.~\ref{fig:bar_structure_unfitted}). Our objective is to impose a Dirichlet boundary condition at the left end of the bar, denoted by $\Gamma_D$. In the body-fitted discretization, $\Gamma_D$ coincides with the left-end of the first element, whereas in the unfitted case, it is located at the midpoint of the first element.

The governing equation for the longitudinal displacement $u(x)$ is given by
\begin{equation}
    \label{eq:bar_equation}
    \frac{d}{dx}\left( EA\, \frac{du}{dx} \right) = -p(x), \quad \text{in } \Omega,
\end{equation}
subject to the boundary conditions
\begin{equation}
    \label{eq:bar_dir_cond}
    u = u_D \quad \text{on } \Gamma_D,
\end{equation}
\begin{equation}
    \label{eq:bar_neum_cond}
    EA\, \frac{du}{dx} = F \quad \text{on } \Gamma_N.
\end{equation}

\noindent
In Eq.~\ref{eq:bar_equation}, $EA$ denotes the axial stiffness of the bar, $p(x)$ represents a longitudinally distributed load, and $u_D$ corresponds to the imposed displacement. For this particular case, we assume $p(x) = 0$ and $u_D = 0$. 

\ignore{
As previously mentioned, imposing the Dirichlet boundary condition can be interpreted as minimizing the following functional, which represents the \( L^2 \)-norm error between the computed displacement \( u(x) \) and the prescribed value \( u_D(x) \)
\begin{equation}
    \label{eq:dirichlet_L2_norm}
    \phi(u) = \frac{1}{2} \int_{\Gamma_D} \left( u(x) - u_D(x) \right)^2 \, d\Gamma 
    \quad \Rightarrow \quad \mathbf{M}_{\Gamma_D} \boldsymbol{u} = \mathbf{b}
\end{equation}
Here, \( \Gamma_D \) corresponds to a single point in the bar structure.
}

We now analyze the resulting linear system used for the imposition of the Dirichlet boundary condition (Eq.~\ref{eq:original_linear_system}), considering both body-fitted and unfitted discretizations. Both in the body-fitted and unfitted cases, the system is given by
\begin{equation}
    \begin{pmatrix}
    N_1(x=0)N_1(x=0) & N_1(x=0)N_2(x=0)\\[0.5em]
    N_2(x=0)N_1(x=0) & N_2(x=0)N_2(x=0)
    \end{pmatrix}
    \begin{pmatrix}
    u_1\\[0.5em]
    u_2
    \end{pmatrix}
    =
    \begin{pmatrix}
    u_D(x=0)N_1(x=0)\\[0.5em]
    u_D(x=0)N_2(x=0)
    \end{pmatrix}
\end{equation}
Substituting the shape function values and the imposed displacement at the \(\Gamma_D\) boundary for the body-fitted case, we obtain
\begin{equation}
    \begin{pmatrix}
    1 & 0\\[0.5em]
    0 & 0
    \end{pmatrix}
    \begin{pmatrix}
    u_1\\[0.5em]
    u_2
    \end{pmatrix}
    =
    \begin{pmatrix}
    0\\[0.5em]
    0
    \end{pmatrix} \quad \Rightarrow \quad u_1 = 0, u_2 \in \mathbb{R}
\end{equation}
Thus, in the body-fitted case, the Dirichlet boundary condition is imposed strongly on the corresponding degree of freedom, enforcing \( u_1 = 0 \). The remaining unknown, \( u_2 \), is not constrained by the boundary condition but will be determined by solving the system of equations arising from the weak form of the governing equation.

Conversely, in the unfitted case, substituting the shape function values and the imposed displacement at the \(\Gamma_D\) boundary leads to the following linear system
\begin{equation}
    \begin{pmatrix}
    \big(\frac{1}{2}\big)^2 & \big(\frac{1}{2}\big)^2\\[0.5em]
    \big(\frac{1}{2}\big)^2 & \big(\frac{1}{2}\big)^2
    \end{pmatrix}
    \begin{pmatrix}
    u_1\\[0.5em]
    u_2
    \end{pmatrix}
    =
    \begin{pmatrix}
    0\\[0.5em]
    0
    \end{pmatrix} \quad \Rightarrow \quad \text{Infinite solutions}.
\end{equation}
Since this system of equations is underdetermined, it admits infinitely many solutions. In other words, attempting to project the Dirichlet boundary condition onto the discretization space of an unfitted boundary leads to an ill-posed problem.

\subsection{Physical interpretation of the singularity problem for unfitted discretizations}

Having presented the body-fitted and unfitted bar structure examples, we now turn our attention to the physical interpretation of this singularity problem observed in unfitted discretizations. 

\begin{figure}[h!]
    \centering
    \begin{tikzpicture}[scale=1.0]
        \centering
        
        \draw[-{Stealth[blue,flex]}, line width = 0.25mm, color=blue] (8.5,0.0)--(9.5,0.0) node[above]{\footnotesize$x$};
        \draw[-{Stealth[blue,flex]}, line width = 0.30mm, color=blue] (0.0,0.0)--(0.0,2.0) node[right]{\footnotesize$u$};
       \draw[line width=0.25mm, color=lightgray, dashed] (7,0.0) -- (8.5,0.0) 
        node[pos=1, anchor=north] {\footnotesize \textcolor{lightgray}{$\Omega$ continues $\rightarrow$}};
        \draw[-{Stealth[gray]}, line width=0.30mm, color=gray]  
        (4.0,1) -- (3.5,0) node[pos=0, anchor=south, color=gray] {\small $u_{\Gamma_D}=0$};

        \draw[ line width = 0.50mm, color=black] (0,0) -- (7,0);
        
        \draw[thick,red, dashed] (0,-0.75) -- (7.0,0.75);
        \draw[-{Stealth[red,flex]}, line width = 0.30mm, color=red, dashed]  (7,0) -- (7,0.75) node[anchor=west] {$u_{2_1}$};
        \draw[-{Stealth[red,flex]}, line width = 0.30mm, color=red, dashed]  (0,0) -- (0,-0.75) node[anchor=east] {$u_{1_1}$}; 

        \draw[thick,cyan, dashed] (0,0.75) -- (7.0,-0.75);
        \draw[-{Stealth[cyan,flex]}, line width = 0.30mm, color=cyan, dashed]  (7,0) -- (7,-0.75) node[anchor=west] {$u_{2_2}$};
        \draw[-{Stealth[cyan,flex]}, line width = 0.30mm, color=cyan, dashed]  (0,0) -- (0,0.75) node[anchor=east] {$u_{1_2}$};

        \draw[thick,green, dashed] (0,1.5) -- (7.0,-1.5);
        \draw[-{Stealth[green,flex]}, line width = 0.30mm, color=green, dashed]  (7,0) -- (7,-1.5) node[anchor=west] {$u_{2_3}$};
        \draw[-{Stealth[green,flex]}, line width = 0.30mm, color=green, dashed]  (0,0) -- (0,1.5) node[anchor=east] {$u_{1_3}$};

        \draw[thick,brown, dashed] (0,-1.5) -- (7.0,1.5);
        \draw[-{Stealth[brown,flex]}, line width = 0.30mm, color=brown, dashed]  (7,0) -- (7,1.5) node[anchor=west] {$u_{2_4}$};
        \draw[-{Stealth[brown,flex]}, line width = 0.30mm, color=brown, dashed]  (0,0) -- (0,-1.5) node[anchor=east] {$u_{1_4}$};
        
        \draw[ line width = 0.20mm, color=lightgray] (3.5,-0.5) -- (3.5,0.5);
         \draw[ line width = 0.20mm, color=lightgray] (3.5,0.0) -- (2.5,0.55);
         \draw[ line width = 0.20mm, color=lightgray] (3.5,-0.5) -- (2.5,0);
         \draw[ line width = 0.20mm, color=lightgray] (3.5,0.5) -- (2.5,1);
         \draw[ line width = 0.20mm, color=lightgray] (3.5,-0.25) -- (2.5,0.25);
         \draw[ line width = 0.20mm, color=lightgray] (3.5,0.25) -- (2.5,0.75);
        
        \fill[red] (0,0) circle (2pt);
        \fill[red] (7,0) circle (2pt);

        \node[anchor=south] at (5.0,2.0) {\textcolor{lightgray}{\small Different $\frac{du}{dx}$}};
    
        \draw[-{Stealth[lightgray,flex]}, line width = 0.30mm, color=lightgray, dashed]  (5.0,2.2) -- (5.3,0.77);
        \draw[-{Stealth[lightgray,flex]}, line width = 0.30mm, color=lightgray, dashed]  (5.0,2.2) -- (5.8,0.5);
        \draw[-{Stealth[lightgray,flex]}, line width = 0.30mm, color=lightgray, dashed]  (5.0,2.2) -- (4.9,-0.25);

        \draw (0.5,0.2) circle (0.15) node {\footnotesize $e_{1}$};
        \draw (7.5,0.2) circle (0.15) node {\footnotesize $e_{2}$};
        
        \node[anchor=north west] at (0,0) {1};
        \node[anchor=north] at (7.0,0) {2};
        
        \fill[red] (0,-2.0) circle (2pt);
        \node[anchor=west] at (0.3,-2.0) {Node};
    \end{tikzpicture}
    \caption{Visualization of an unfitted discretization for a bar structure, illustrating how different displacement gradients (\( \frac{du}{dx} \)) emerge from various combinations of nodal displacements in the unfitted boundary that satisfy the Dirichlet boundary condition \( u_{\Gamma_D} = 0 \). The dashed lines represent possible displacement solutions \( u(x) \), and the domain \( \Omega \) continues to the right, as indicated by the gray dashed line. Nodes are marked in red, and the structured mesh elements are annotated as \( e_1 \) and \( e_2 \).}
    \label{fig:bar_structure_unfitted_diff_grad}
\end{figure}
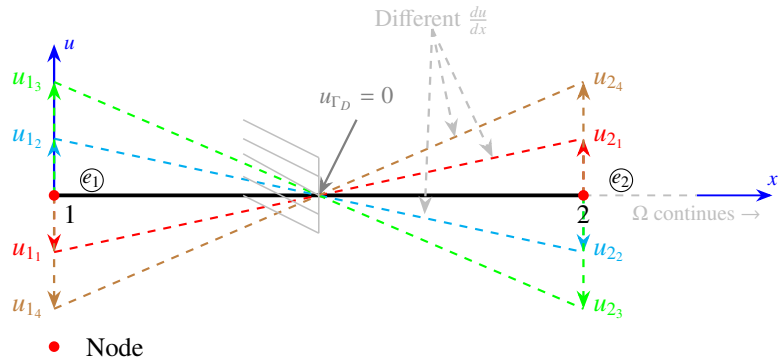

Focusing on the first element of the bar structure, we can observe from Fig.~\ref{fig:bar_structure_unfitted_diff_grad} that there are infinite possible assignments for  $u_1$ and $u_2$ that still satisfy the condition $u_{\Gamma_D} = 0$.
As observed in Fig.~\ref{fig:bar_structure_unfitted_diff_grad}, the infinite solutions differ only in the gradient of the solution field within the trimmed element. Moreover, it is crucial to note that many of these solutions, such as those with a negative gradient inside the trimmed element, do not correspond with the physical behaviour of the problem. This raises a fundamental question whether this issue could be resolved by imposing an additional condition on the gradient within the first element (the trimmed element). This concept \emph{serves as the foundation for the methodology proposed in this work.}

\section{Stabilization of the $L^2$-norm error functional for unfitted meshes}
\label{sec:stab_L2_functional_unfitted}

In this section, we present a novel approach to stabilize the $L^2$-norm error functional for unfitted meshes, enabling a strong-like, external imposition of Dirichlet boundary conditions without altering the underlying weak form of the partial differential equation. This approach results in a non-intrusive iterative method suitable for body-fitted and unfitted discretizations, effectively addressing the challenges associated with trimmed elements.

Before presenting the method, we clarify the concept of strong-like imposition of Dirichlet boundary conditions as used in this work. By strong-like, we refer to the explicit modification of both the left-hand side (LHS) and right-hand side (RHS) of the linear system arising from the weak form of the physical problem to enforce the boundary condition. The nodal values used for this enforcement are computed by solving an auxiliary algebraic system that is independent of the weak form governing the physics.

We introduce a novel functional $\psi(\phi)$ to define strong-like Dirichlet boundary conditions in unfitted meshes. This functional consists of two key components and is given by

\begin{equation}
    \psi(\phi^{k+1}) =
    \frac{1}{2} \int_{\Gamma_D} \left( \phi^{k+1}(\mathbf{x}) - \phi_D(\mathbf{x}) \right)^2 \, d\Gamma +
    \frac{1}{2} \int_{\Omega_\text{trim}} \left( \nabla(\phi^{k+1})\cdot \mathbf{n} - \widetilde{\nabla}(\phi^k) \cdot \mathbf{n} \right)^2 \, d\Omega
    \label{eq:new_functional_stabilised}
\end{equation}
In Eq.~\ref{eq:new_functional_stabilised}, $\phi^{k+1}$ represents the solution at the Dirichlet boundary at iteration $k+1$, $\Omega_{\text{trim}}$ denotes the intersected elements (or knot spans in the case of Isogeometric Analysis, $\nabla(\phi^{k+1})$ is the gradient inside an intersected element at iteration $k+1$, and $\widetilde{\nabla}(\phi_k)$ is an approximation of the gradient inside the intersected element, obtained using information computed at iteration $k$.

The second term in Eq.~\ref{eq:new_functional_stabilised} acts as a regularization mechanism that addresses the underdetermined nature of the projection problem in unfitted discretizations. This stabilization term introduces an additional constraint to the projection that guides the gradient within the trimmed region to be compatible with the gradient field in adjacent, fully-resolved elements. The term involves the normal component of the gradient, where $\widetilde{\nabla} (\phi^k)$  is an approximation computed at iteration $k$ using information from neighboring non-intersected elements.  Notably, the term is integrated over the entire trimmed element (not just the active portion) and has shown to produce accurate and stable results in practice.

\ignore{
The idea of regularizing the boundary condition imposition using the normal gradient was inspired by the Shifted Boundary Method (SBM) \cite{ATALLAH2020113341}. In the initial SBM formulations, boundary conditions were applied on a surrogate boundary using a Taylor expansion from the true boundary, which inherently involves the gradient in the normal direction. This emphasizes the importance of controlling the normal variation of the solution near the boundary. In our method, we adopt a similar principle by incorporating the normal gradient into the stabilization term.
}

\ignore{
The purpose of each term in Eq.~\ref{eq:new_functional_stabilised}
\begin{itemize}
    \item \textbf{First term:} This boundary term calculated along $\Gamma_D$ is exactly the same as in the original $L^2$-norm error formulation, enforcing the Dirichlet boundary condition
    \item \textbf{Second term (Regularization or Stabilization Term):} This term ensures the consistency of the normal gradient of $\phi$ within trimmed elements. The gradient $\widetilde{\nabla_n} (\phi^k)$ is an approximation of the actual normal gradient at iteration $k$, computed using gradient information from the non-intersected neighbour elements.  Notably, this term is computed by numerically integrating over the entire trimmed element, not just the active portion, and has proven to yield very good results
\end{itemize}
}

\begin{rem}
    To approximate the gradient inside intersected elements, \( \widetilde{\nabla} (\phi) \), the gradient in non-intersected elements must first be determined. However, since the gradient in these elements depends on the nodal values, unknown prior to imposing boundary conditions, the procedure is inherently iterative. Notably, if the exact gradient at the boundary were known, the algorithm would converge in a single iteration.
\end{rem}

\begin{rem}
    It is important to mention that at the first iteration ($k=0$), the normal gradient within the intersected elements is initialized as
    \begin{equation}
        \nabla (\phi^0) \cdot \mathbf{n} = 0
        \label{eq:initial_normal_gradient}
    \end{equation}
\end{rem}

Now, we would like to introduce the new discrete form of the problem based on the proposed functional in Eq.~\ref{eq:new_functional_stabilised}. Taking the derivative of Eq.~\ref{eq:new_functional_stabilised} with respect to each $\phi_i^{k+1}$ and setting it to zero yields
\begin{equation}
    \label{eq:new_variation}
    \begin{split}
        \frac{\partial \psi}{\partial \phi_i^{k+1}} = \int_{\Gamma_D} \left( \sum_{j=1}^{n} N_j(\mathbf{x}) \phi_j^{k+1} - \phi_D(\mathbf{x}) \right) N_i(x) \, d\Gamma +
        \int_{\Omega_{\text{trim}}} \left( \sum_{j=1}^{n} (\nabla N_j(\mathbf{x}) \cdot \mathbf{n}) \, \phi_j^{k+1} - \widetilde{\nabla}(\phi^k) \cdot \mathbf{n} \right) (\nabla N_i(\mathbf{x}) \cdot \mathbf{n}) \, d\Omega = 0, \\ \quad i = 1, \dots, n.
    \end{split}
\end{equation}

\ignore{
This results in the following linear system expressed in terms of the nodal degrees of freedom at the Dirichlet boundary $\Gamma_D$
\begin{equation}
    \label{eq:new_linear_system}
    \begin{split}
        \sum_{i=1}^{n} \left( \int_{\Gamma_D} N_i(\mathbf{x}) N_j(\mathbf{x}) \, d\Gamma +
        \int_{\Omega_{\text{trim}}} (\nabla N_i(\mathbf{x}) \cdot \mathbf{n}) \, (\nabla N_j(\mathbf{x}) \cdot \mathbf{n}) \, d\Omega \right) \phi_i^{k+1}
        =
        \int_{\Gamma_D} N_j(\mathbf{x}) \phi_D(\mathbf{x}) \, d\Gamma 
        + \int_{\Omega_{\text{trim}}} \, (\nabla N_j(\mathbf{x}) \cdot \mathbf{n})(\widetilde{\nabla}(\phi^k) \cdot \mathbf{n}) \, d\Omega, \\
        \quad j = 1, \dots, n.
    \end{split}
\end{equation}
}

\noindent
In matrix form, the algebraic system can be expressed as a residual equation  
\begin{equation}
    \mathbf{A} \boldsymbol{\phi}^{k+1} - \mathbf{f}^k = \mathbf{0},
    \label{eq:new_matrix_form}
\end{equation}  
where the matrix \(\mathbf{A}\) and the right-hand-side vector \(\mathbf{f}\) are defined as  

\begin{equation}
    \mathbf{A}_{ij} = \int_{\Gamma_D} N_i(\mathbf{x}) N_j(\mathbf{x}) \, d\Gamma +
    \int_{\Omega_{\text{trim}}} (\nabla N_i(\mathbf{x}) \cdot \mathbf{n}) (\nabla N_j(\mathbf{x}) \cdot \mathbf{n}) \, d\Omega=(A_{{ij}})_{\gamma}+(A_{{ij}})_{\Omega_{trim}},
    \label{eq:matrix_A}
\end{equation}

\begin{equation}
    \mathbf{f}_i^k = \int_{\Gamma_D} \phi_D(\mathbf{x}) N_i(\mathbf{x}) \, d\Gamma +
    \int_{\Omega_{\text{trim}}} (\widetilde{\nabla}(\phi^k) \cdot \mathbf{n}) (\nabla N_i(\mathbf{x}) \cdot \mathbf{n}) \, d\Omega =(f_{i})_{\gamma}+(f^k_{i})_{\Omega_{trim}}
    \label{eq:vector_f}
\end{equation}

\noindent
This formulation extends the classical discrete system in Eq.~\ref{eq:original_linear_system} by incorporating a regularization (or augmentation) term within the intersected elements, thereby ensuring a well-posed problem. Notably, each individual contribution to the left-hand side is inherently ill-posed; only their combined effect yields a stable and solvable linear system.

\begin{rem}
    It is important to emphasize that Equations~\ref{eq:matrix_A} and \ref{eq:vector_f} are \emph{not incorporated into the weak form of the physical problem}. Instead, Equation~\ref{eq:new_matrix_form} defines an independent algebraic problem that is solved iteratively to determine the boundary conditions to be imposed on the nodes of the cut elements. This procedure remains decoupled from the weak formulation given in Equation~\ref{eq:final_weak_form}.
\end{rem}

\begin{rem}
    In Equations~\ref{eq:matrix_A} and \ref{eq:vector_f}, only the contribution 
    $(f^k_{i})_{\Omega_{\text{trim}}}$, associated with the intersected (or trimmed) elements needs to be updated at each fixed-point iteration. 
    The boundary-related terms are computed only once at the beginning of the time step, as they remain constant throughout the iterative process.
\end{rem}

\textcolor{black}{
\begin{rem}
    Although the focus of this work is on the strong-like imposition of Dirichlet boundary conditions in unfitted meshes, the proposed framework can accommodate Neumann conditions in a fully black-box fashion. Given the weak form contribution
    \begin{equation}
        \int_{\Gamma_N} t_N w \, d\Gamma,
    \end{equation}
    the associated force vector is \emph{computed externally}, using the geometry of the Neumann boundary, and then \emph{passed to the solver} as an additional forcing term in its global right-hand side vector. The stiffness matrix assembled by the solver is left completely unchanged.
    For each cut element $K$, a boundary quadrature is constructed on $\Gamma_N \cap K$, the outward unit normal is computed from the exact geometry, and the contribution
    \begin{equation}
    f_{N,i} = \int_{\Gamma_N} t_N N_i(x) \, d\Gamma
    \end{equation}
    is evaluated externally, where $t_N$ is the prescribed Neumann traction, $N_i(x)$ is the $i$-th basis function of the host solver, and $f_{N,i}$ is the corresponding entry in the global load vector. The vector $\mathbf{f}_N = \{f_{N,i}\}$ is then added to the solver's existing load vector via its public API. 
    \label{rem:treatment_neumann_bcs}
\end{rem}
}

\begin{figure}[h!]
    \centering
    \begin{tikzpicture}[scale=1.2]
        \centering
        
        \draw[-{Stealth[blue,flex]}, line width = 0.25mm, color=blue] (-0.5357,0.0)--(8.5,0.0) node[above]{\footnotesize$x$};
        \draw[-{Stealth[blue,flex]}, line width = 0.30mm, color=blue] (0.0,0.0)--(0.0,1.5) node[right]{\footnotesize$y$};
        
        \draw[-{Stealth[black,flex]}, line width = 0.30mm, color=black];
        
        \draw[ line width = 0.50mm, color=black] (0,0) -- (7,0);
        
          \draw[-{Stealth[cyan,flex]}, line width = 0.35mm, color=cyan]  (7,0) -- (8.0,0) node[anchor=south east] {$F$};
        
        \draw[thick,red, dashed] (0,0) -- (7.0,2.5);
        \draw[-{Stealth[red,flex]}, line width = 0.30mm, color=red, dashed]  (7,0) -- (7,2.5) node[anchor=west] {$u(x) = \alpha x + \beta, \quad \text{where } \alpha, \beta \in \mathbb{R}$
        };

        \draw[thick,green, dashed] (-0.5357,0) -- (0.6,0.0);
        \draw[thick,green, dashed] (0.6,0) -- (7.0,1.9) 
    node[pos=0.5, below, sloped] {\tiny Iteration 1};

        \draw[thick,cyan, dashed] (-0.5357,-0.1) -- (0.6,0.1);
        \draw[thick,cyan, dashed] (0.6,0.1) -- (7.0,2.3) 
    node[pos=0.7, below, sloped] {\tiny Iteration 2};

        \foreach \i in {0,...,7} {
            \pgfmathsetmacro{\xval}{-0.5357 + \i * (6.9617 - (-0.5357)) / 7}
            \fill[red] (\xval,0) circle (2pt);
        }

        \foreach \i in {1,...,7} {
            \pgfmathsetmacro{\xval}{-0.5357 + (\i - 0.5) * (6.9617 - (-0.5357)) / 7}
            \draw (\xval,-0.3) circle (0.15) node {\footnotesize $e_{\i}$};
        }
        
        \node[anchor=north west] at (-0.5357,0) {1};
        \node[anchor=north] at (0.5357,0) {2};
        \node[anchor=north] at (1.6067,0) {3};
        \node[anchor=north] at (2.6777,0) {4};
        \node[anchor=north] at (3.7487,0) {5};
        \node[anchor=north] at (4.8197,0) {6};
        \node[anchor=north] at (5.8907,0) {7};
        \node[anchor=north] at (6.9617,0) {8};
        
        \node[anchor=north, color=gray] at (7.4,0.0) {\small $\Omega$};
    \end{tikzpicture}
    \caption{Illustration of the iterative correction process for enforcing a zero-displacement ($u=0$) boundary condition at the midpoint of the first element in an unfitted bar structure. The bar, fixed at the left end, is subjected to an external force $F$ at its free end. The structured mesh discretization is represented by nodal points (marked in red).}
    \label{fig:iterative_algorithm_example}
\end{figure}
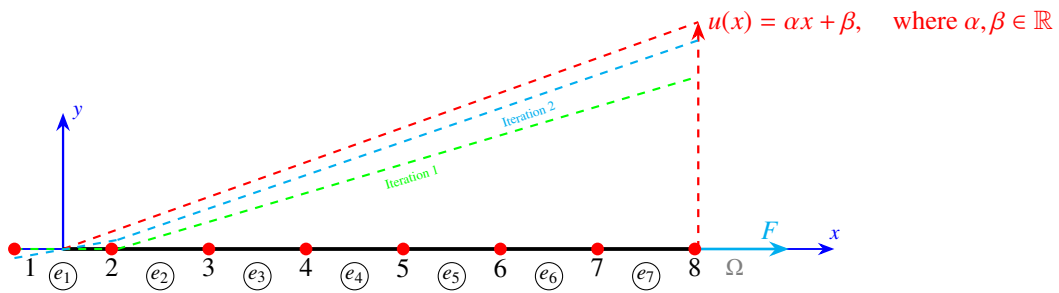

At this stage, it would be highly beneficial for the reader to re-examine a simple example based on the unfitted bar structure shown in Fig.~\ref{fig:bar_structure_unfitted} to better understand the iterative algorithm. This illustrative example is presented in Fig.~\ref{fig:iterative_algorithm_example}, where we attempt to enforce a zero-displacement ($u=0$) boundary condition at the midpoint of the first element. In the first iteration (green dashed line), the normal gradient inside the trimmed element is initially set to zero, leading the solution of Eq.~\ref{eq:new_matrix_form} to yield $u_1 = u_2 = 0$. In the second iteration (cyan dashed line), information from the previous step regarding the normal gradients within the domain is incorporated, enabling a correction of the boundary condition values by approximating the normal gradient inside the trimmed element. The algorithm continues iterating until a specified absolute or relative tolerance is met.

To further clarify the discussion, we now outline the general solution algorithm (Alg.~\ref{alg:strong_dir_bc}) used to non-intrusively impose Dirichlet boundary conditions on both body-fitted and unfitted discretizations. The proposed method follows an iterative procedure that is applied at each time step for transient problems. First, an auxiliary problem is solved to determine the nodal values on the Dirichlet boundary. These values are then imposed explicitly as boundary conditions in the main equations system. The physical problem is subsequently solved for the remaining degrees of freedom, the relative error between iterations is evaluated, and the process is repeated until convergence is achieved.

\ignore{
To further clarify the discussion, this section presents the algorithm for the proposed method, which enables a non-intrusive application of Dirichlet BCs in unfitted discretizations. 
We outline the simulation strategy for transient problems, where, at each time step, the procedure consists of the following steps: solving the auxiliary problem for Dirichlet BCs imposition, enforcing the obtained values as strong Dirichlet BCs, solving for the DOFs of the physical problem, computing the relative error between iterations, and repeating these steps until convergence is achieved.
}


\begin{algorithm}[h!]
\footnotesize
\SetAlgoLined
    \KwData{\\
        $k_{max}$ : Max. iterations \\
        $tol$ : Residual tolerance\\
        $T$ : Simulation end time\\
        $\Delta t$ : Time step increment\\
        $\mathcal{M}$ : Background mesh\\
        $\mathcal{EMB}$ : Embedded geometry\\
    }
    \texttt{\\}
    \tcp{Classify the elements (or knot spans) in the background mesh as active, inactive or intersected (note that this is a preprocessing step)}
    \texttt{ClassifyElements}($\mathcal{M, EMB}$)$\to \mathcal{A, I}$ \tcp*[r]{$\mathcal{A}$: subset of active elements, $\mathcal{I}$: subset of intersected elements}

    \texttt{\\}
    \While{$t \leq T$}{

        \tcp{Assemble the system matrices for the physical problem, considering only the contribution from active elements}
        \texttt{AssembleSystemMatrices($\mathcal{A}$)}
        
        \texttt{\\}
        \tcp{Non-linear solution strategy loop}
        \While{$k \leq k_{max}$}{
            \tcp{Solve the auxiliary algebraic problem to compute nodal boundary values in the Dirichlet boundary $\Gamma_D$}
            \texttt{SolveAuxiliaryProblem($\mathcal{A, I}$)} \tcp*[r]{Assemble and solve Eq.\ref{eq:new_matrix_form}}

            \texttt{\\}
            \tcp{Enforce the computed boundary values explicitly by altering the system matrix and right-hand side}
            \texttt{ModifySystemWithComputedBoundaryValues($\mathcal{I}$)}\;

            \texttt{\\}
            \tcp{Solve for the remaining DOFs of the original problem with updated BCs}
            \texttt{Solve($\mathcal{A}$)}\;

            \texttt{\\}
            \tcp{Compute the relative error between iterations $e_{rel}$}
            \texttt{ComputeRelativeErrorBetweenIterations($\mathcal{A, I}$)}\;

            \texttt{\\}
            \tcp{Check convergence}
            \eIf{$e_{rel} \leq tol$}{
                $\mathtt{break}$\;
            }{
                $k \mathrel{+}= 1$
            }
        }

        \texttt{\\}
        \tcp{Advance in time}
        $t \mathrel{+}= \Delta t$\;
     }
     \caption{Algorithm for the strong imposition of Dirichlet BCs in unfitted meshes}
     \label{alg:strong_dir_bc}
\end{algorithm}

The most critical preprocessing step in the algorithm is the \textit{elements classification}, which determines their role in the computation and assembly of the global system. Elements are categorized into three types: \textit{active elements, intersected elements, and inactive elements}.

\begin{itemize}
    \item \textit{Active elements:} These elements are fully assembled into the global system of equations, contributing to both the left-hand side (LHS) and right-hand side (RHS) of the system.
    \item \textit{Intersected elements:} These elements participate in the auxiliary algebraic problem required to compute the coefficients \( \phi_{\Gamma_{Di}} \), but they are not included in the assembly of the global system. This exclusion arises because the nodal values of the active nodes in these elements are explicitly imposed.
    \item \textit{Inactive elements:} These elements do not contribute to the global system and are excluded from the assembly process.
\end{itemize}

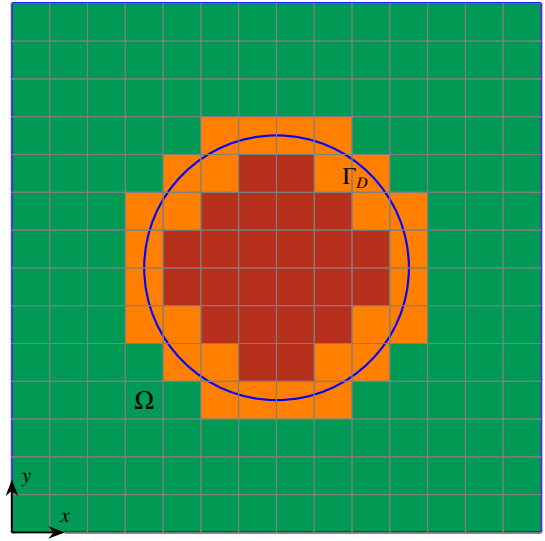
\begin{figure}[h!]
    \centering
    \begin{tikzpicture}[scale=7] 

        \def\step{1/14} 

        \foreach \x/\y in {
            0/0, 1/0, 2/0, 3/0, 4/0, 5/0, 6/0, 7/0, 8/0, 9/0, 10/0, 11/0, 12/0, 13/0,
            0/1, 1/1, 2/1, 3/1, 4/1, 5/1, 6/1, 7/1, 8/1, 9/1, 10/1, 11/1, 12/1, 13/1,
            0/2, 1/2, 2/2, 3/2, 4/2, 5/2, 6/2, 7/2, 8/2, 9/2, 10/2, 11/2, 12/2, 13/2,
            0/3, 1/3, 2/3, 3/3, 4/3, 9/3,10/3,11/3, 12/3, 13/3,
            0/4, 1/4, 2/4, 3/4, 10/4, 11/4, 12/4, 13/4,
            0/5, 1/5, 2/5, 11/5, 12/5, 13/5,
            0/6, 1/6, 2/6,11/6, 12/6, 13/6,
            0/7, 1/7, 2/7, 11/7, 12/7, 13/7,
            0/8, 1/8, 2/8, 11/8, 12/8, 13/8,
            0/9, 1/9, 2/9, 3/9, 10/9, 11/9, 12/9, 13/9,
            0/10, 1/10, 2/10, 3/10, 4/10, 8/10, 9/10, 10/10, 11/10,12/10, 13/10,
            0/11, 1/11, 2/11, 3/11, 4/11, 5/11, 6/11, 7/11, 8/11, 9/11, 10/11, 11/11, 12/11, 13/11,
            0/12, 1/12, 2/12, 3/12, 4/12, 5/12, 6/12, 7/12, 8/12, 9/12, 10/12, 11/12, 12/12, 13/12,
            0/13, 1/13, 2/13, 3/13, 4/13, 5/13, 6/13, 7/13, 8/13, 9/13, 10/13, 11/13, 12/13, 13/13
        } {
            \fill[ForestGreen] (\x*\step, \y*\step) rectangle (\x*\step+\step, \y*\step+\step);
        }

        \foreach \x/\y in {
             5/3, 6/3, 7/3, 8/3, 4/4, 5/4, 6/4, 7/4, 8/4, 9/4,
            3/5, 4/5, 9/5,
            3/6,
            3/7, 4/7, 5/7, 8/7, 9/7,
            3/8, 4/8, 5/8, 8/8, 9/8,
            4/9, 5/9, 6/9, 7/9, 8/9, 9/9,
            5/10, 6/10, 7/10, 8/10,
            10/5, 10/6, 10/7, 10/8
        } {
            \fill[orange] (\x*\step, \y*\step) rectangle (\x*\step+\step, \y*\step+\step);
        }

        \foreach \x/\y in {
            5/5, 8/5, 6/4, 7/4, 6/5, 7/5, 6/6, 7/6, 6/7, 7/7, 6/8, 7/8, 6/9, 7/9, 8/6, 9/6, 4/6, 5/6, 4/7, 5/7, 8/7, 9/7, 5/8, 8/8
        } {
            \fill[BrickRed] (\x*\step, \y*\step) rectangle (\x*\step+\step, \y*\step+\step);
        }

        \draw[thick,blue] (0.5,0.5) circle (0.25);
        \draw[thick,blue] (0.1,0.0)--(1.0,0.0);
        \draw[thick,blue] (1.0,0.0)--(1.0,1.0);
        \draw[thick,blue] (1.0,1.0)--(0.0,1.0);
        \draw[thick,blue] (0.0,1.0)--(0.0,0.0);

            \foreach \x in {0,0.0714,...,0.9286}
                \foreach \y in {0,0.0714,...,0.9286}
                    \draw[gray,very thin] (\x,\y) rectangle (\x+0.0714,\y+0.0714);

        \draw[-{Stealth[black]}, line width = 0.25mm, color=black] (0,0.0)--(0.1,0.0) node[above]{\footnotesize$x$};
        \draw[-{Stealth[black]}, line width = 0.25mm, color=black] (0.0,0.0)--(0.0,0.1) node[right]{\footnotesize$y$};
        \node at (0.25, 0.25) {$\Omega$};
        \node at (0.65, 0.67) {\small $\Gamma_D$};

    \end{tikzpicture}
    \caption{Classification of elements within the Cartesian grid: \textcolor{ForestGreen}{Active} (inside), \textcolor{orange}{Intersected} (crossing the circle), and \textcolor{BrickRed}{Inactive} (outside).}
    \label{fig:classification_cartesian_grid}
\end{figure}

Figure (\ref{fig:classification_cartesian_grid}) illustrates an example of element classification applied to a plate with a hole.


\section{Normal gradient approximation inside trimmed elements}
\label{sec:normal_gradient_approximation}

The key aspect of the proposed iterative algorithm for the strong-like imposition of Dirichlet BCs is the gradient approximation within the intersected elements. In this section, we present the fundamental principles of this approximation.

As previously mentioned, the gradient within intersected elements is approximated using the computed gradients from neighboring non-intersected elements. In finite element discretizations, interpolation points are typically selected at the element centers of the $N$ closest unperturbed elements (Def.\ref{def:unperturbed_element}). Conversely, in isogeometric discretizations, they are chosen as the $N$ nearest integration points within unperturbed knot spans. The number of interpolation points, 
$N$, is specified by the user and can be adapted based on the discretization density or problem setup.

\begin{definition}
    \label{def:unperturbed_element}
    An \textbf{unperturbed element or knot span} is an element in which no degrees of freedom are influenced or modified by the auxiliary algebraic problem \ref{eq:new_matrix_form}.
\end{definition}

\begin{figure}[h!]
    \centering
    \begin{tikzpicture}[scale=7] 

        \def\step{1/14} 

        \foreach \x/\y in {
            0/0, 1/0, 2/0, 3/0, 4/0, 5/0, 6/0, 7/0, 8/0, 9/0, 10/0, 11/0, 12/0, 13/0,
            0/1, 1/1, 2/1, 3/1, 4/1, 5/1, 6/1, 7/1, 8/1, 9/1, 10/1, 11/1, 12/1, 13/1,
            0/2, 1/2, 2/2, 3/2, 4/2, 5/2, 6/2, 7/2, 8/2, 9/2, 10/2, 11/2, 12/2, 13/2,
            0/3, 1/3, 2/3, 3/3, 4/3, 9/3,10/3,11/3, 12/3, 13/3,
            0/4, 1/4, 2/4, 3/4, 10/4, 11/4, 12/4, 13/4,
            0/5, 1/5, 2/5, 11/5, 12/5, 13/5,
            0/6, 1/6, 2/6,11/6, 12/6, 13/6,
            0/7, 1/7, 2/7, 11/7, 12/7, 13/7,
            0/8, 1/8, 2/8, 11/8, 12/8, 13/8,
            0/9, 1/9, 2/9, 3/9, 10/9, 11/9, 12/9, 13/9,
            0/10, 1/10, 2/10, 3/10, 4/10, 8/10, 9/10, 10/10, 11/10,12/10, 13/10,
            0/11, 1/11, 2/11, 3/11, 4/11, 5/11, 6/11, 7/11, 8/11, 9/11, 10/11, 11/11, 12/11, 13/11,
            0/12, 1/12, 2/12, 3/12, 4/12, 5/12, 6/12, 7/12, 8/12, 9/12, 10/12, 11/12, 12/12, 13/12,
            0/13, 1/13, 2/13, 3/13, 4/13, 5/13, 6/13, 7/13, 8/13, 9/13, 10/13, 11/13, 12/13, 13/13
        } {
            \fill[ForestGreen] (\x*\step, \y*\step) rectangle (\x*\step+\step, \y*\step+\step);
        }

        \foreach \x/\y in {
             5/3, 6/3, 7/3, 8/3, 4/4, 5/4, 6/4, 7/4, 8/4, 9/4,
            3/5, 4/5, 9/5,
            3/6,
            3/7, 4/7, 5/7, 8/7, 9/7,
            3/8, 4/8, 5/8, 8/8, 9/8,
            4/9, 5/9, 6/9, 7/9, 8/9, 9/9,
            5/10, 6/10, 7/10, 8/10,
            10/5, 10/6, 10/7, 10/8
        } {
            \fill[orange] (\x*\step, \y*\step) rectangle (\x*\step+\step, \y*\step+\step);
        }

        \foreach \x/\y in {
            5/5, 8/5, 6/4, 7/4, 6/5, 7/5, 6/6, 7/6, 6/7, 7/7, 6/8, 7/8, 6/9, 7/9, 8/6, 9/6, 4/6, 5/6, 4/7, 5/7, 8/7, 9/7, 5/8, 8/8
        } {
            \fill[BrickRed] (\x*\step, \y*\step) rectangle (\x*\step+\step, \y*\step+\step);
        }

        \draw[thick,blue] (0.5,0.5) circle (0.25);
        \draw[thick,blue] (0.1,0.0)--(1.0,0.0);
        \draw[thick,blue] (1.0,0.0)--(1.0,1.0);
        \draw[thick,blue] (1.0,1.0)--(0.0,1.0);
        \draw[thick,blue] (0.0,1.0)--(0.0,0.0);

            \foreach \x in {0,0.0714,...,0.9286}
                \foreach \y in {0,0.0714,...,0.9286}
                    \draw[gray,very thin] (\x,\y) rectangle (\x+0.0714,\y+0.0714);

        \draw[-{Stealth[black]}, line width = 0.25mm, color=black] (0,0.0)--(0.1,0.0) node[above]{\footnotesize$x$};
        \draw[-{Stealth[black]}, line width = 0.25mm, color=black] (0.0,0.0)--(0.0,0.1) node[right]{\footnotesize$y$};
        \node at (0.25, 0.25) {$\Omega$};
        \node at (0.65, 0.67) {\small $\Gamma_D$};

        \fill[Purple] (0.45, 0.233) circle (0.25pt);
        \fill[blue] (0.47, 0.105) circle (0.25pt);
        \fill[blue] (0.541, 0.105) circle (0.25pt);
        \fill[blue] (0.399, 0.105) circle (0.25pt);
        \fill[blue] (0.47, 0.034) circle (0.25pt);
        \fill[blue] (0.541, 0.034) circle (0.25pt);
        \fill[blue] (0.399, 0.034) circle (0.25pt);

    \end{tikzpicture}
    \caption{Illustration of the unperturbed element concept: The integration point where the gradient is approximated is shown in \textcolor{Purple}{purple}, while the unperturbed elements used for the gradient approximation are highlighted in \textcolor{black}{blue}.}
    \label{fig:unperturbed_elements}
\end{figure}
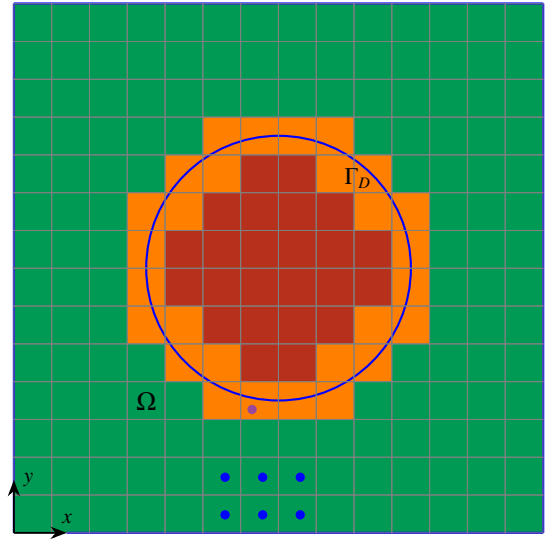

Fig.~\ref{fig:unperturbed_elements} illustrates the concept of an unperturbed element within a Cartesian finite element background mesh. In this case, the physical domain consists of a square plate with a hole in the center.

Through extensive numerical experiments, it has been observed that for the algorithm to achieve convergence, the gradient must be approximated using information exclusively from these unperturbed elements. These tests confirmed that including data from perturbed elements introduces inaccuracies that hinder the iterative process, reinforcing the necessity of relying solely on unperturbed regions for stable and accurate gradient reconstruction.

\begin{rem}
    Gradient reconstruction inside intersected elements should be performed exclusively using information from unperturbed elements, as contributions from perturbed elements may hinder convergence
\end{rem}

\subsection{Radial basis function interpolation with polynomial extension}

{\color{black} 
Radial Basis Function (RBF) interpolation is a well-established technique for scattered data approximation and field transfer, widely used in computational mechanics. In immersed and embedded settings, it has been employed, for example, to transfer solution fields in the Finite Cell Method (FCM) after remeshing~\cite{Sartorti2024}. In the present work, RBF interpolation is adopted to approximate the gradient within intersected elements, using gradient information from the surrounding unperturbed active elements.

Given $N$ data points $\mathbf{x}_i$, $i=1, \dots, N$, with known gradient values $\nabla \phi_i$, the RBF interpolation approximates the gradient field $\nabla \phi(\mathbf{x})$ as
\begin{equation}
    \label{EQ:rbf_interpolation}
    \nabla \phi (\mathbf{x}) = \sum_{i=1}^{N} w_i \, \beta( \| \mathbf{x} - \mathbf{x}_i \| ) + \sum_{k=1}^{N_p} \lambda_k q_k(\mathbf{x)},
\end{equation}
\noindent
where $\beta(r)$ is a chosen radial basis function, $w_i$ are the RBF weights, and $\mathbf{p}(\mathbf{x}) = \sum_{i=1}^{N_p} \lambda_j q_j(\mathbf{x)}$ is an optional polynomial term used to improve approximation accuracy and ensure the reproduction of polynomial fields up to a given degree. 
\medskip

Two conditions are imposed to determine the unknowns:
\begin{enumerate}
    \item \textbf{Interpolation condition:} At each known point $\mathbf{x}_j$,
    \begin{equation}
        \sum_{i=1}^{N} w_i \, \beta( \| \mathbf{x}_j - \mathbf{x}_i \| ) + \sum_{k=1}^{N_p} \lambda_k q_k(\mathbf{x_j)} = \nabla \phi (\mathbf{x}_j).
    \end{equation}
    \item \textbf{Polynomial orthogonality condition:} The RBF component must be orthogonal to the chosen polynomial space,
    \begin{equation}
        \sum_{i=1}^{N} w_i \, q(\mathbf{x}_i) = 0 \quad \text{for all basis functions } q(\mathbf{x}) \text{ of } \mathbf{p}(\mathbf{x}),
    \end{equation}
    which ensures that the polynomial term is exactly reproduced, removes rank deficiencies in the system, and guarantees uniqueness of the solution.
\end{enumerate}

\noindent
Combining these two conditions yields the augmented linear system
\begin{equation}
\begin{bmatrix}
    \mathbf{B} & \mathbf{P} \\
    \mathbf{P}^T & \mathbf{0}
\end{bmatrix}
\begin{bmatrix}
    \mathbf{w} \\
    \boldsymbol{\lambda}
\end{bmatrix}
=
\begin{bmatrix}
    \mathbf{g} \\
    \mathbf{0}
\end{bmatrix},
\end{equation}

\noindent
where $\mathbf{B}$ is the interpolation matrix with entries $B_{ji} = \beta(\| \mathbf{x}_j - \mathbf{x}_i \|)$,  
$\mathbf{P}$ is the polynomial matrix with rows $\mathbf{p}(\mathbf{x}_j)$,  
$\mathbf{w} = (w_1, \dots, w_N)^T$ contains the RBF weights,  
$\boldsymbol{\lambda}$ are the polynomial coefficients,  
and $\mathbf{g} = (\nabla \phi (\mathbf{x}_1), \dots, \nabla \phi (\mathbf{x}_N))^T$ contains the known gradients.
\medskip

The choice of $\beta(r)$ influences accuracy and stability. Common options include:
\begin{itemize}
    \item Gaussian: $\beta(r) = e^{-(\epsilon r)^2}$,
    \item Multiquadric (MQ): $\beta(r) = \sqrt{r^2 + \epsilon^2}$,
    \item Inverse Multiquadric (IMQ): $\beta(r) = 1 / \sqrt{r^2 + \epsilon^2}$,
    \item Thin-Plate Spline (TPS): $\beta(r) = r^2 \log r$,
    \item Cubic: $\beta(r) = r^3$.
\end{itemize}
\noindent
It is important to note that the shape parameter $\epsilon$ governs the flatness or sharpness of the radial basis function, directly influencing both the interpolation accuracy and the numerical stability of the solution. However, selecting an appropriate value for $\epsilon$ typically requires careful fine-tuning, which we aim to avoid in the present method. This motivates the investigation of alternative interpolation techniques that require no parameter tuning, such as the Moving Least Squares (MLS) method, which will be discussed later in this work. For an in-depth discussion of RBF interpolation and its polynomial extension, the reader is referred to \cite{powell1992rbf} and \cite{buhmann2000rbf}.}

\subsection{MLS interpolation}

The \textit{Moving Least Squares} (MLS) method is a well-known numerical technique used for function approximation, surface reconstruction, and meshless methods in computational mechanics. Unlike traditional least squares approximation, MLS provides a localized and adaptive approach to fitting a function to scattered data points.

The primary goal of MLS is to construct a smooth function approximation $P(\mathbf{x})$ that closely follows a given set of data points while being flexible enough to adapt to local variations. This is achieved by minimizing a weighted least squares error that emphasizes nearby points more than distant ones.

Given $N$ a set of scattered data points $\mathbf{x}_i$, $i=1, \dots, N$, with known gradient values $\nabla \phi_i$, the MLS method constructs a \textit{local polynomial approximation} at each evaluation point $\mathbf{x}$. The polynomial is defined as
\begin{equation}
    P(\mathbf{x}) = \sum_{j=0}^{m} a_j(\mathbf{x}) \beta_j(\mathbf{x}) = \mathbf{a}(\mathbf{x}) \cdot \boldsymbol{\beta}(\mathbf{x})
\end{equation}
where $\beta_j(\mathbf{x})$ are basis functions (typically polynomials), and $a_j(\mathbf{x})$ are the unknown coefficients.
The coefficients $a_j(\mathbf{x})$ are obtained by minimizing the weighted least squares error
\begin{equation}
    J(\mathbf{a}) = \frac{1}{2}\sum_{i} W(\mathbf{x} - \mathbf{x}_i) \left( P(\mathbf{x}_i) - \nabla\phi_i\right)^2
\end{equation}
 $W(\mathbf{x} - \mathbf{x}_i)$ is a \textit{weight or kernel function} introduced to ensure that points closer to $\mathbf{x}$ have greater influence. A common choice is the Gaussian weight function
\begin{equation}
    W(\mathbf{x} - \mathbf{x}_i) = \exp\left(-\frac{|\mathbf{x} - \mathbf{x}_i|^2}{h^2}\right)
\end{equation}
where $h$ is a smoothing parameter controlling the influence range of each point.

\noindent
This leads to a system of linear equations, which is solved to determine the optimal values of $a_j(\mathbf{x})$. The system is given by
\begin{equation}
    \mathbf{M}(\mathbf{x}) \mathbf{a}(\mathbf{x}) = \mathbf{H}(\mathbf{x})
\end{equation}
being
\begin{equation}
    \mathbf{M}(\mathbf{x}) = \mathbf{B}^T(\mathbf{x}) \mathbf{W}(\mathbf{x}) \mathbf{B}(\mathbf{x})
\end{equation}
and
\begin{equation}
    \mathbf{H}(\mathbf{x}) = \mathbf{B}^T(\mathbf{x}) \mathbf{W}(\mathbf{x}) \mathbf{\nabla \phi} = 
    \begin{bmatrix}
        \sum\limits_{i=1}^{N} W_i \nabla \phi_i \\
        \sum\limits_{i=1}^{N} x_i W_i \nabla \phi_i \\
        \sum\limits_{i=1}^{N} y_i W_i \nabla \phi_i
    \end{bmatrix}
\end{equation}

\noindent
Here, $\mathbf{M}(\mathbf{x})$ is the weighted moment matrix and $\mathbf{H}(\mathbf{x})$ is defined as $\mathbf{B}^T(\mathbf{x}) \mathbf{W}(\mathbf{x}) \mathbf{\nabla\Phi}$, being $\mathbf{B}(\mathbf{x})$ the matrix of basis functions and $\mathbf{\nabla\Phi}$ a vector containing the function values to be interpolated. The term $\mathbf{W}(\mathbf{x})$ is a diagonal weighting matrix, with diagonal entries $W_{ii}=W(\mathbf{x} - \mathbf{x}_i)$.

Once the coefficients are determined, the final approximation function is given by:
\begin{equation}
    \nabla \phi(\mathbf{x}) \approx P(\mathbf{x})
\end{equation}
which smoothly adapts to local data variations. For a comprehensive mathematical review of MLS interpolation, the reader is referred to \cite{Levin1998} and \cite{Afshar2021}.

\section{Application of the method to the numerical solution of the Poisson problem for FEM and IGA discretizations}
\label{sec:numerical_solution_poisson_problem}

In this section, we assess the proposed algorithm for the Poisson problem \ref{eq:poisson} across various scenarios, considering different discretization methods (low- and high-order FEM basis functions, as well as B-Spline basis), different element geometries (triangular and quadrilateral elements), and different gradient approximation techniques within trimmed elements. The error convergence is evaluated using the following manufactured solution for the Poisson problem:
\begin{equation*}
    \label{eq:man_solution}
    \phi(x,y) = \sin(\pi x) \cos(\pi y)
\end{equation*}

\noindent
For the given manufactured solution, the source term $\sigma(x,y)$ is defined as:
\begin{equation*}
    \label{eq:source_term}
    \sigma(x,y) = -2\pi^2 \sin(\pi x) \cos(\pi y)
\end{equation*}

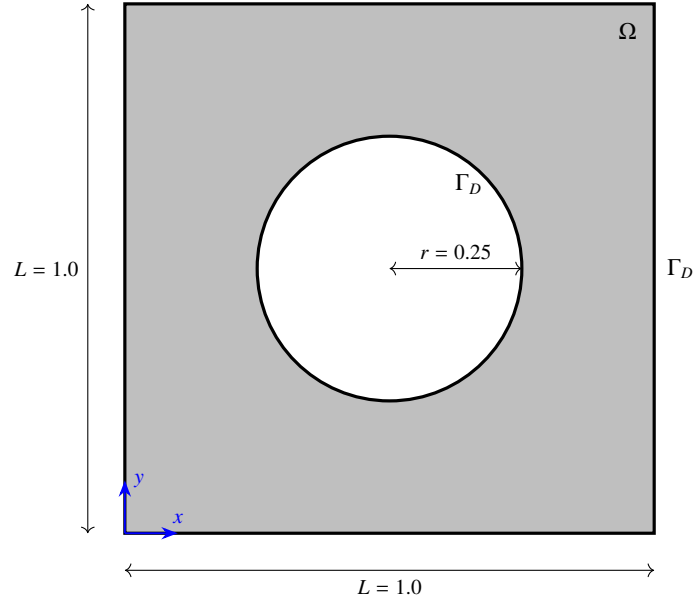
\begin{figure}[h!]
    \centering
    \begin{tikzpicture}[scale=7] 

        \fill[gray!50] (0,0) rectangle (1,1);  
        \fill[white] (0.5,0.5) circle (0.25);  

        \draw[very thick,black] (0,0) rectangle (1,1); 
        \draw[very thick,black] (0.5,0.5) circle (0.25); 

        \draw[-{Stealth[blue]}, line width = 0.25mm, color=blue] (0,0.0)--(0.1,0.0) node[above]{\footnotesize$x$};
        \draw[-{Stealth[blue]}, line width = 0.25mm, color=blue] (0.0,0.0)--(0.0,0.1) node[right]{\footnotesize$y$};

        \node at (0.95, 0.95) {\small $\Omega$};
        \node at (0.65, 0.66) {\small $\Gamma_D$};
        \node at (1.05, 0.5) {\small $\Gamma_D$};

        \draw[thin,<->] (0,-0.07) -- (1,-0.07) node[midway,below]{\footnotesize $L=1.0$};
        \draw[thin,<->] (-0.07,0) -- (-0.07,1) node[midway,left]{\footnotesize $L=1.0$};

        \draw[thin,<->] (0.5,0.5) -- (0.75,0.5) node[midway,above]{\footnotesize $r=0.25$};

    \end{tikzpicture}
    \caption{Illustration of the computational domain, consisting of a 
[1×1] square with an embedded circular hole of radius $r=0.25$, centered at $C=(0.5,0.5)$.}

    \label{fig:domain_geometry}
\end{figure}

The domain consists of a $[1 \times 1]$ square with a circular hole centered at $C = (0.5, 0.5)$ and a radius of $r = 0.25$ (Fig.~\ref{fig:domain_geometry}).We impose Dirichlet boundary conditions across the entire boundary, $\Gamma_D = \Gamma$, by explicitly enforcing the manufactured solution. When the boundary is embedded (trimmed), the proposed algorithm strongly enforces these conditions. For body-fitted boundaries, where the geometry aligns with the underlying discretization, the same enforcement approach is used. 

The error is measured using the $L^2$-norm of the difference between the analytical solution and the numerical approximation:
\begin{equation*}
    \label{eq:L2_norm}
    \| \phi - \phi_h \|_{L^2(\Omega)} =
    \sqrt{\int_{\Omega} (\phi - \phi_h)^2 \, d\Omega}
\end{equation*}

The proposed strategy was implemented using the Kratos Multiphysics API, based on the release version \texttt{v10.1} of the open-source framework \href{https://github.com/KratosMultiphysics/Kratos}{Kratos Multiphysics} \cite{dadvand2010, dadvand2013}. This implementation demonstrates that it is possible to adapt existing body-fitted solvers to operate in unfitted mesh scenarios, provided that the solver satisfies four key conditions: (i) it supports user scripting, (ii) it allows Dirichlet boundary conditions to be imposed at the node level through scripting, (iii) it permits deactivation of elements outside the physical domain, and (iv) it provides access to the solution gradient within active elements.

In Kratos, the structure and flow of a simulation are typically managed by a class called \texttt{AnalysisStage}. This class governs the full simulation lifecycle, including model initialization, time stepping, and solution loop execution. The proposed algorithm was implemented by extending \texttt{AnalysisStage}, overriding only the necessary member functions of this class to inject the Dirichlet enforcement step at the appropriate stage of the loop.

To illustrate this, we present in Listing~\ref{lst:algorithm_implementation} a minimal but complete implementation of this algorithm using the Kratos API. The class shown integrates seamlessly with Kratos' solver infrastructure and provides a reusable mechanism for solving unfitted problems without modifying the internals of the solver.

\begin{lstlisting}[style=kratos, caption={Core implementation of the proposed algorithm in Kratos.}, label={lst:algorithm_implementation}]
# --- Import Kratos core and the convection-diffusion solver wrapper ---
import KratosMultiphysics
from KratosMultiphysics.ConvectionDiffusionApplication import python_solvers_wrapper_convection_diffusion as solver_wrapper
from KratosMultiphysics.analysis_stage import AnalysisStage
from scipy.interpolate import Rbf  # Used to intra/extrapolate gradients from the physical domain

# --- Main analysis class inheriting from Kratos' AnalysisStage infrastructure ---
# This class wraps the entire solution process and integrates the proposed strategy.
class ExtendedGradientMethodConvectionDiffusionAnalysis(AnalysisStage):

    def RunSolutionLoop(self):
        # Retrieve the primary unknown (e.g. temperature or concentration) from solver settings
        unknown_variable = self._GetUnknownVariable()

        # Initialize the solution field from the previous time step (used for gradient extrapolation)
        self._InitializeOldSolutionField(unknown_variable)

        # Loop over time steps
        while self.KeepAdvancingSolutionLoop():
            self._AdvanceTime()
            self.InitializeSolutionStep()

            # Begin fixed-point iterations for applying Dirichlet BCs strongly on an unfitted mesh
            self.iteration_number = 0
            while self._NotConverged():
                self._GetSolver().Predict()  # Kratos predictor step

                # Proposed method: apply Dirichlet BCs using auxiliary system
                self.ApplyDirichletBoundaryConditions()

                # Solve the physical problem after applying boundary conditions
                self._GetSolver().SolveSolutionStep()

                # Check error between iterations to assess convergence
                self._UpdateConvergenceStatus()

            # Update stored field and finalize current step
            self._UpdateOldSolutionField(unknown_variable)
            self.FinalizeSolutionStep()
            self.OutputSolutionStep()

    def ApplyDirichletBoundaryConditions(self):
        # Extract relevant submodel parts from Kratos' data structures
        self.model_part = self._GetSolver().GetComputingModelPart()
        self.intersected_elements_sub_model_part = self.model_part.GetSubModelPart("intersected_elements")
        self.active_elements_sub_model_part = self.model_part.GetSubModelPart("active_elements")
        self.embedded_body_boundary_model_part = self.model_part.GetModel().GetModelPart("embedded_body_boundary")

        # Assemble the LHS matrix and the RHS vector from the boundary only once per time step
        if self.iteration_number == 0:
            self.LHS = self.CalculateLHS()  # Compute the A matrix (Includes both boundary and intersected-elements terms)
            self.RHSBoundaryContribution = self.CalculateRHSBoundaryContribution() # Compute f_gamma

        # Compute the RHS contribution from the intersected elements (updated at each iteration)
        self.RHSTrimmedElementsContribution = self.CalculateRHSTrimmedElementsContribution() # Compute (f^k)_trim
        self.RHS = self.RHSBoundaryContribution + self.RHSTrimmedElementsContribution

        # Solve the auxiliary linear system to compute nodal Dirichlet values
        solution_dir = self._SolveLinearSystem(self.LHS, self.RHS)

        # Apply the computed Dirichlet values to the nodes belonging to the intersected elements
        # (This typically calls node.Fix(VARIABLE) and node.SetSolutionStepValue(VARIABLE, value)
        #  for each node in the intersected elements sub-model part)
        self._ApplyBoundaryValues(solution_dir)

    # Assemble the global LHS matrix from both boundary and intersected element contributions
    def CalculateLHS(self):
        LHS_bc = self.CalculateLHSBoundaryContribution()
        LHS_trimmed = self.CalculateLHSTrimmedElementsContribution()
        return self._AssembleLHS(LHS_bc, LHS_trimmed)

    # Compute the RHS vector corresponding to boundary integration (fixed per time step)
    def CalculateRHSBoundaryContribution(self):
        return self._IntegrateBoundaryBCs()

    # Use gradient extrapolation (via RBF or MLS) to compute trimmed-element RHS at each iteration
    def CalculateRHSTrimmedElementsContribution(self):
        gradient_interpolator = self._BuildRBFGradientInterpolator()
        return self._IntegrateTrimmedElementContributions(gradient_interpolator)

# --- Main script execution ---
# Loads parameters, constructs the analysis class, and runs the simulation
if __name__ == "__main__":
    project_parameters_file_name = "ProjectParameters.json"

    # Read solver and problem configuration from file.
    # This JSON file contains geometry and mesh definition, material properties, boundary conditions, solver settings and output configuration
    with open(project_parameters_file_name, 'r') as parameter_file:
        parameters = KratosMultiphysics.Parameters(parameter_file.read())

    # Create model and run the simulation
    model = KratosMultiphysics.Model()
    simulation = ExtendedGradientMethodConvectionDiffusionAnalysis(model, parameters)
    simulation.Initialize() # The elements classification and deactivation (element.Set(ACTIVE, False)) 
    # is performed here
    simulation.RunSolutionLoop()
    simulation.Finalize()
\end{lstlisting}

\subsection{FEM discretizations}

In this section, we present numerical results for the application of the proposed method to the Poisson problem using different Finite Element Method (FEM) discretizations. The primary objective is to evaluate the effectiveness of the strong Dirichlet boundary condition enforcement technique for \textit{unfitted meshes} and compare its performance across different element types and polynomial orders.

We consider the \textit{manufactured solution} already defined above for the Poisson problem described in Eq.~\ref{eq:poisson}, which allows us to evaluate the \textit{$L^2$-norm error} for different discretizations. As described in Section~\ref{sec:numerical_solution_poisson_problem}, the computational domain consists of a \textit{square region with an embedded circular hole}, with Dirichlet BCs applied to the entire boundary.

The numerical experiments explore the performance of the method under \textit{different FEM discretizations}, specifically considering linear triangular elements, as well as quadratic quadratic elements.

For each case, we assess the \textit{convergence behaviour} and \textit{accuracy} of the numerical solution. Additionally, we analyze the impact of \textit{gradient approximation techniques} (Radial Basis Functions (RBF) and Moving Least Squares (MLS)) on the accuracy of the strong Dirichlet BCs enforcement.

\subsubsection{FEM Discretization with linear triangular elements}

This section examines the accuracy and convergence behaviour of the proposed strong Dirichlet BCs enforcement approach in unfitted triangular meshes. 

\begin{figure}[h!]
    \centering
    \includegraphics[width=0.8\textwidth]{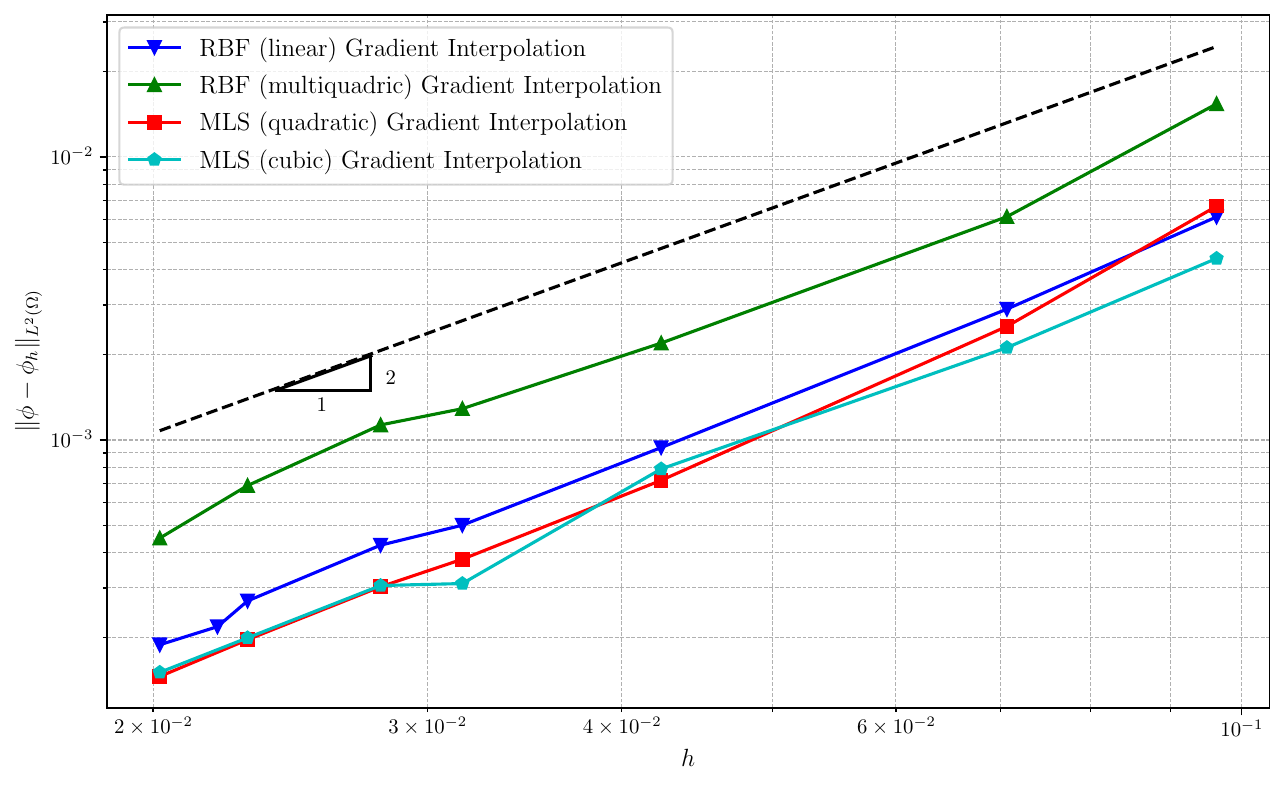}
    \caption{Comparison of $L^2$-norm error convergence for linear triangular elements under different gradient approximation techniques. The error is computed for the manufactured solution $\phi(x,y) = \sin(\pi x) \cos(\pi y)$ using an unfitted boundary mesh. The study evaluates four gradient interpolation methods: RBF with linear basis (blue triangles), RBF with multiquadric basis (green triangles), MLS with quadratic basis (red squares), and MLS with cubic basis (cyan circles). The dashed black line represents the theoretical second-order convergence rate ($\mathcal{O}(h^2)$).}
    \label{fig:L2_error_linear_triangle}
\end{figure}

\begin{figure}[h!]
    \centering
    \includegraphics[width=0.8\textwidth]{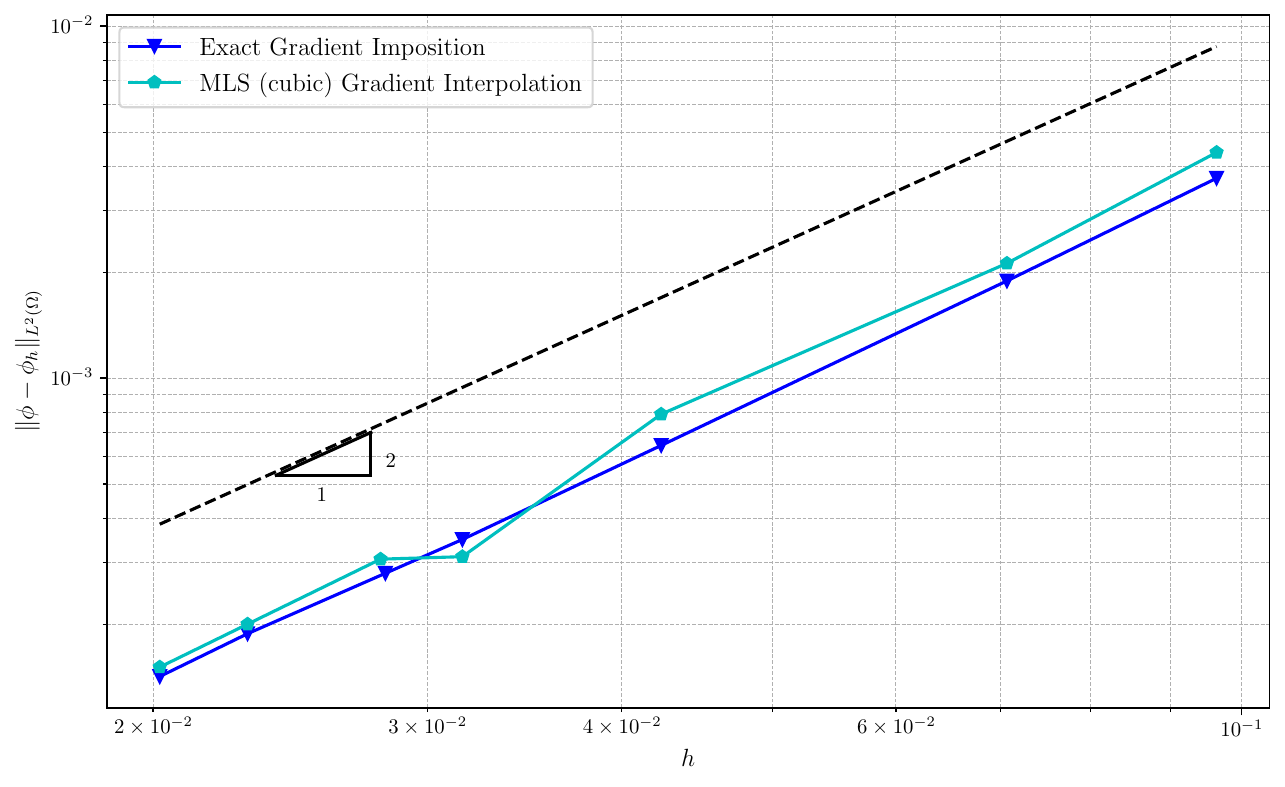}
    \caption{Comparison of $L^2$-norm error convergence for linear triangular elements discretization. The study compares the $L^2$-norm error using MLS with cubic basis functions gradient approximation (cyan pentagons) and and the $L^2$-norm error imposing the exact gradient in the trimmed elements (blue triangles). The dashed black line represents the theoretical second-order convergence rate ($\mathcal{O}(h^2)$).}
    \label{fig:L2_error_linear_triangle_exact_grad}
\end{figure}

Figs.(\ref{fig:L2_error_linear_triangle}) and (\ref{fig:L2_error_linear_triangle_exact_grad}) present the $L^2$-norm error convergence for different gradient approximation strategies. The analysis considers two interpolation methods: Radial Basis Function (RBF) interpolation and Moving Least Squares (MLS) approximation. The RBF method is evaluated with both a linear and a multiquadric basis, while the MLS approach is tested with quadratic and cubic polynomial orders. Additionally, the performance of enforcing the exact gradient is compared against the MLS-based approximations (Fig.~\ref{fig:L2_error_linear_triangle_exact_grad}).

All methods exhibit the expected second-order convergence rate ($\mathcal{O}(h^2)$) in the $L^2$-norm error, confirming theoretical predictions for first-order FEM discretizations. However, multiquadric RBF interpolation consistently produces higher errors, emphasizing the impact of basis function selection on accuracy. In contrast, MLS cubic interpolation closely matches exact gradient imposition (Fig.~\ref{fig:L2_error_linear_triangle_exact_grad}), demonstrating its effectiveness in gradient approximation. The shown results confirm that MLS interpolation, both the quadratic and cubic variant, represent a robust and accurate approach for the gradient approximation in this algorithm.

\begin{figure}[h!]
    \centering
    \begin{subfigure}{0.47\textwidth} 
        \centering
        \includegraphics[width=\linewidth]{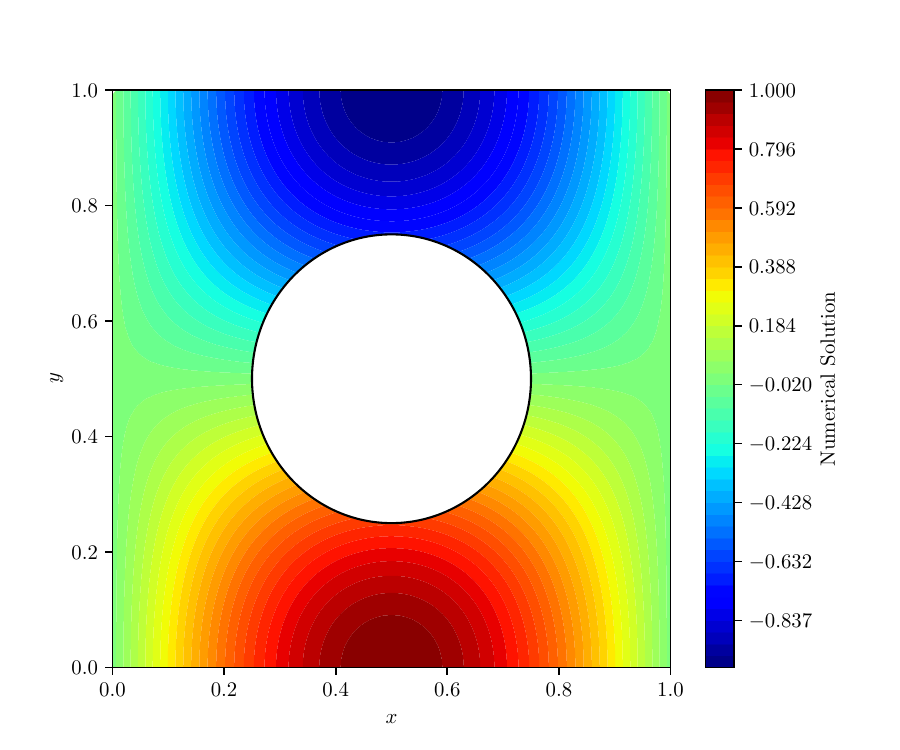}
        \caption{Numerical solution}
        \label{fig:plate_hole_numerical_triangle}
    \end{subfigure}
    \hfill 
    \begin{subfigure}{0.47\textwidth} 
        \centering
        \includegraphics[width=\linewidth]{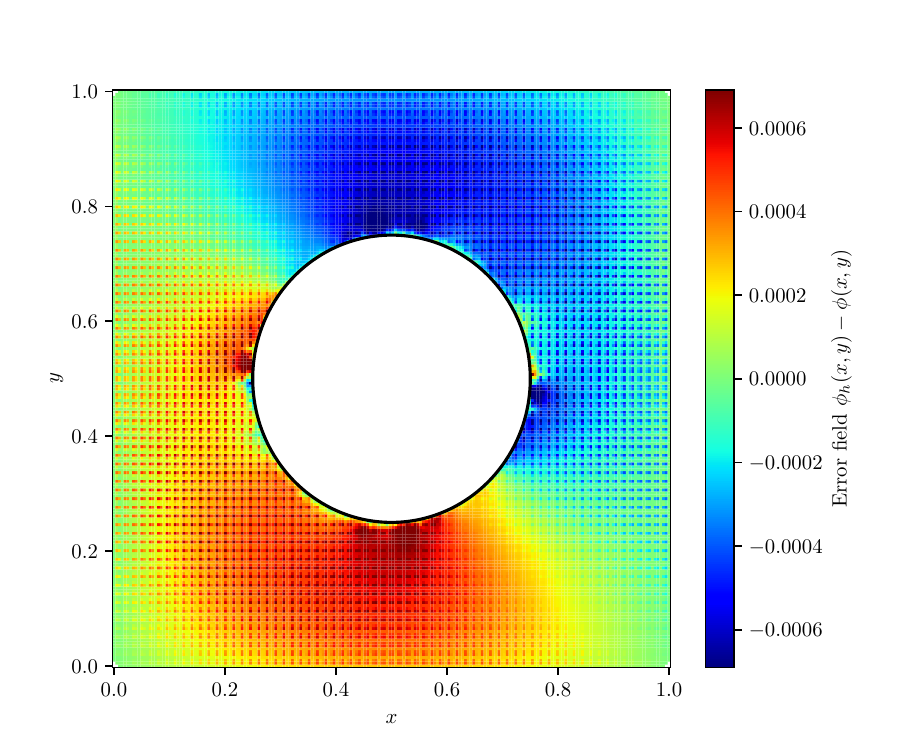}
        \caption{Error distribution}
        \label{fig:plate_hole_error_triangle}
    \end{subfigure}
    \caption{Comparison of the numerical solution and the corresponding error distribution for the plate with a hole problem using linear triangular elements discretization. (a) Computed numerical solution. (b) Error distribution, illustrating the deviation of the numerical solution from the exact solution.}
    \label{fig:plate_hole_comparison}
\end{figure}

Finally, Fig.~\ref{fig:plate_hole_comparison} provides a visualization of the numerical solution and the corresponding error distribution for the Poisson problem. The results correspond to an average element size of $h=0.0317$, using MLS cubic interpolation for gradient approximation inside trimmed elements.

\subsubsection{FEM Discretization with quadratic quadrilateral elements}

The performance of the strong-like Dirichlet BCs enforcement is analyzed for unfitted quadratic quadrilateral discretizations. The discretization employs 9-node quadratic quadrilateral elements.

\begin{figure}[h!]
    \centering
    \includegraphics[width=0.8\textwidth]{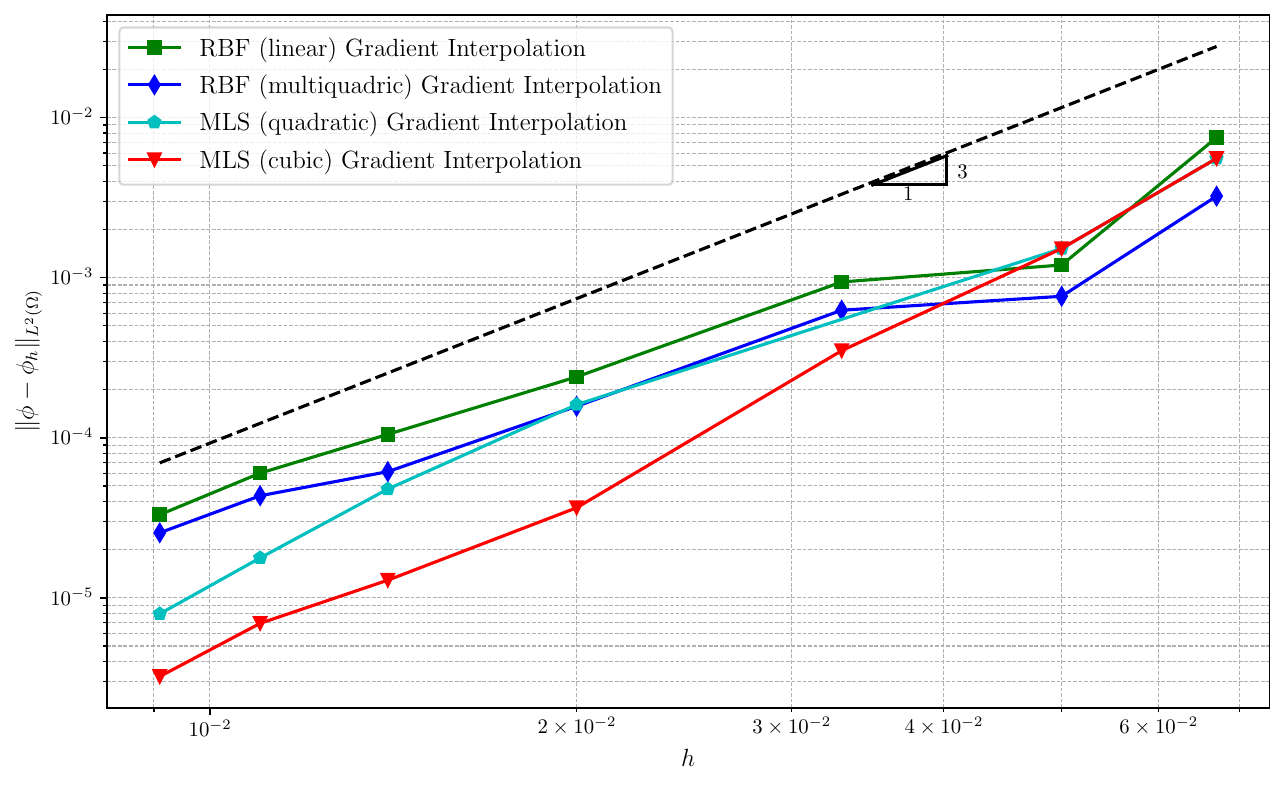}
    \caption{Comparison of $L^2$-norm error convergence for 9-node quadratic quadrilateral elements under different gradient approximation techniques. The error is computed for the manufactured solution $\phi(x,y) = \sin(\pi x) \cos(\pi y)$ using an unfitted boundary mesh. The study evaluates four gradient interpolation methods: RBF with linear basis (blue triangles), RBF with multiquadric basis (green triangles), MLS with quadratic basis (red squares), and MLS with cubic basis (cyan circles). The dashed black line represents the theoretical third-order convergence rate ($\mathcal{O}(h^2)$).}
    \label{fig:L2_error_quadratic_quad}
\end{figure}

\begin{figure}[h!]
    \centering
    \includegraphics[width=0.8\textwidth]{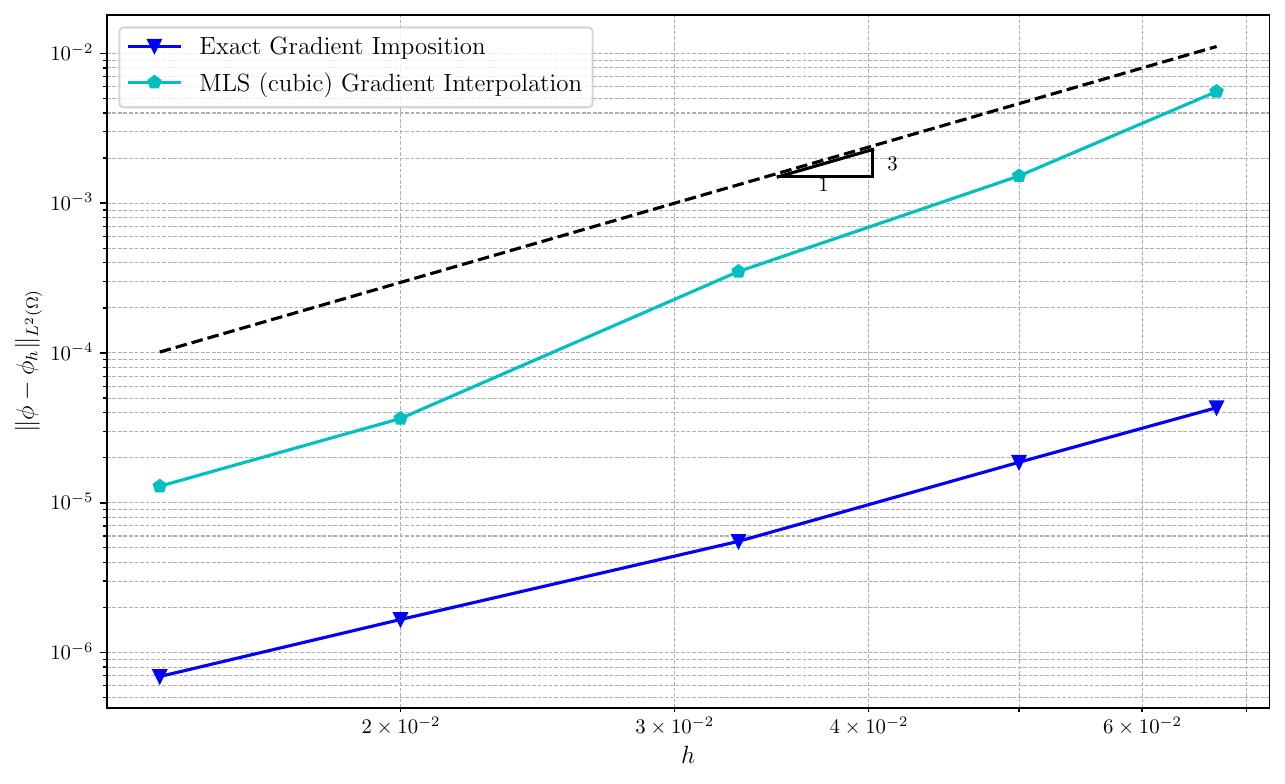}
    \caption{Comparison of $L^2$-norm error convergence for 9-node quadratic quadrilateral elements discretization. The study compares the $L^2$-norm error using MLS with cubic basis functions gradient approximation (cyan pentagons) and and the $L^2$-norm error imposing the exact gradient in the trimmed elements (blue triangles). The dashed black line represents the theoretical third-order convergence rate ($\mathcal{O}(h^3)$).}
    \label{fig:L2_error_quadratic_quad_exact_grad}
\end{figure}

Figs.~(\ref{fig:L2_error_quadratic_quad}) and (\ref{fig:L2_error_quadratic_quad_exact_grad}) present the $L^2$-norm error convergence for different gradient approximation strategies. The study evaluates again MLS interpolation, RBF-based methods with linear and multiquadric basis functions, and direct exact gradient enforcement. The convergence behaviour differs among the tested methods. MLS cubic interpolation exhibits optimal third-order convergence ($\mathcal{O}(h^3)$), consistent with theoretical expectations for quadratic FEM discretizations. In contrast, the RBF-based approaches show a slower error decay, indicating slightly suboptimal convergence. This suggests that while MLS effectively reconstructs the gradient field, the accuracy of RBF interpolation is more sensitive to the choice of basis function. These results highlight the superior performance of MLS, which stems from its local minimization-based formulation. 

The comparison between MLS cubic interpolation and exact gradient imposition (Fig.~\ref{fig:L2_error_quadratic_quad_exact_grad}) reveals a greater discrepancy than in the previously analyzed cases. While MLS maintains the expected third-order convergence, its absolute error remains higher than that of direct gradient enforcement. This suggests that for quadratic quadrilateral elements, the MLS approach introduces a larger numerical error, possibly due to increased sensitivity in gradient reconstruction when higher-order approximations are used.

\begin{figure}[h!]
    \centering
    \begin{subfigure}{0.47\textwidth} 
        \centering
        \includegraphics[width=\linewidth]{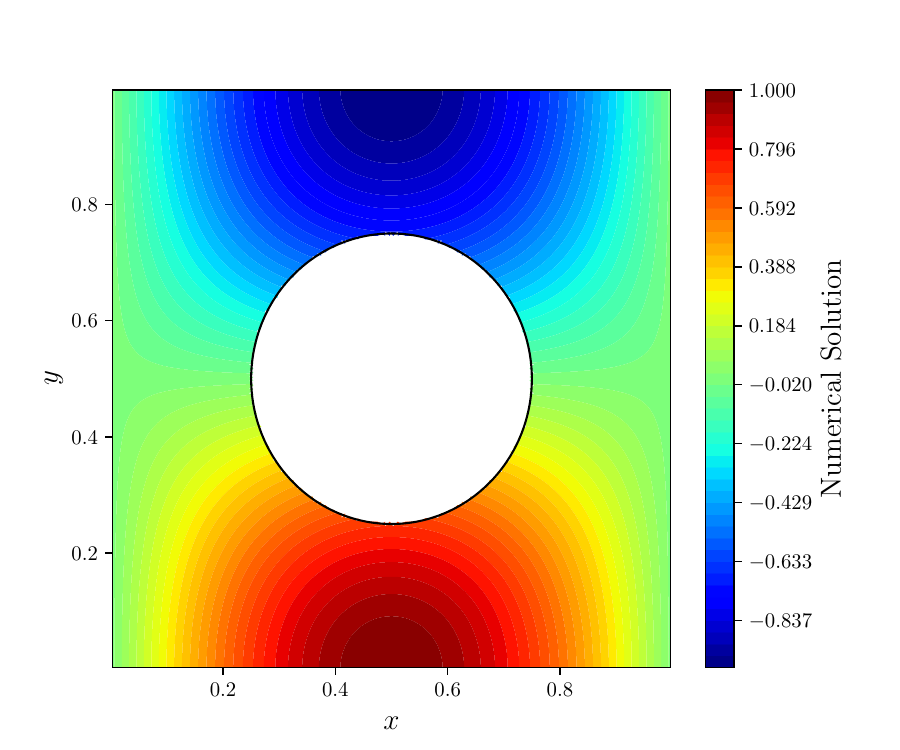}
        \caption{Numerical solution}
        \label{fig:plate_hole_numerical_quad}
    \end{subfigure}
    \hfill 
    \begin{subfigure}{0.47\textwidth} 
        \centering
        \includegraphics[width=\linewidth]{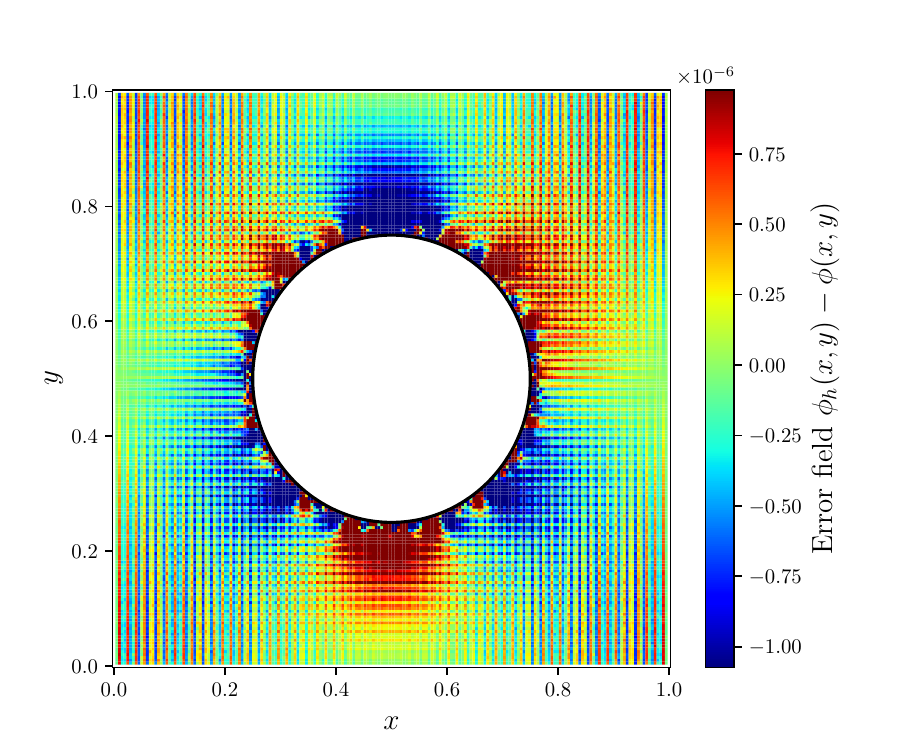}
        \caption{Error distribution}
        \label{fig:plate_hole_error_quad}
    \end{subfigure}
    \caption{Comparison of the numerical solution and the corresponding error distribution for the plate with a hole problem using 9-node quadratic quadrilateral elements discretization. (a) Computed numerical solution. (b) Error distribution, illustrating the deviation of the numerical solution from the exact solution.}
    \label{fig:plate_hole_comparison_quadratic_quad}
\end{figure}

Finally, Fig.~\ref{fig:plate_hole_comparison_quadratic_quad} provides a visualization of the numerical solution and error distribution for the Poisson problem, computed with an average element size of $h=0.014$.

\subsection{IGA discretizations}

In this section, we present the $L^2$-norm error convergence study for B-Spline basis functions, which form the foundation of Isogeometric Analysis (IGA) \cite{hughes2005isogeometric}. Rather than revisiting the mathematical formulation of B-Splines, typically defined via the Cox–de Boor recursion formula \cite{cox1972numerical, deboor1972calculating}, we refer the reader to the standard reference The NURBS Book by Piegl and Tiller \cite{piegl2012nurbs}, which offers a thorough and accessible introduction to the theory and practical aspects of B-Splines and NURBS.

\subsubsection{Quadratic IGA discretizations}

This section studies the accuracy and convergence properties of the proposed strong Dirichlet BCs enforcement technique in unfitted (trimmed) quadratic B-Spline discretizations.

\begin{figure}[h!]
    \centering
    \includegraphics[width=0.8\textwidth]{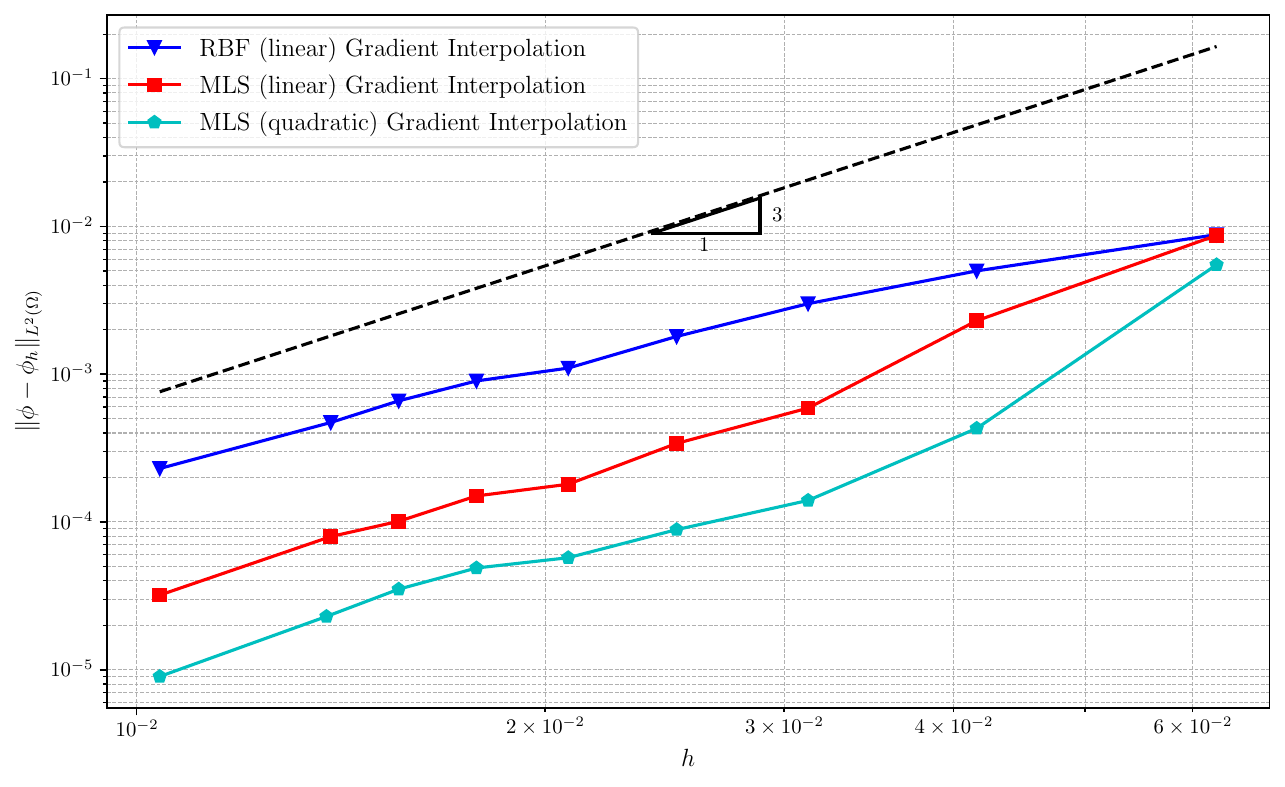}
    \caption{Comparison of $L^2$-norm error convergence for a quadratic IGA discretization under different gradient approximation techniques. The error is computed for the manufactured solution $\phi(x,y) = \sin(\pi x) \cos(\pi y)$ using an unfitted boundary mesh. The study evaluates three gradient interpolation methods: RBF with linear basis (blue triangles), MLS with linear basis (red squares), and MLS with quadratic basis (cyan circles). The dashed black line represents the theoretical third-order convergence rate ($\mathcal{O}(h^3)$).}
    \label{fig:L2_error_quadratic_IGA}
\end{figure}

\begin{figure}[h!]
    \centering
    \includegraphics[width=0.8\textwidth]{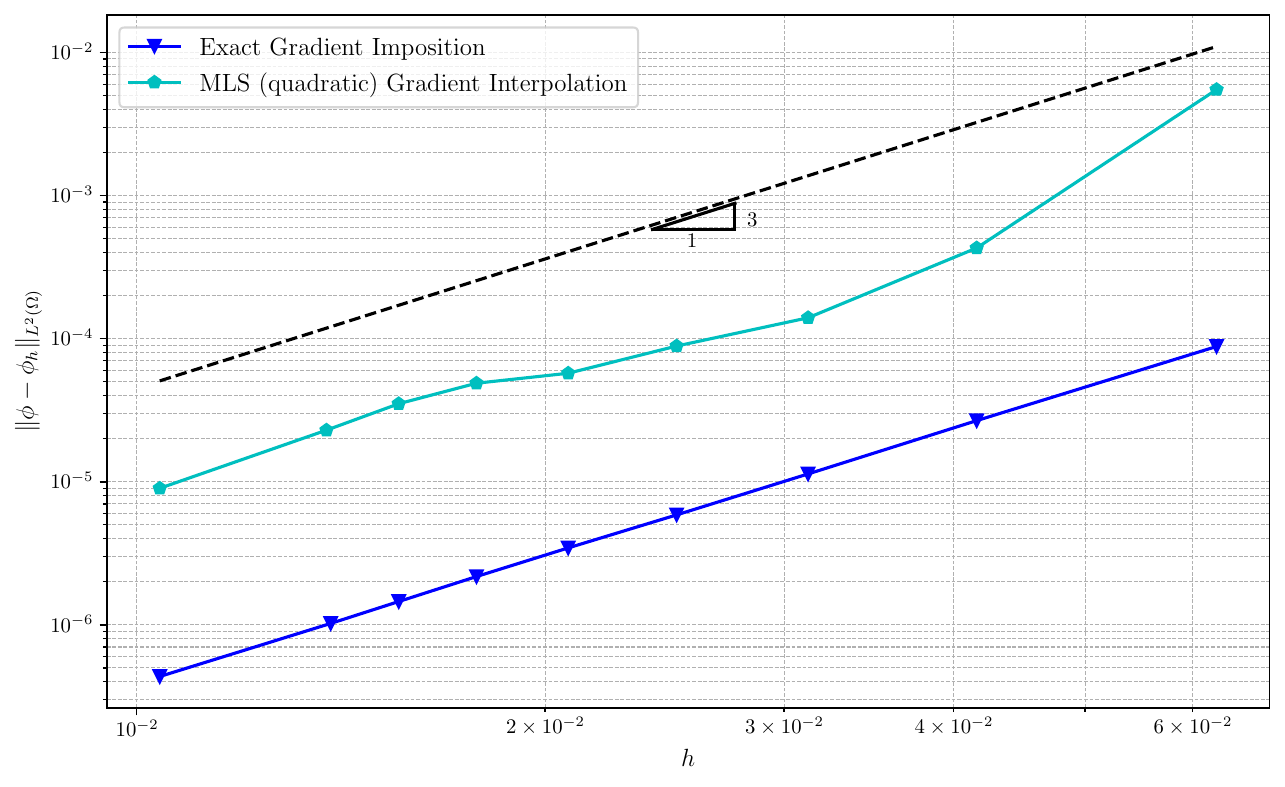}
    \caption{Comparison of $L^2$-norm error convergence for a quadratic B-Spline discretization. The study compares the $L^2$-norm error using MLS with quadratic basis functions gradient approximation (cyan pentagons) and and the $L^2$-norm error imposing the exact gradient in the trimmed elements (blue triangles). The dashed black line represents the theoretical third-order convergence rate ($\mathcal{O}(h^3)$).}
    \label{fig:L2_error_quadratic_IGA_exact_grad}
\end{figure}

Figs. (\ref{fig:L2_error_quadratic_IGA}) and (\ref{fig:L2_error_quadratic_IGA_exact_grad}) present the $L^2$-norm error convergence for different gradient approximation strategies.
In the section discussing FEM discretizations, we demonstrated that the accuracy gain from using a cubic basis in MLS, as opposed to a quadratic basis, is negligible, while the computational cost of cubic basis functions is significantly higher. Given this observation, in this section, we reassess MLS interpolation using linear and quadratic basis, RBF-based methods with linear basis, and direct exact gradient enforcement. 

The convergence behavior varies among the tested methods. Both MLS quadratic and MLS linear interpolations exhibit the expected third-order convergence rate ($\mathcal{O}(h^3)$), in agreement with theoretical predictions for quadratic IGA discretizations. In contrast, the RBF-based approach shows a slower error decay, indicating suboptimal convergence. This suggests that while MLS effectively reconstructs the gradient field, the reduced accuracy of RBF interpolation is likely due to the global nature of the interpolation, which can introduce approximation errors and reduced stability, particularly in regions with non-uniform point distributions. It is worth to mention that in this case, among the available RBF methods, only the linear basis ($\phi(r)=r$) demonstrated proper accuracy in gradient reconstruction inside trimmed elements, highlighting its relative robustness in this specific setting.

Analyzing the results of gradient interpolation using quadratic MLS and exact gradient enforcement (Fig.~\ref{fig:L2_error_quadratic_IGA_exact_grad}) reveals some difference in accuracy. While MLS achieves the expected third-order convergence, its error remains  larger than that of enforcing the exact gradient. This suggests that for quadratic IGA elements, the gradient approximation process in MLS introduces greater inaccuracies compared to lower-order discretizations.  

It is important to highlight that the interpolation points used for gradient reconstruction belong to the so-called unperturbed elements (Def. \ref{def:unperturbed_element}),  where no degrees of freedom are modified by this method. In the case of IGA with a quadratic discretization ($p=2$), where a basis function spans over $p+1=3$ knot spans, this implies that the gradient information is extracted from interpolation points located at a distance of approximately $\mathcal{O}(3h)$, with $h$ being the knot span size. Conversely, in the FEM case, where shape function support is strictly local within each element, the gradient information used for interpolation is, in the worst-case scenario, at a distance of only $\mathcal{O}(2h)$ from the evaluation point. This fundamental difference in gradient information clearly contributes to the observed accuracy discrepancies between MLS-based and exact gradient enforcement methods.

\begin{figure}[h!]
    \centering
    \begin{subfigure}{0.47\textwidth} 
        \centering
        \includegraphics[width=\linewidth]{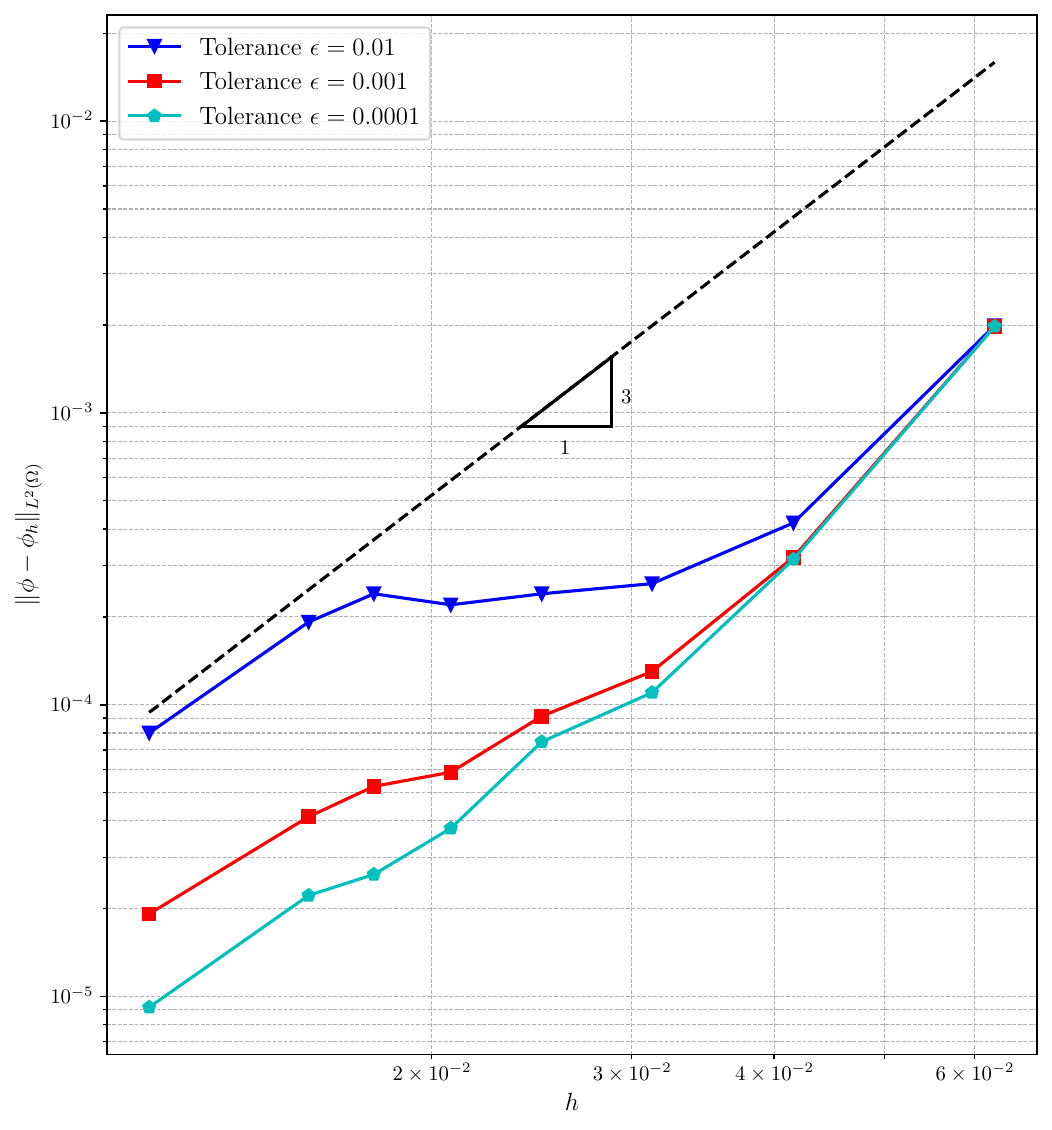}
        \caption{$L^2$-norm error convergence plot for different algorithm tolerances}
        \label{fig:plate_hole_quad_IGA_L2_diff_tol}
    \end{subfigure}
    \hfill 
    \begin{subfigure}{0.455\textwidth} 
        \centering
        \includegraphics[width=\linewidth]{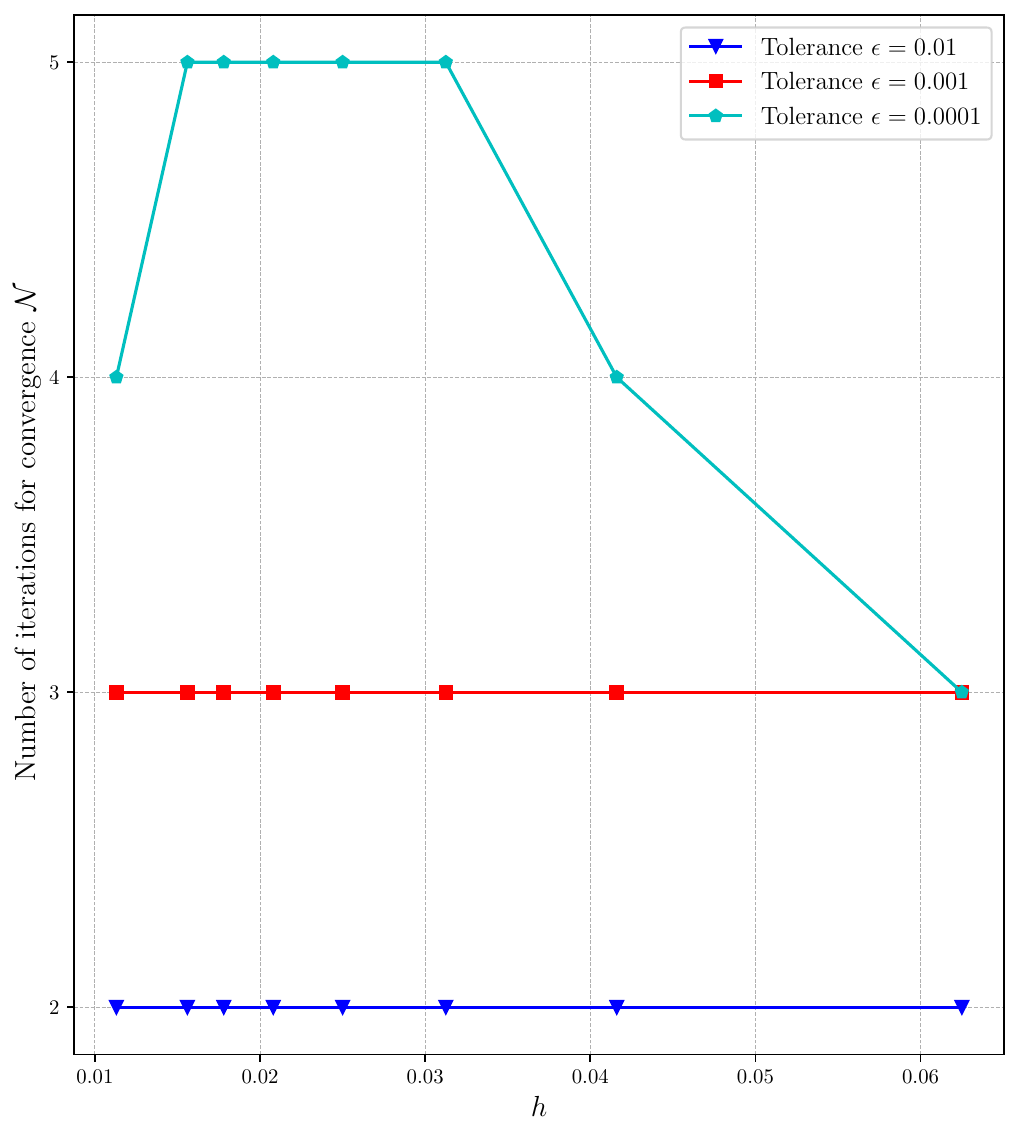}
        \caption{Number of iterations $\mathcal{N}$ for convergence}
        \label{fig:plate_hole_quad_IGA_iterations}
    \end{subfigure}
   \caption{Analysis of solver accuracy and convergence behavior for the embedded circle geometry using quadratic IGA discretization. (a) Convergence of the $L^2$-norm error for different iterative solver tolerances. (b) Required number of iterations $\mathcal{N}$ to achieve convergence for varying solver tolerances.}
    \label{fig:plate_with_a_hole_diff_tolerances}
\end{figure}

An important aspect to analyze is how the $L^2$-norm error varies with the iterative solver tolerance 
$\epsilon$ and how the number of iterations needed for convergence depends on the mesh size $h$ for different target accuracies. The corresponding results are presented in Fig.~\ref{fig:plate_with_a_hole_diff_tolerances} for the embedded circle case. From these plots, it can be observed that reducing the iterative solver tolerance \( \epsilon \) leads to a decrease in the $L^2$-norm error, as expected. However, this reduction in tolerance also results in a higher number of iterations \( \mathcal{N} \) required for convergence. As in most engineering problems, an optimal balance must be struck between computational cost and the required accuracy for a given application. In this particular case, a tolerance of \( \epsilon = 0.001 \) provides satisfactory results. It should also be noted that only a few iterations are typically needed to reach convergence, highlighting the efficiency of the proposed algorithm. Additionally, the first iteration is always performed under the assumption of a zero normal gradient within each trimmed element ($\nabla\phi \cdot \mathbf{n}=0$), so there is no associated computational cost for the gradient approximation inside the trimmed elements during this step.

\begin{figure}[h!]
    \centering
    \begin{subfigure}{0.47\textwidth} 
        \centering
        \includegraphics[width=\linewidth]{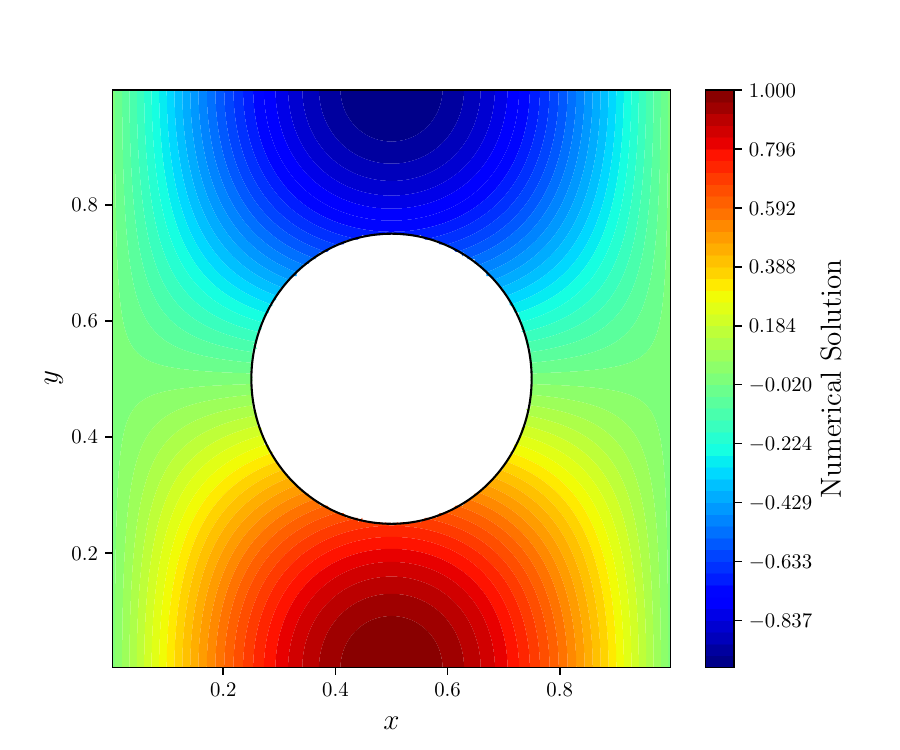}
        \caption{Numerical solution}
        \label{fig:plate_hole_quad_IGA_numerical_solution}
    \end{subfigure}
    \hfill 
    \begin{subfigure}{0.47\textwidth} 
        \centering
        \includegraphics[width=\linewidth]{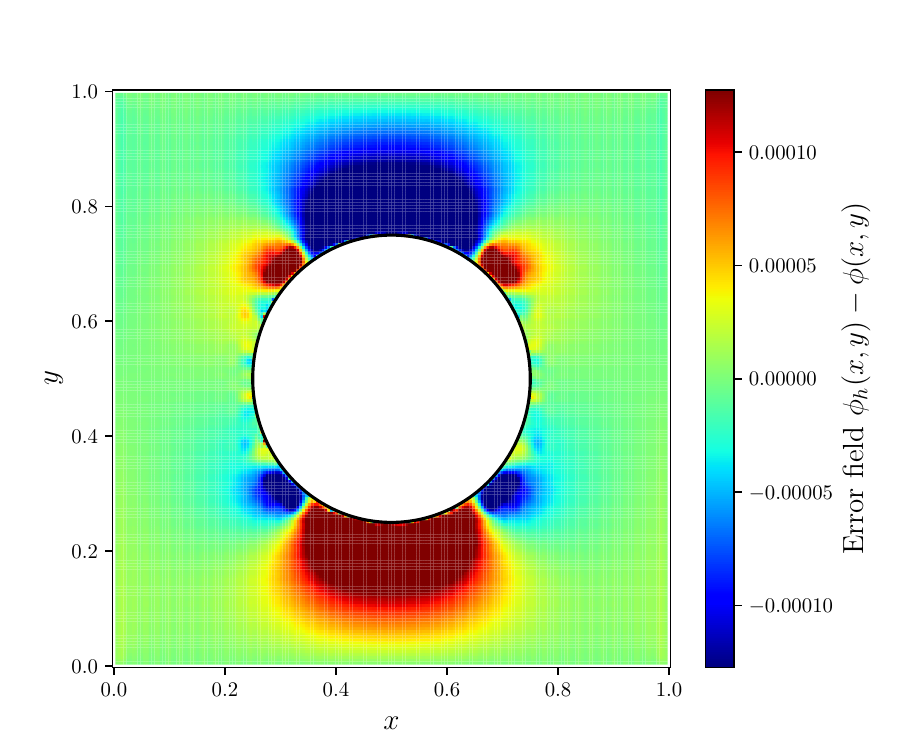}
        \caption{Error distribution}
        \label{fig:plate_hole_quad_IGA_error_dist_plot}
    \end{subfigure}
    \caption{Comparison of the numerical solution and the corresponding error distribution for the plate with a hole problem using quadratic IGA discretization. (a) Computed numerical solution. (b) Error distribution, illustrating the deviation of the numerical solution from the exact solution.}
    \label{fig:plate_hole_comparison_quad_IGA}
\end{figure}

Finally, Fig.~\ref{fig:plate_hole_comparison_quad_IGA} provides a visualization of the numerical solution and error distribution for the Poisson problem, computed with an  element size of $h=0.021$.

\subsubsection{Cubic IGA discretizations}

In this section, we present an evaluation of our developed method for strongly imposing Dirichlet BCs, focusing on its performance in terms of accuracy and convergence when applied to unfitted (trimmed) cubic B-Spline discretizations.

\begin{figure}[h!]
    \centering
    \includegraphics[width=0.80\textwidth]{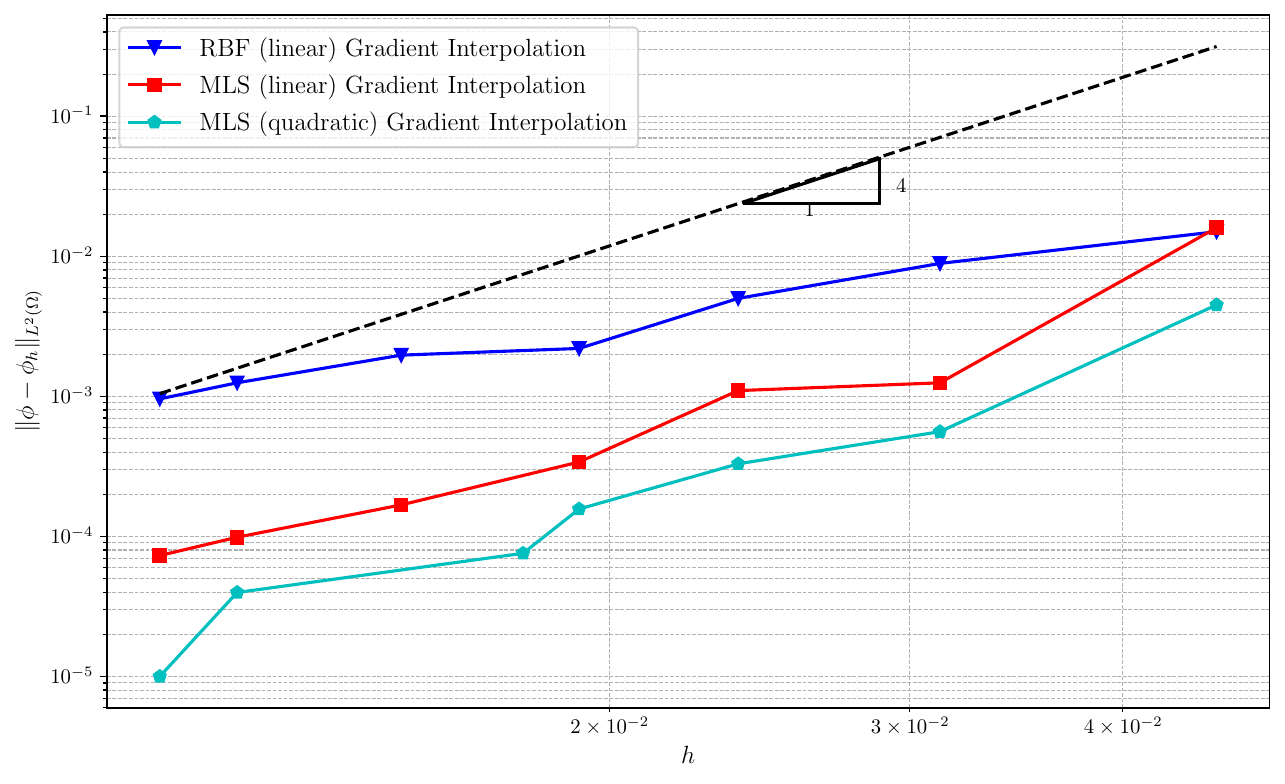}
    \caption{Comparison of $L^2$-norm error convergence for a cubic IGA discretization under different gradient approximation techniques. The error is computed for the manufactured solution $\phi(x,y) = \sin(\pi x) \cos(\pi y)$ using an unfitted boundary mesh. The study evaluates three gradient interpolation methods: RBF with linear basis (blue triangles), MLS with linear basis (red squares), and MLS with quadratic basis (cyan circles). The dashed black line represents the theoretical fourth-order convergence rate ($\mathcal{O}(h^4)$).}
    \label{fig:L2_error_cubic_IGA}
\end{figure}

\begin{figure}[h!]
    \centering
    \includegraphics[width=0.8\textwidth]{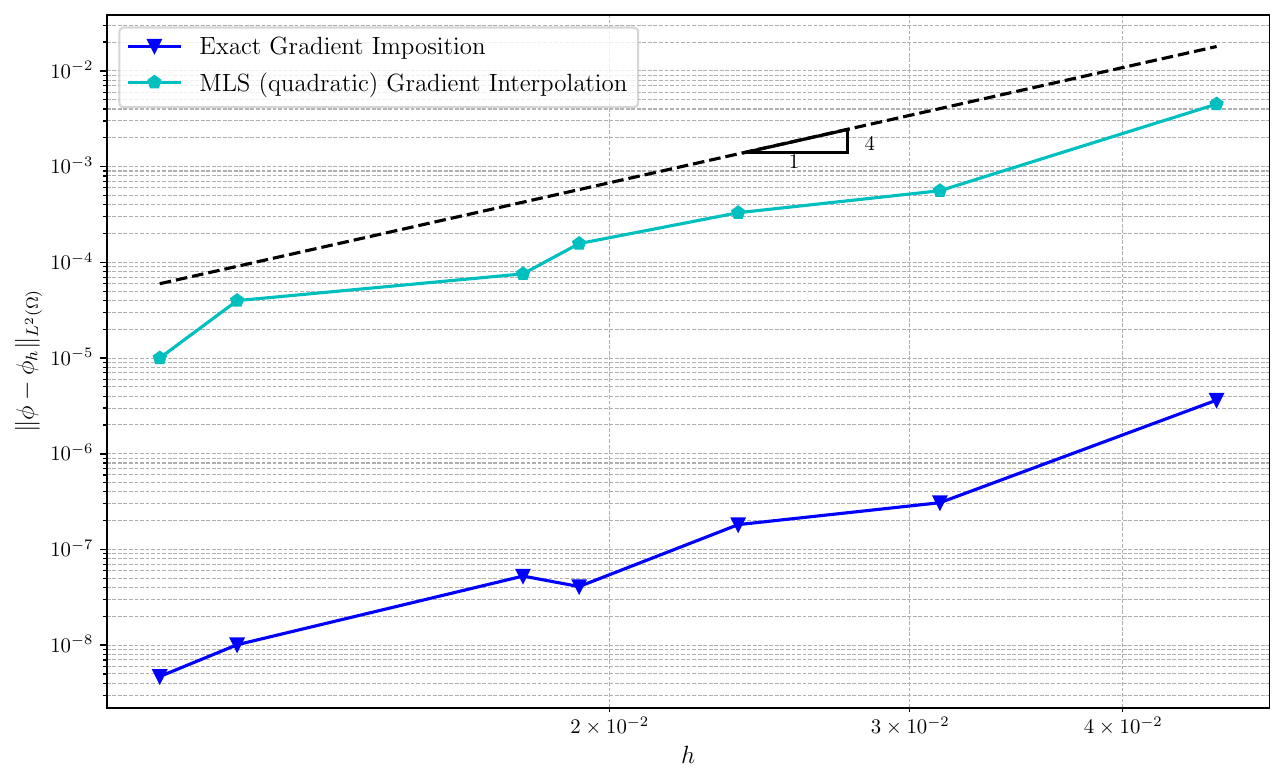}
    \caption{Comparison of $L^2$-norm error convergence for a cubic B-Spline discretization. The study compares the $L^2$-norm error using MLS with quadratic basis functions gradient approximation (cyan pentagons) and and the $L^2$-norm error imposing the exact gradient in the trimmed elements (blue triangles). The dashed black line represents the theoretical third-order convergence rate ($\mathcal{O}(h^4)$).}
    \label{fig:L2_error_cubic_IGA_exact_grad}
\end{figure}

Figs. (\ref{fig:L2_error_cubic_IGA}) and (\ref{fig:L2_error_cubic_IGA_exact_grad}) present the $L^2$-norm error convergence for different gradient approximation strategies. Among the methods evaluated, convergence characteristics differ noticeably. Both the quadratic MLS and linear MLS interpolation techniques achieve a fourth-order convergence rate ($\mathcal{O}(h^4)$), aligning with theoretical expectations for cubic IGA discretizations. In contrast, the RBF-based method demonstrates a slower reduction in error, suggesting that its convergence performance is suboptimal. Once again, the quadratic MLS method delivers the best accuracy, underscoring its robustness as the most reliable approach for gradient approximation.

An analysis of the gradient interpolation results using quadratic MLS versus exact gradient enforcement (see Fig.~\ref{fig:L2_error_cubic_IGA_exact_grad}) indicates a noticeable accuracy difference. Although the MLS method achieves the anticipated 4th-order convergence ($\mathcal{O}(h^4)$), its error level remains higher than that obtained with exact gradient enforcement. This is once again attributed to the fact that the gradient information for interpolation is sourced from the so-called unperturbed elements, which, for cubic discretizations, are located approximately $4h$ away from the interpolation point.

{\color{black} An analysis of the gradient interpolation results using quadratic MLS versus exact gradient enforcement (Fig.~\ref{fig:L2_error_cubic_IGA_exact_grad}) indicates a noticeable accuracy gap. Although the MLS method achieves the anticipated $\mathcal{O}(h^4)$ convergence rate, its absolute error remains consistently higher than that obtained with exact gradient enforcement. This is attributed to the fact that the gradient information for interpolation is sourced exclusively from the so-called unperturbed elements, which, for cubic discretizations, are located approximately $4h$ away from the interpolation point (see \ref{def:unperturbed_element}). This distance amplifies the approximation error in trimmed regions.

At present, no specific countermeasure has been incorporated to mitigate this loss of accuracy for high-order B-Splines discretizations; addressing this will be part of future work. We note, however, that the impact is significantly reduced for lower-order elements (including quadratic), for which our method achieves accuracy comparable to established techniques.
}

\begin{figure}[!tb]
    \centering
    \begin{subfigure}{0.47\textwidth} 
        \centering
        \includegraphics[width=\linewidth]{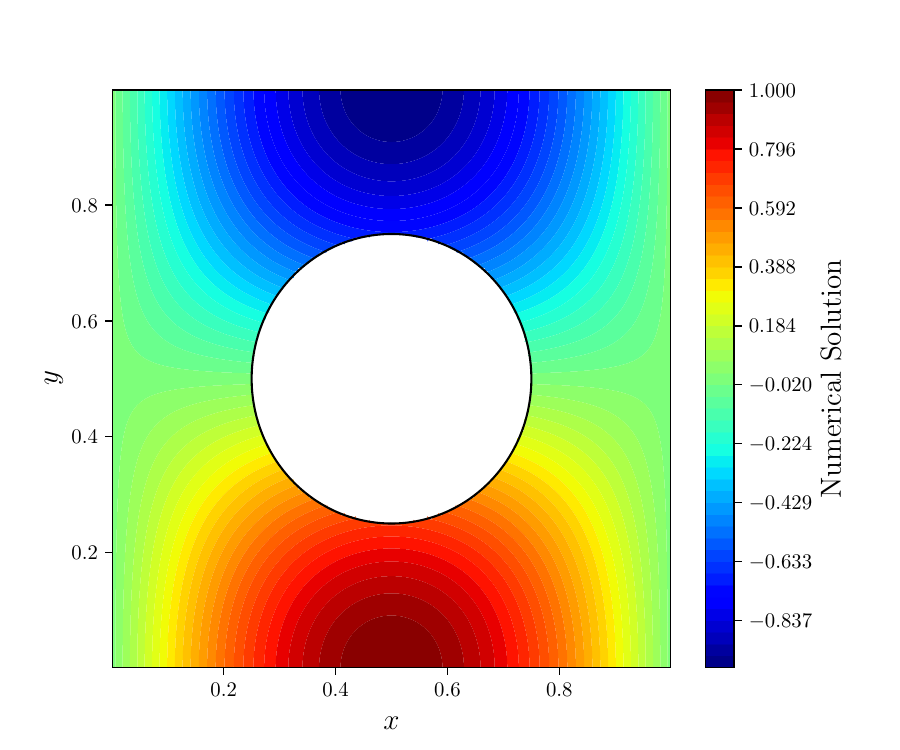}
        \caption{Numerical solution}
        \label{fig:plate_hole_cubic_IGA_numerical_solution}
    \end{subfigure}
    \hfill 
    \begin{subfigure}{0.47\textwidth} 
        \centering
        \includegraphics[width=\linewidth]{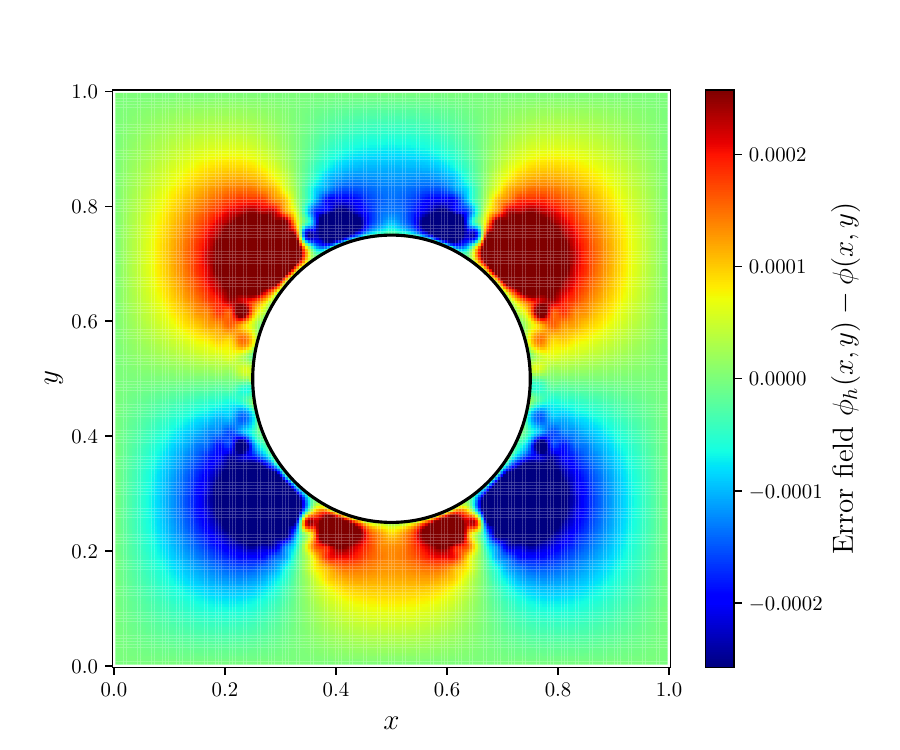}
        \caption{Error distribution quadratic MLS as the gradient approximator}
        \label{fig:plate_hole_cubic_IGA_error_plot}
    \end{subfigure}

    \begin{subfigure}{0.5\textwidth}
        \centering
        \includegraphics[width=\linewidth]{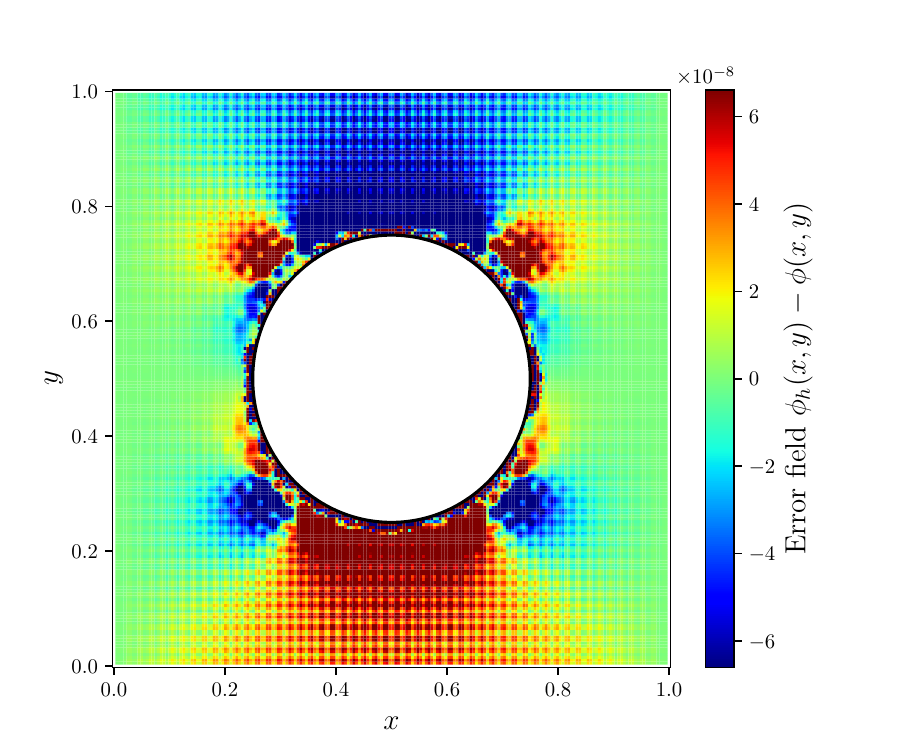}
        \caption{Error distribution for exact gradient imposition}
        \label{fig:plate_hole_cubic_IGA_error_plot_exact_grad}
    \end{subfigure}
    
    \caption{Comparison of the numerical solution and the corresponding error distribution for the plate with a hole problem using cubic IGA discretization. (a) Computed numerical solution. (b) Error distribution, illustrating the deviation of the numerical solution from the exact solution. c) Error distribution, for exact gradient imposition}
    \label{fig:plate_hole_comparison_cubic_IGA}
\end{figure}

As a final illustration, Fig.~\ref{fig:plate_hole_comparison_cubic_IGA} presents the numerical solution and error distribution for the Poisson problem using quadratic MLS gradient approximation, alongside the error distribution plot for exact gradient enforcement, computed with an element size of $h=0.019$.

\subsection{Comparison of the proposed strategy, IBRA and SBM}

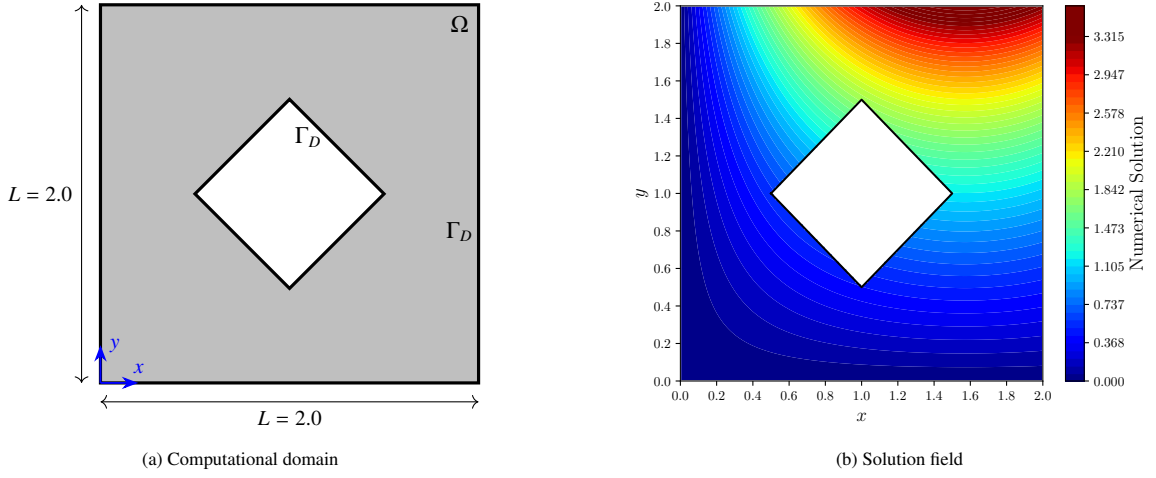
\begin{figure}[h]
    \centering
    \begin{subfigure}{0.47\textwidth}
        \centering
        \begin{tikzpicture}[scale=2.5] 

            \fill[gray!50] (0,0) rectangle (2,2);  
            \fill[white] (0.5,1) -- (1,0.5) -- (1.5,1) -- (1,1.5) -- cycle;  
        
            \draw[very thick,black] (0,0) rectangle (2,2); 
            \draw[very thick,black] (0.5,1) -- (1,0.5) -- (1.5,1) -- (1,1.5) -- cycle; 
        
            \draw[-{Stealth[blue]}, line width = 0.25mm, color=blue] (0,0.0)--(0.2,0.0) node[above]{\footnotesize$x$};
            \draw[-{Stealth[blue]}, line width = 0.25mm, color=blue] (0.0,0.0)--(0.0,0.2) node[right]{\footnotesize$y$};
        
            \node at (1.9, 1.9) {\small $\Omega$};
            \node at (1.1, 1.3) {\small $\Gamma_D$};
            \node at (1.9, 0.8) {\small $\Gamma_D$};
        
            \draw[thin,<->] (0,-0.1) -- (2,-0.1) node[midway,below]{\footnotesize $L=2.0$};
            \draw[thin,<->] (-0.1,0) -- (-0.1,2) node[midway,left]{\footnotesize $L=2.0$};
        
        \end{tikzpicture}
        \caption{Computational domain}
        \label{fig:comp_domain_rotated_square}
    \end{subfigure}
    \hfill
    \begin{subfigure}{0.47\textwidth}
        \centering
        \includegraphics[width=\linewidth]{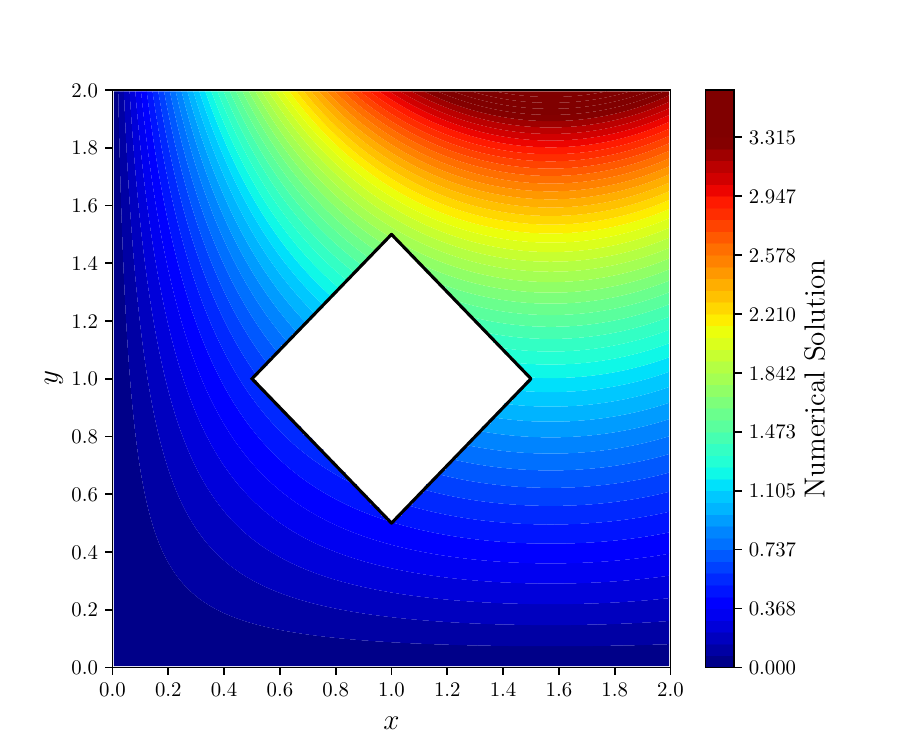}
        \caption{Solution field}
        \label{fig:solution_field_rotated_square}
    \end{subfigure}
    \caption{Computational setup and solution field for the problem defined by the field 
    \(\phi(x,y) = \sin(x) \sinh(y)\). (a) The computational domain consists of a square domain 
    \(\Omega = [0,2] \times [0,2]\) with an embedded quadrilateral hole where Dirichlet BCs
    are imposed. (b) Solution field \(\phi(x,y)\).}
    \label{fig:embedded_square_problem_definition}
\end{figure}

In this section, we compare our proposed method with the Isogeometric B-Rep Analysis (IBRA) methodology \cite{Breitenberger2015, Teschemacher2018, Teschemacher2022} and the Shifted Boundary Method (SBM) \cite{Main2018a, Main2018b, Antonelli2024}, evaluating both the accuracy in terms of the $L^2$-norm error and the conditioning of the resulting linear system. To this end, we consider the geometry shown in Fig.~\ref{fig:comp_domain_rotated_square} and prescribe the following manufactured solution of the Poisson problem as a Dirichlet boundary condition on the entire boundary, $\Gamma = \Gamma_D$:

\begin{equation}
    \phi(x,y) = \sin(x) \sinh(y)
    \label{eq:manufactured_solution_2}
\end{equation}

\noindent where the corresponding source term is identically zero throughout the domain, i.e., $\sigma(x,y) = 0$.

The primary distinction between the methods lies in how each handles the imposition of boundary conditions. Our approach employs a non-intrusive framework that enforces Dirichlet boundary conditions strongly by solving an auxiliary algebraic problem, independent from the physics-based weak form. This strategy not only simplifies implementation but also avoids the need for integrating over trimmed knot spans, as the stabilization term is computed across the entire intersected element.

\begin{figure}[!tb]
    \centering
    \includegraphics[width=0.4\linewidth]{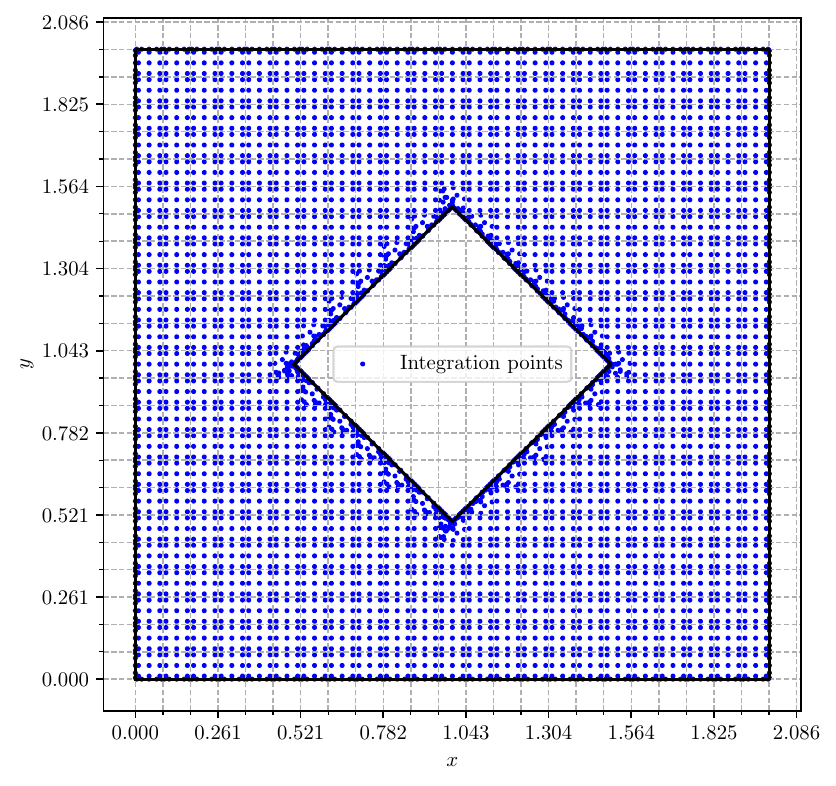}
  \caption{Distribution of IBRA integration points for quadratic B-splines, with 9 integration points per knot span. The plot is shown in the parametric space $(\xi, \eta)$, which in this case coincides with the physical domain $(x, y)$. A high concentration of integration points is observed near the trimmed boundary, where triangulation is applied to enforce accurate integration in cut elements.}
    \label{fig:integration_points_IBRA}
\end{figure}

In contrast, the Isogeometric B-Rep Analysis (IBRA) imposes Dirichlet boundary conditions weakly using techniques such as penalty methods, Lagrange multipliers, or Nitsche’s method. IBRA maintains optimality by triangulating and integrating the active portions of cut knot spans, as illustrated in Fig.~\ref{fig:integration_points_IBRA}. This often leads to a high concentration of integration points near trimmed boundaries, where accurate numerical integration is achieved through fine triangulation. To address the computational burden and potential over-integration in these regions, reduced integration techniques using fewer Gauss points have been proposed \cite{Messmer2022}. However, the presence of very small trimmed elements can introduce the so-called “small-cut cell instability”, which may significantly degrade the conditioning of the resulting linear systems. In the following examples, IBRA employs a penalty-free Nitsche method for boundary condition enforcement, as introduced in \cite{Collins2023} and further validated in \cite{Antonelli2024}.

The Shifted Boundary Method (SBM), on the other hand, avoids direct integration over trimmed elements by shifting the boundary conditions from the true boundary to a surrogate boundary located entirely within the computational domain. This enables the use of standard integration schemes, sidestepping the geometric complexities of cut elements. Dirichlet conditions are weakly enforced on the surrogate boundary using techniques like Nitsche's method.

\begin{figure}[!tb]
    \centering
    \begin{subfigure}{0.47\textwidth} 
        \centering
        \includegraphics[width=\linewidth]{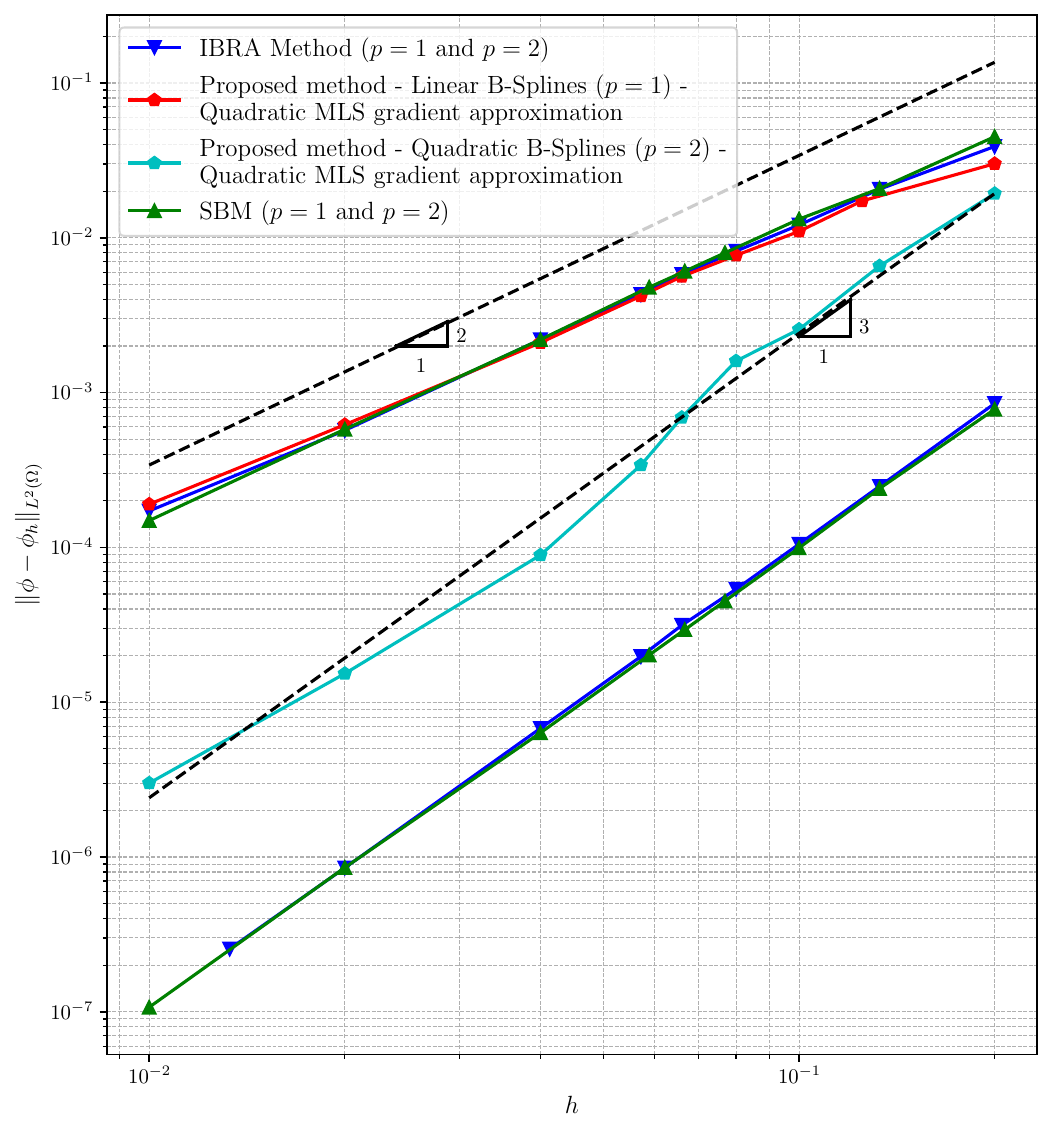}
        \caption{Comparison of the $L^2$-norm error between IBRA, SBM and the proposed method}
        \label{fig:plate_square_hole_p1_p2_error_IBRA_comparison}
    \end{subfigure}
    \hfill 
    \begin{subfigure}{0.47\textwidth} 
        \centering
        \includegraphics[width=\linewidth]{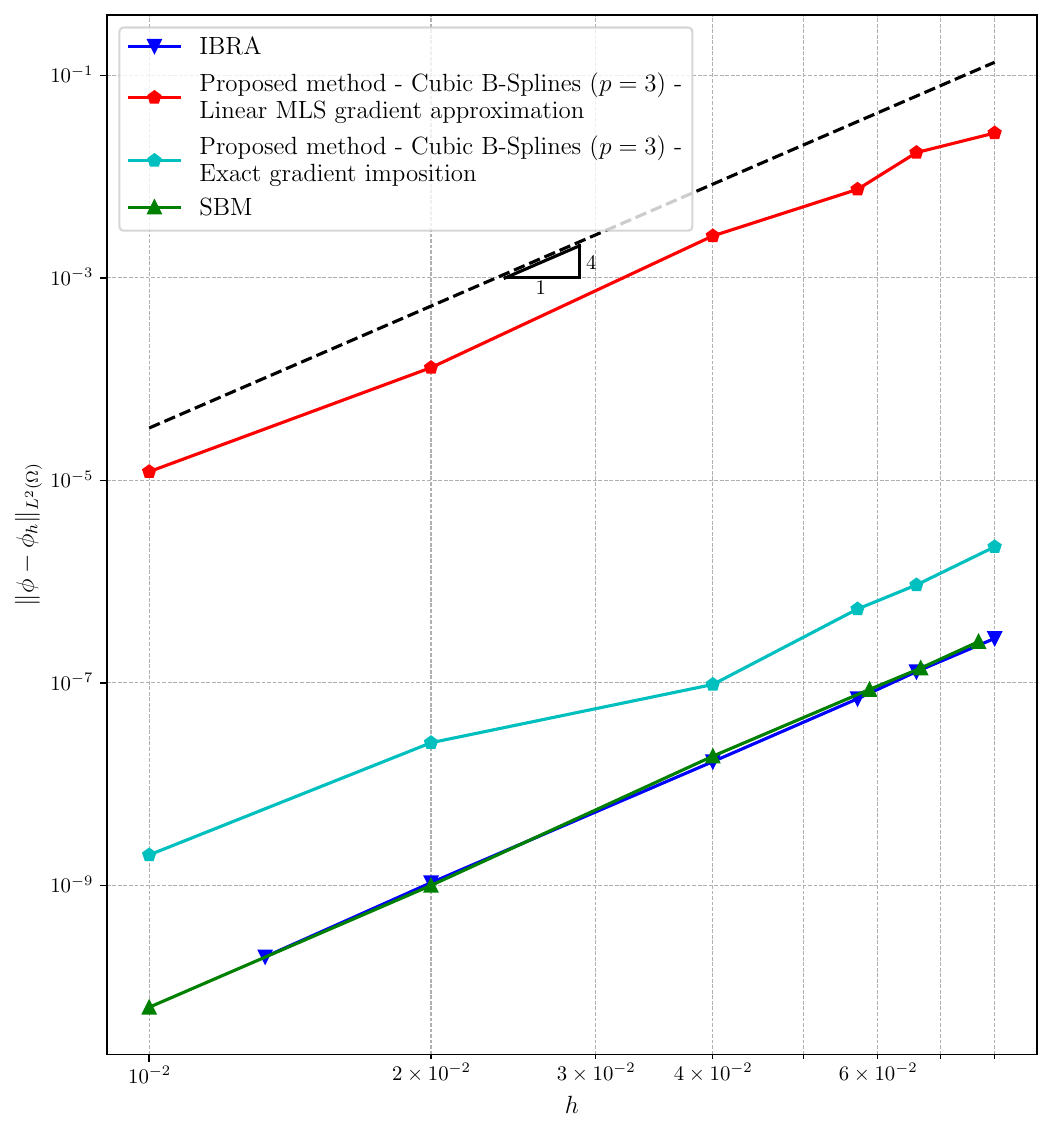}
        \caption{Comparison of the $L^2$-norm error between IBRA, SBM and the proposed method}
        \label{fig:plate_square_hole_p3_error_IBRA_comparison}
    \end{subfigure}

   \caption{Error convergence comparison between IBRA, SBM and the proposed approach. 
    (a) Convergence results for IBRA, SBM and the proposed method using linear (\(p=1\)) and quadratic (\(p=2\)) B-Spline discretizations. In the proposed method, the gradient is approximated using a quadratic MLS scheme.  
    (b) Convergence results for IBRA, SBM and the proposed method using cubic B-Spline discretizations (\(p=3\)). The proposed method is evaluated using both a linear MLS gradient approximation and exact gradient imposition.}
    \label{fig:plate_square_hole_p1_p2_p3_error_IBRA_comparison}
\end{figure}

In Figs. (\ref{fig:plate_square_hole_p1_p2_error_IBRA_comparison}) and (\ref{fig:plate_square_hole_p3_error_IBRA_comparison}), we compare the three approaches for the embedded square computational domain in terms of the $L^2$-norm error.  
For degree $p=1$, the three methods yield very similar results in terms of the $L^2$-norm error. However, for degree $p=2$, IBRA and SBM exhibit better accuracy than the proposed approach, likely due to inaccuracies introduced by the gradient approximation within the trimmed elements. A similar trend is observed for degree $p=3$.  
Interestingly, as shown in Fig.~\ref{fig:plate_square_hole_p3_error_IBRA_comparison}, when the exact gradient is imposed in the trimmed elements, the accuracy of the proposed method remains slightly lower than that of IBRA and SBM, but the difference in accuracy is significantly reduced. This suggests that the method remains optimal in terms of the $L^2$-norm error, with the loss of accuracy primarily originating from the gradient reconstruction in the trimmed elements. This issue becomes increasingly complex for higher polynomial degrees.  

\begin{figure}[!tb]
    \centering
    \begin{subfigure}{0.455\textwidth} 
        \centering
        \includegraphics[width=\linewidth]{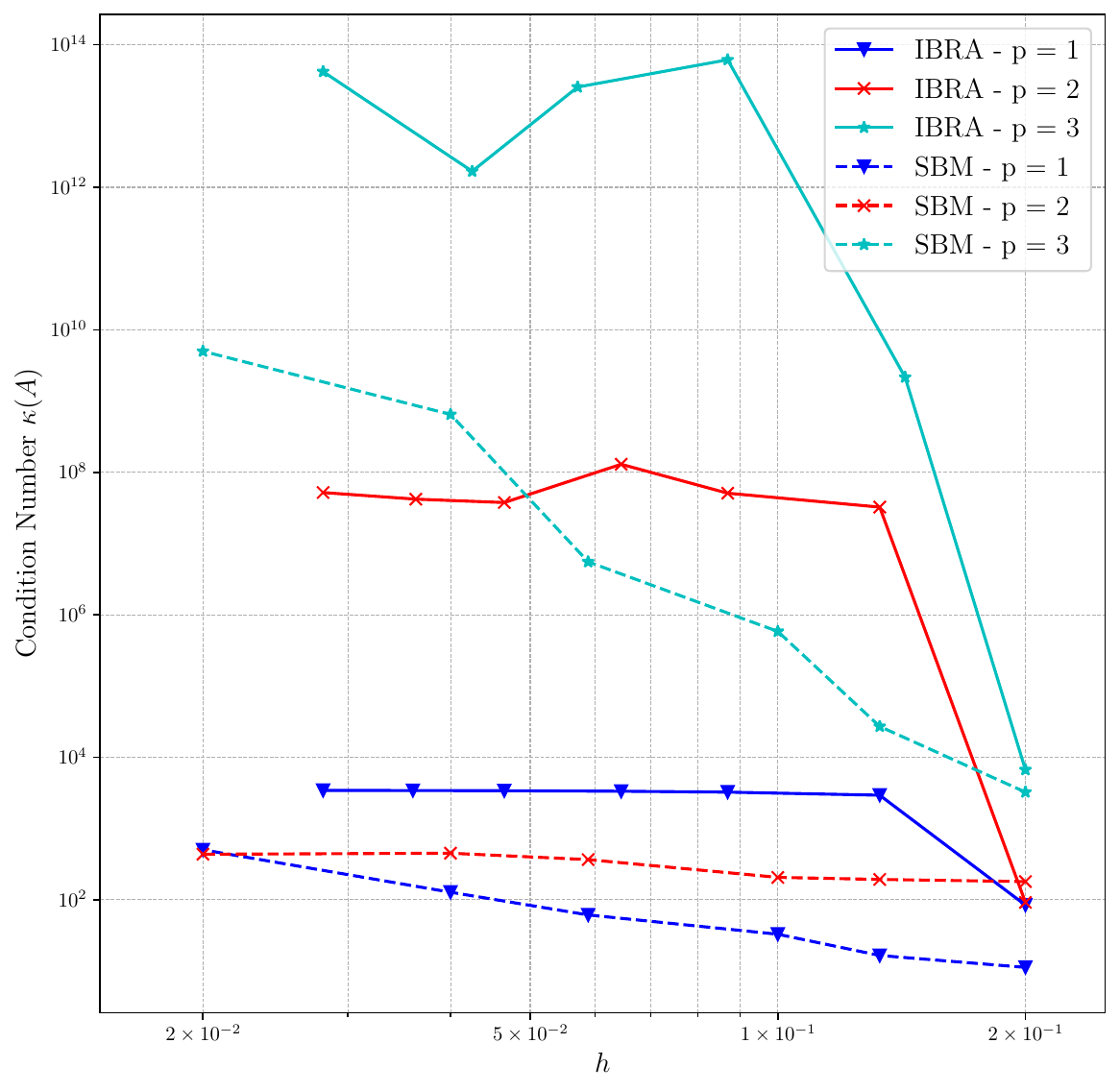}
        \caption{Condition number $\kappa(A)$ of IBRA and SBM system matrix}
        \label{fig:plate_square_hole_IBRA_cond_number}
    \end{subfigure}
    \hfill 
    \begin{subfigure}{0.42\textwidth} 
        \centering
        \includegraphics[width=\linewidth]{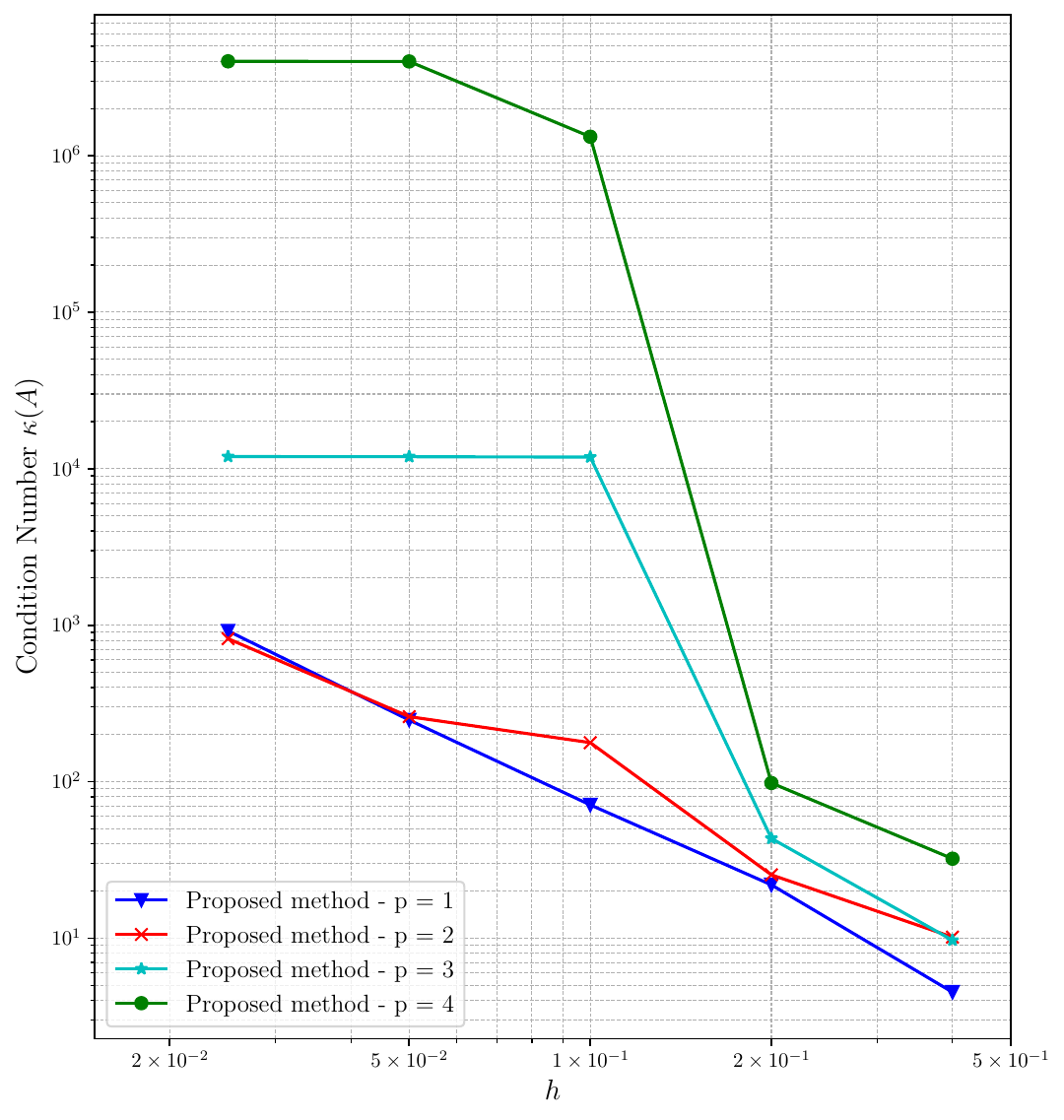}
        \caption{Condition number $\kappa(A)$ of the proposed approach system matrix}
        \label{fig:plate_square_hole_proposed_approach_cond_number}
    \end{subfigure}

   \caption{Comparison of the condition number behavior for IBRA (left panel) and our proposed method (right panel), considering different orders of basis functions for the embedded square case in Fig.~\ref{fig:comp_domain_rotated_square}. The order of the basis functions is represented by the following symbols: triangles (\( p = 1 \)), crosses (\( p = 2 \)), stars (\( p = 3 \)) and circles (\( p = 4 \)).}
    \label{fig:plate_square_hole_cond_number_comparison}
\end{figure}

The condition number of the system matrix $\kappa(A)$ is crucial in FEM-like simulations, as it affects numerical stability and iterative solver performance. In unfitted methods, small cut-cell instabilities and the weak imposition of boundary conditions can significantly increase the condition number, sometimes impairing the use of iterative solvers. 
Fig.~\ref{fig:plate_square_hole_cond_number_comparison} examines the condition number behavior for the embedded square geometry shown in Fig. ~\ref{fig:comp_domain_rotated_square}, where the proposed approach, SBM and IBRA are used to impose Dirichlet BCs on the complete boundary. Compared to IBRA, both the formulation presented in this paper and SBM consistently produce condition numbers $\kappa(\mathbf{A})$ that are several orders of magnitude lower.

To conclude, we emphasize that our proposed method, whether combined with low order or higher order discretizations, remains stable and unaffected by small-cut cell instability, a frequent challenge in CutFEM-like approaches and other methods that require integration over trimmed elements such as IBRA. IBRA-based discretizations tend to encounter this issue, often leading to numerical difficulties that hinder the performance of iterative solvers. In our case, the resulting system remains well-conditioned, ensuring efficient and reliable solutions with both direct and iterative solvers.

\subsection{Capabilities of the proposed approach}

\textcolor{black}{In this section, we investigate two representative examples in which the proposed method is applied to domains with complex boundary geometries. Both cases involve solving the Poisson problem introduced in Eq.~\ref{eq:strong_form}, using a Cartesian mesh composed of linear and quadratic B-Spline discretizations. To enable a controlled evaluation of the method's accuracy, we adopt again the manufactured solution \(\phi(x, y) = \sin(\pi x) \cos(\pi y)\), which is also prescribed as a Dirichlet boundary condition along both the inner and outer boundaries of the domain. In addition, a quantitative comparison with the shifted boundary method (SBM) is provided for the first example to assess the relative accuracy of the proposed approach in handling more complex immersed geometries.}

\begin{figure}[h!]
    \centering
    \begin{subfigure}{0.47\textwidth} 
        \centering
        \includegraphics[width=0.97\linewidth]{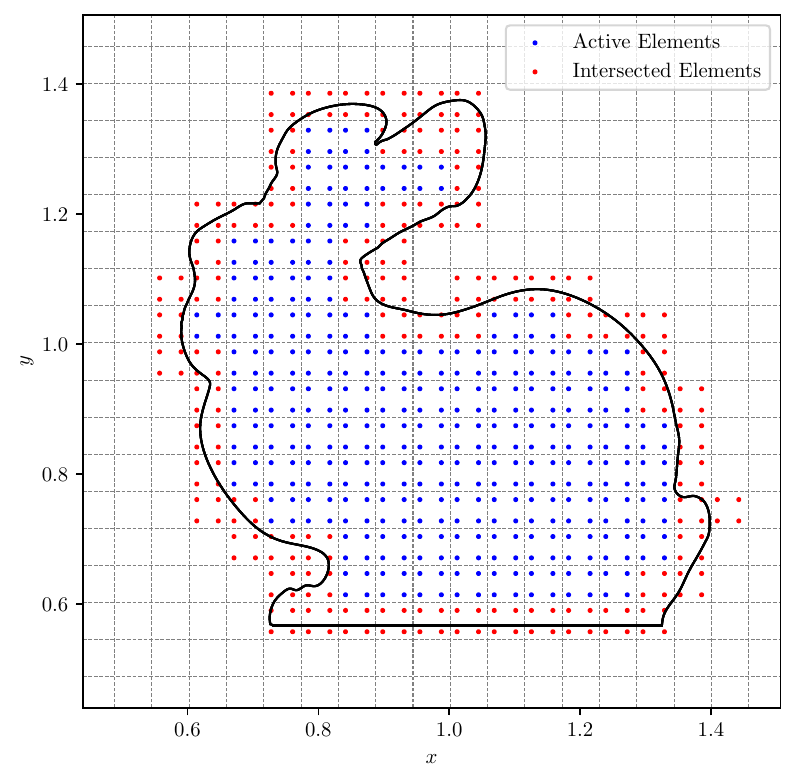}
        \caption{\color{black} Computational set-up, active and intersected elements}
        \label{fig:stanford_bunny_geometry}
    \end{subfigure}
    \hfill 
    \begin{subfigure}{0.47\textwidth} 
        \centering
        \includegraphics[width=\linewidth]{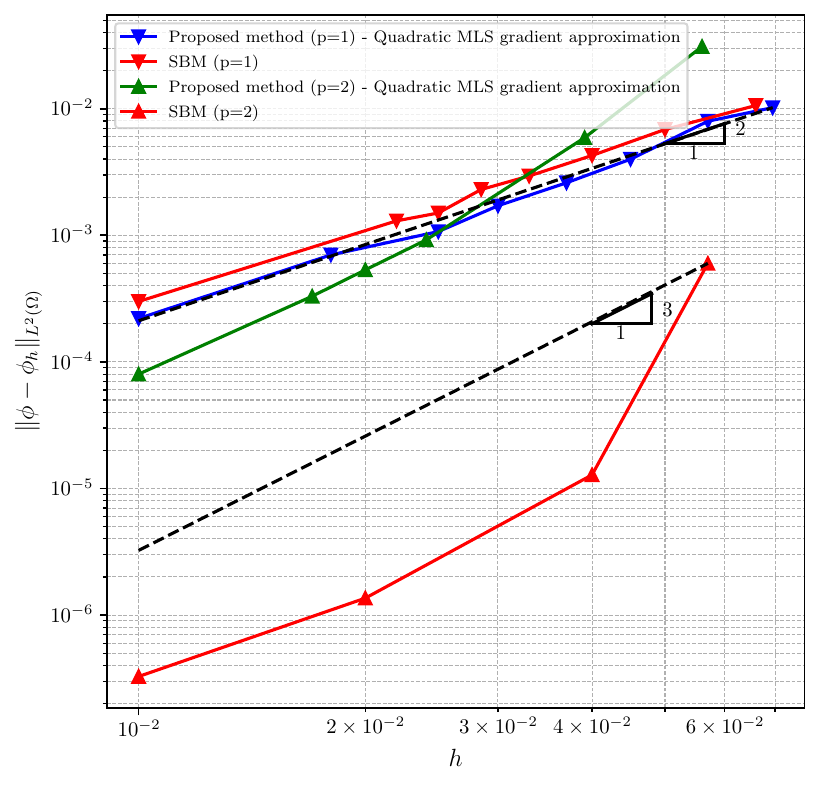}
        \caption{\color{black} Convergence study for the $L^2$-norm error and comparison with SBM}
        \label{fig:stanford_bunny_convergence}
    \end{subfigure}

   \caption{\color{black} Capabilities of the proposed approach. a) the two-dimensional Stanford Bunny is used as the external embedded boundary, and the blue and red dots are the active and intersected integration points respectively. b) the convergence study for linear and quadratic B-Spline discretizations, demonstrating optimal convergence and a comparison with the SBM, demonstrating similar accuracy for low-order elements
study for linear quadrilateral elements}
    \label{fig:stanford_bunny}
\end{figure}

The first case, shown in Fig.~\ref{fig:stanford_bunny_geometry}, applies the method to the external boundary of the two-dimensional Stanford Bunny. In the figure, blue dots represent the active integration points (located within active elements), which contribute to assembling both the left- and right-hand sides of the governing equations. Red dots indicate the intersected integration points (located within intersected elements), which contribute to the stabilization term in the additional algebraic problem. The convergence results in Fig.~\ref{fig:stanford_bunny_convergence} confirm that the method achieves the expected optimal convergence rate, and for low-order elements, its accuracy is comparable to that of the Shifted Boundary Method.

\begin{figure}[h!]
    \centering
    \begin{subfigure}{0.47\textwidth} 
        \centering
        \includegraphics[width=0.97\linewidth]{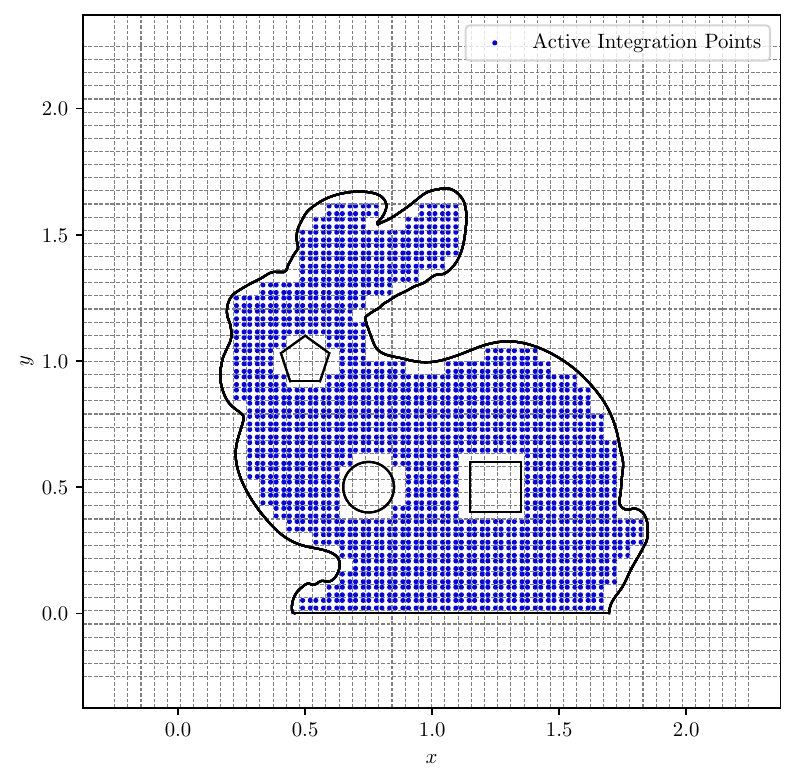}
        \caption{Computational set-up and active integration points}
        \label{fig:stanford_bunny_inner_loops_geometry}
    \end{subfigure}
    \hfill 
    \begin{subfigure}{0.47\textwidth} 
        \centering
        \includegraphics[width=\linewidth]{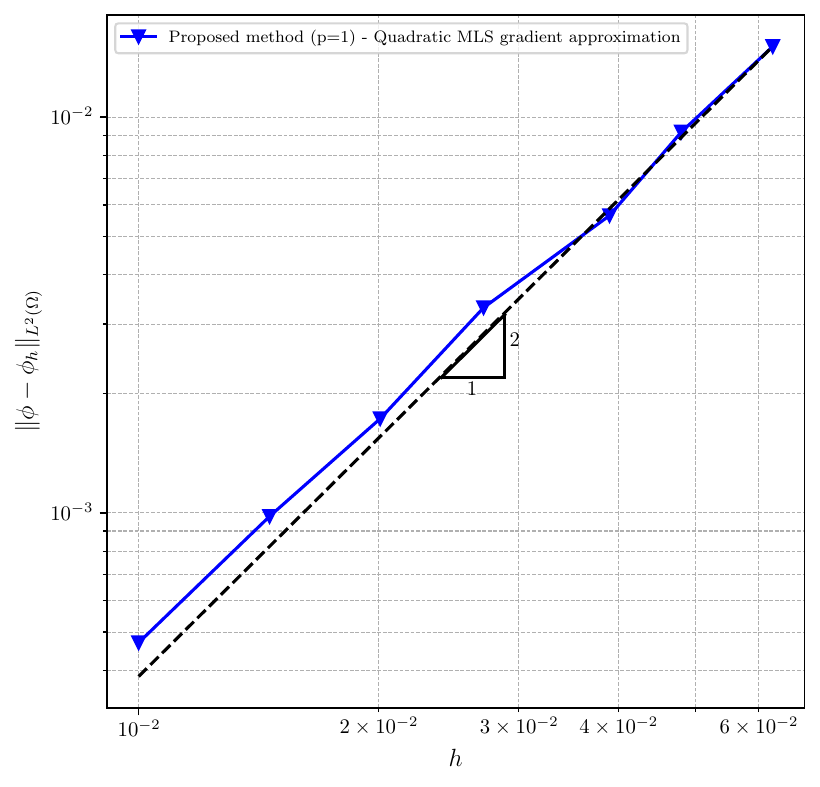}
        \caption{Convergence study for the $L^2$-norm error}
        \label{fig:stanford_bunny_inner_loops_convergence}
    \end{subfigure}

   \caption{Capabilities of the proposed approach. a) a complex example composed of an external and three internal embedded boundaries treated with the proposed method b) the convergence study for linear quadrilateral elements, demonstrating optimal convergence
study for linear quadrilateral elements}
    \label{fig:stanford_bunny_inner_loops}
\end{figure}

A second, more challenging case is shown in Fig.~\ref{fig:stanford_bunny_inner_loops_geometry}, where the domain contains multiple inner loops. The corresponding convergence study in Fig.~\ref{fig:stanford_bunny_inner_loops_convergence} again confirms the robustness of the method and its ability to maintain optimal convergence properties.

{\color{black} While all examples in the present work are two-dimensional, the proposed formulation is dimension-independent and can be directly extended to three-dimensional problems without modification to its mathematical structure. In 3D, the additional challenges are mainly related to the naturally higher computational effort required for element classification and gradient reconstruction, which stem from the increased problem size rather than from any limitation of the method itself.}

\subsection{Explicit transient heat diffusion}

This section evaluates the performance of the proposed method for the explicit simulation of transient heat diffusion, governed by the following partial differential equation (PDE)

\begin{equation}
    \rho c_p \frac{\partial \phi}{\partial t} = \nabla \cdot (k\nabla\phi) + Q(\mathbf{x},t)
    \label{eq:explicit_heat_diff}
\end{equation}

where $\rho$ is the density of the material, $c_p$ is the specific heat capacity, $k$ is the thermal conductivity, and $Q(\mathbf{x}, t)$ represents a heat source that varies with position $\mathbf{x}$ and time $t$. For simplicity, we set $\rho = c_p = k = 1$ in all subsequent derivations.

{\color{black}To numerically solve this equation, we employ an explicit time integration scheme, specifically the explicit Euler method \cite{hairer1993solving}, combined with a finite element spatial discretization with linear triangular elements. }

Since explicit methods require the inversion of the consistent mass matrix $M$, a common simplification is to use a lumped mass matrix approximation, replacing $M^{-1}$ with a diagonal matrix to improve computational efficiency. However, as with any explicit scheme, the stability of the method is conditionally restricted by the Courant–Friedrichs–Lewy (CFL) condition, which imposes an upper bound on the time step $\Delta t$ to prevent numerical instability (any explicit time scheme is inherently conditionally stable).

With this example, we aim to demonstrate that for explicit dynamic simulations with smooth and non-smooth solution fields, good accuracy can be achieved without the need of iterating at each time step. This is because, in explicit simulations, time steps are typically very small, resulting in negligible gradient differences between successive steps. Consequently, at each time step, the normal gradient for the first (and only) iteration can be assumed to be the same as that from the previous time step.

As an initial example, we examine a smooth manufactured solution for the transient Poisson problem (\ref{eq:explicit_heat_diff}) within the embedded circle domain defined in Fig.~(\ref{fig:domain_geometry} which is discretized with a background mesh of linear triangular elements. The manufactured solution $\phi(x,y,t)$ is given by

\begin{equation*}
    \phi(x,y,t) = \sin{(\pi x)}\cos{(\pi y)} \sin{(50\pi t)}
    \label{eq:smooth_man_solution_transient_poisson}
\end{equation*}

\noindent The corresponding source term, $Q(\mathbf{x},t)$ , is given by

\begin{equation*}
Q(x,y,t) = \sin{(\pi x)}\cos{(\pi y)}[50\pi \cos{(50\pi t)}+2 \pi^2 \sin{(50 \pi t)]}
\end{equation*}

\begin{figure}[!tb]
    \centering
    \begin{subfigure}{0.47\textwidth} 
        \centering
        \includegraphics[width=\linewidth]{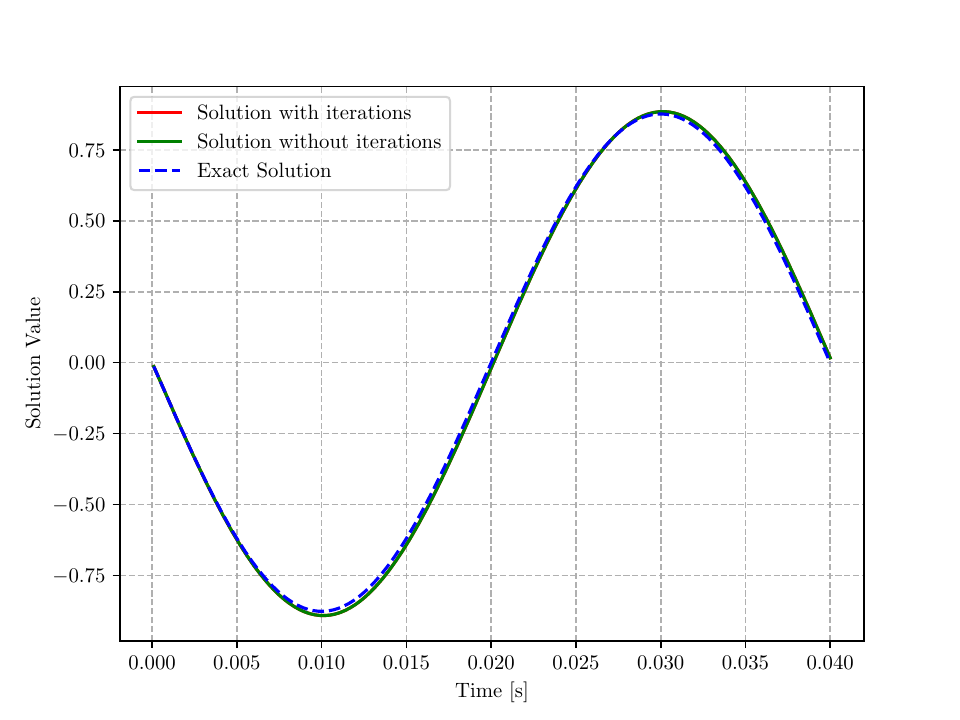}
        \caption{Exact solution vs. Numerical solution with and without iterations}
        \label{fig:plate_hole_num_sol_smooth_transient}
    \end{subfigure}
    \hfill 
    \begin{subfigure}{0.47\textwidth} 
        \centering
        \includegraphics[width=\linewidth]{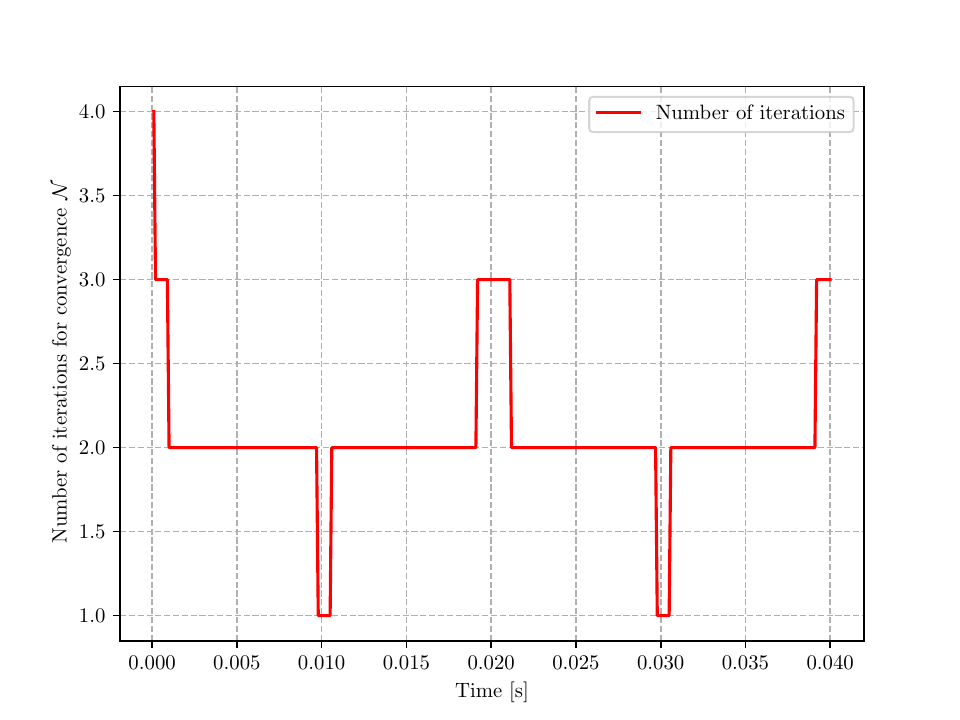}
        \caption{Iterations vs. time for algorithm convergence}
        \label{fig:plate_hole_smooth_transient_num_iterations}
    \end{subfigure}

   \caption{(a) Time-dependent numerical solution at the fixed point $(x,y)=(0.5,0.84)$ using a time step $\Delta t = 0.00005\;s$. The plot compares three different solutions: the solution obtained with iterations (solid red line), the solution obtained without iterations (solid green line over the red one), and the exact solution (dashed blue line).The results demonstrate excellent accuracy for this transient smooth problem, even without iterations at each time step. (b) Number of iterations required for convergence when the algorithm tolerance is set to $\epsilon=0.001$. The peak number of iterations occurs at the initial stage of the transient simulation and during instances where the solution's concavity changes.}
    \label{fig:plate_hole_smooth_transient}
\end{figure}

In Fig.~\ref{fig:plate_hole_num_sol_smooth_transient}, we present the time-dependent numerical solution at a specific point in the domain with coordinates $(x,y)=(0.5,0.84)$, using a time step $\Delta t =0.00005\;s$. The numerical solution is obtained using the forward Euler scheme with a lumped mass matrix.
The results clearly demonstrate that, for this transient  problem, very good accuracy is achieved without the need for iteration at each time step. Additionally, Fig.~\ref{fig:plate_hole_smooth_transient_num_iterations} illustrates the number of iterations required for convergence when the algorithm tolerance is set to $\epsilon=0.001$.  As depicted in the plot, the maximum number of iterations required for convergence occurs primarily at the initial stage of the transient simulation and during instances where the solution's concavity changes (around $t=0.02\;s$).

As a second example, we consider a manufactured solution for the transient Poisson problem (\ref{eq:explicit_heat_diff}) exhibiting a shock-like pattern in time around $t=0.01$ within the embedded circle domain shown in Fig.~\ref{fig:domain_geometry} which is discretized again with a background mesh of linear triangular elements. The manufactured solution $\phi(x,y,t)$ is given by,

\begin{equation*}
    \phi(x,y,t) = \sin{(\pi x)}\cos{(\pi y)} \tanh{[10000(t-0.01)]}
    \label{eq:non_smooth_man_solution_transient_poisson}
\end{equation*}

\noindent The corresponding source term, $Q(\mathbf{x},t)$ , is given by

\begin{equation*}
Q(x,y,t) = \sin(\pi x) \cos(\pi y) \left( \frac{10000}{\cosh^2(10000(t - 0.01))} + 2\pi^2 \tanh(10000(t - 0.01)) \right)
\end{equation*}

\begin{figure}[!tb]
    \centering
    \begin{subfigure}{0.47\textwidth} 
        \centering
        \includegraphics[width=\linewidth]{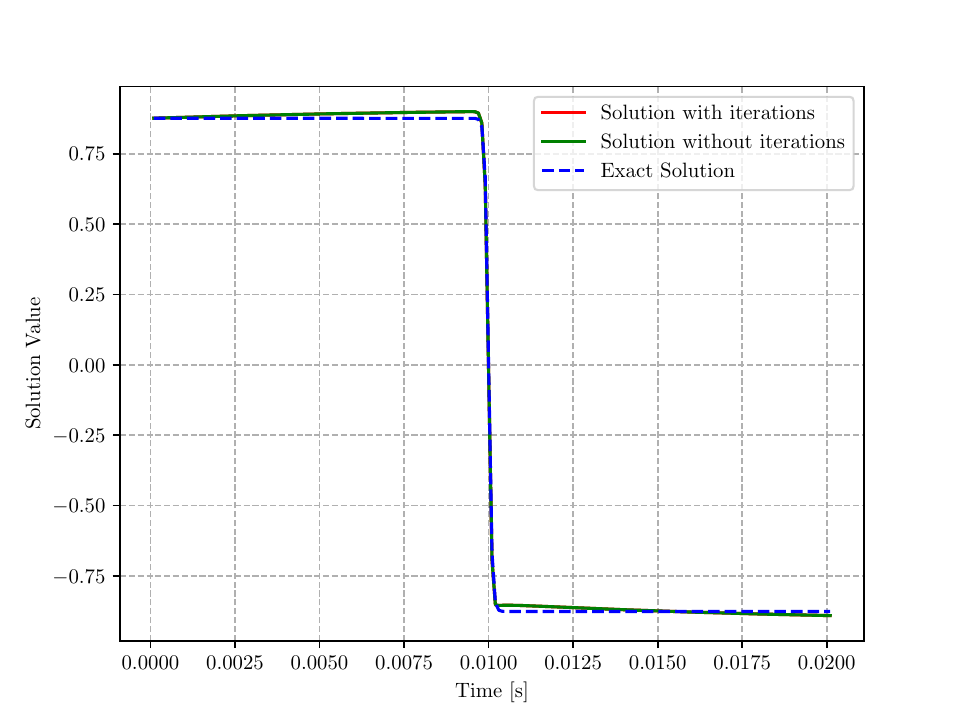}
        \caption{Exact solution vs. Numerical solution with and without iterations}
        \label{fig:plate_hole_num_sol_non_smooth_transient}
    \end{subfigure}
    \hfill 
    \begin{subfigure}{0.47\textwidth} 
        \centering
        \includegraphics[width=\linewidth]{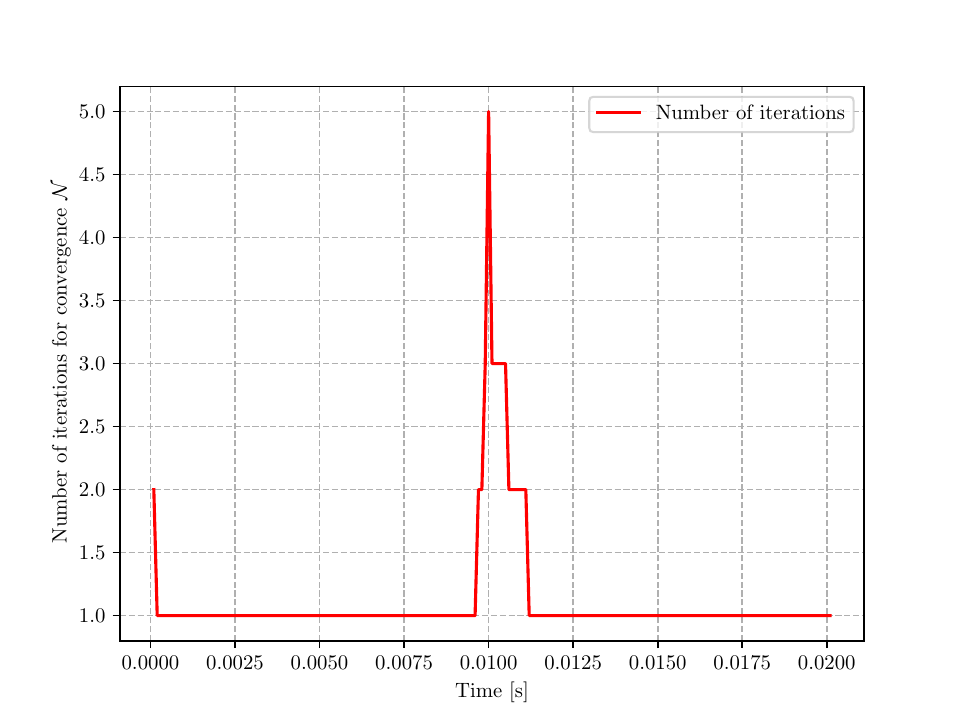}
        \caption{Iterations vs. time for algorithm convergence}
        \label{fig:plate_hole_non_smooth_transient_num_iterations}
    \end{subfigure}

   \caption{(a) Time-dependent numerical solution at the fixed point $(x,y)=(0.5,0.84)$ using a time step $\Delta t = 0.00005\;s$. The plot compares three different solutions: the solution obtained with iterations (solid red line), the solution obtained without iterations (solid green line over the red one), and the exact solution (dashed blue line). The results demonstrate that even without iterations at each time step, the numerical solution maintains excellent accuracy for this transient non-smooth problem. (b) Number of iterations required for convergence when the algorithm tolerance is set to $\epsilon=0.001$. The peak number of iterations occurs at the initial stage of the transient simulation and near the shock position.}
    \label{fig:plate_hole_non_smooth_transient}
\end{figure}

In Fig.~\ref{fig:plate_hole_num_sol_non_smooth_transient}, we depict the time-dependent numerical solution at a specific location within the domain, given by the coordinates $(x,y) = (0.5,0.84)$, with a time step of $\Delta t = 0.00005\;s$. The numerical solution is obtained using again the forward Euler scheme with a lumped mass matrix. The results indicate that, despite the transient and non-smooth nature of the problem, high accuracy is maintained without requiring iterative corrections at each time step. 
Furthermore, Fig.~\ref{fig:plate_hole_non_smooth_transient_num_iterations} presents the number of iterations necessary for convergence when the algorithm's tolerance is set to $\epsilon = 0.001$. As illustrated in the figure, the peak number of iterations occurs predominantly at the initial stage of the transient simulation and around the shock location, approximately at $t = 0.01\;s$.

These results demonstrate that the proposed algorithm is well-suited for explicit dynamic simulations, as it enables the strong enforcement of Dirichlet boundary conditions with virtually no additional computational cost, even in the presence of non-smooth solution fields.

\section{Conclusions}
 \label{sec:conclusions}

In this work, we have presented a physics-agnostic and non-intrusive iterative approach for the strong-like imposition of Dirichlet boundary conditions in unfitted meshes. The method enforces the prescribed conditions by reformulating the problem as an $L^2$-norm error minimization, enhanced with a stabilization term that approximates the normal gradient within trimmed elements. This minimization defines an auxiliary algebraic problem, independent from the weak form of the governing equations. A key feature of this approach is that it preserves the original variational formulation, avoiding modifications to the physical system and maintaining its intrinsic properties.

The method has been implemented using the Kratos Multiphysics API (release \texttt{v10.1}), and a central contribution of this work is to demonstrate that such a strategy can enable already validated, body-fitted, black-box solvers to operate in unfitted meshes. This is possible as long as four conditions are fulfilled: (i) scripting support, (ii) allowance for the imposition of Dirichlet boundary conditions at the node level, (iii) element deactivation, and (iv) access to the solution gradient in active elements. \textcolor{black}{It is worth noting that the last condition can also be satisfied by externally reconstructing the gradient from nodal values and connectivity information, provided the element formulation is known, making it optional in practice. The treatment of Neumann boundary conditions in this black-box context has also been detailed earlier in the manuscript (see Rem.~\ref{rem:treatment_neumann_bcs}). This greatly expands the flexibility of traditional solvers, allowing them to handle complex geometries without the need for body-fitted meshes or intrusive code changes.}

The proposed method has been thoroughly validated through the numerical solution of the Poisson problem, using both Finite Element Method (FEM) and Isogeometric Analysis (IGA) discretizations. The results confirm optimal $L^2$-norm error convergence, demonstrating the effectiveness of the approach. A detailed comparison of different gradient approximation strategies has shown that Moving Least Squares (MLS) interpolation yields the most accurate reconstruction inside intersected elements, likely due to its local minimization formulation.

Additionally, the method eliminates the need for penalty parameter tuning and improves system conditioning compared to Isogeometric B-Rep Analysis (IBRA), while maintaining comparable accuracy. It is especially suitable for explicit dynamic simulations, where small time steps lead to minor changes in the gradient field between iterations. In such cases, the method enforces strong Dirichlet conditions with virtually no additional computational cost, even for non-smooth solutions.

{\color{black}
From a computational cost standpoint, although the proposed strategy is iterative and therefore entails a higher total runtime than any single-pass method, the additional cost per iteration is small compared to solving the main system. This is because the auxiliary problem for updating Dirichlet values involves only a small fraction of the total degrees of freedom, and in each iteration only the gradient-approximation term needs to be recomputed.

In terms of accuracy, the proposed method has shown results comparable to established approaches such as IBRA and SBM for low-order discretizations, but for higher-order discretizations its accuracy is generally lower, primarily due to the gradient reconstruction procedure.

It is therefore important to emphasize that the main strength of the proposed approach lies neither in outperforming alternative methods in accuracy nor in minimizing computational cost, but in enabling a solver originally designed for body-fitted solution strategies to be adapted, without intrusive modifications, to work in unfitted scenarios.}

While the method has shown promising results, further improvements in gradient approximation could enhance its accuracy, particularly in high-order discretizations. Extending the method to additional physical problems is a natural next step, including its application to embedded structures, fluid domains, and fluid–structure interaction. Integration into high-performance computing environments will also enable its use in large-scale simulations.

\section*{Declarations}
\paragraph{\textbf{Conflict of interest}} 
The authors have no competing interests to declare that are relevant to the content of this article.

\section*{Data availability}
\noindent Data will be made available on request.

\section*{Code availability}
The complete implementation of the proposed method, as used to produce all results in this paper, 
is publicly available at 
\href{https://github.com/KratosMultiphysics/Kratos/tree/iga/embedded_extended_gradient_method}
{https://github.com/KratosMultiphysics/Kratos/tree/iga/embedded\_extended\_gradient\_method}. 
This repository contains the Kratos Multiphysics (v10.1) implementation of the algorithm. 
While the present work employs Kratos as the host solver, the formulation is solver-agnostic and 
can be implemented in any programming language that provides an interface to impose Dirichlet boundary conditions at the node level, deactivate elements, and calculate the solution gradient in active elements.

\section*{Acknowledgments}
\noindent The authors gratefully acknowledge the \emph{Design for IGA-type discretization workflows (GECKO)} project. The Design for IGA-type discretization workflows has received funding from the European Union’s Horizon Europe research and Innovation programme under grant agreement No.~101073106, Call: HORIZON-MSCA-2021-DN-01.


 \bibliographystyle{elsarticle-num} 
 \bibliography{camarotti_paper-refs}





\end{document}